\documentclass[12pt]{article}
\usepackage{epsfig,amsfonts,amssymb}
\usepackage{hyperref}
\usepackage{cite}
\input epsf.sty
\usepackage{cite}

\newcommand{\pii}{\pi}

\def\ZZZ{{\hbox{ Z\kern-1.6mm Z}}}
\def\RRR{{\hbox{ R\kern-2.4mm R}}}
\def\CCC{{\hbox{ C\kern-2.0mm C}}}
\def\zzz{{\hbox{z\kern-1mm z}}}

\newcommand{\vt}{\vartheta}

\newcommand{\qeq}{{\hbox{=\kern-2.3mm ? \kern.5mm }}}
\renewcommand{\qeq}{=}

\newcommand{\eps}{\epsilon}

\newcommand{\vp}{\varphi}
\newcommand{\ve}{\varepsilon}

\newcommand{\VV}{{\cal V}}

\newcommand{\AAA}{{\cal A}}
\newcommand{\GG}{{\cal G}}

\newcommand{\FF}{{\cal F}}
\newcommand{\JJ}{{\cal J}}
\newcommand{\HH}{{\cal H}}

\newcommand{\OO}{{\cal O}}

\newcommand{\EE}{{\cal E}}
\newcommand{\LL}{{\cal L}}

\newcommand{\XX}{{\cal X}}

\newcommand{\wt}{\widetilde}
\newcommand{\wh}{\widehat}

\newcommand{\NN}{{\cal N}}

\newcommand{\be}{\begin{equation}}
\newcommand{\ee}{\end{equation}}
\newcommand{\ben}{\begin{eqnarray}\displaystyle}
\newcommand{\een}{\end{eqnarray}}

\newcommand{\refb}[1]{(\ref{#1})}
\newcommand{\p}{\partial}
\newcommand{\sectiono}[1]{\section{#1}\setcounter{equation}{0}}

\newcommand{\gsim}{\stackrel{>}{\sim}}

\def\one{{\hbox{ 1\kern-.8mm l}}}
\def\zero{{\hbox{ 0\kern-1.5mm 0}}}

\newcommand{\bea}[1]{\begin{eqnarray}\label{#1} }
\newcommand{\eea}{\end{eqnarray}}

\newcommand{\eqref}{\refb}

\newcommand{\dotalpha}{{\dot{\alpha}}}

\newcommand{\dotbeta}{{\dot{\beta}}}

\newcommand{\dotgamma}{{\dot{\gamma}}}

\newcommand{\Vm}{V}

\newcommand{\gb}{G}

\newcommand{\q}{e}

\def\figwtgam{
\def\JPicScale{0.8}
\ifx\JPicScale\undefined\def\JPicScale{1}\fi
\unitlength \JPicScale mm
\begin{picture}(140,80)(0,0)
\linethickness{0.3mm}
\put(65.46,53.77){\line(0,1){0.5}}
\multiput(65.45,54.77)(0.01,-0.5){1}{\line(0,-1){0.5}}
\multiput(65.42,55.27)(0.03,-0.5){1}{\line(0,-1){0.5}}
\multiput(65.38,55.76)(0.04,-0.5){1}{\line(0,-1){0.5}}
\multiput(65.32,56.26)(0.06,-0.49){1}{\line(0,-1){0.49}}
\multiput(65.25,56.75)(0.07,-0.49){1}{\line(0,-1){0.49}}
\multiput(65.17,57.24)(0.08,-0.49){1}{\line(0,-1){0.49}}
\multiput(65.07,57.73)(0.1,-0.49){1}{\line(0,-1){0.49}}
\multiput(64.96,58.21)(0.11,-0.49){1}{\line(0,-1){0.49}}
\multiput(64.84,58.69)(0.12,-0.48){1}{\line(0,-1){0.48}}
\multiput(64.7,59.17)(0.14,-0.48){1}{\line(0,-1){0.48}}
\multiput(64.55,59.65)(0.15,-0.47){1}{\line(0,-1){0.47}}
\multiput(64.38,60.12)(0.16,-0.47){1}{\line(0,-1){0.47}}
\multiput(64.2,60.58)(0.18,-0.46){1}{\line(0,-1){0.46}}
\multiput(64.01,61.04)(0.1,-0.23){2}{\line(0,-1){0.23}}
\multiput(63.81,61.5)(0.1,-0.23){2}{\line(0,-1){0.23}}
\multiput(63.59,61.94)(0.11,-0.22){2}{\line(0,-1){0.22}}
\multiput(63.37,62.39)(0.11,-0.22){2}{\line(0,-1){0.22}}
\multiput(63.13,62.82)(0.12,-0.22){2}{\line(0,-1){0.22}}
\multiput(62.87,63.25)(0.13,-0.21){2}{\line(0,-1){0.21}}
\multiput(62.61,63.67)(0.13,-0.21){2}{\line(0,-1){0.21}}
\multiput(62.33,64.09)(0.14,-0.21){2}{\line(0,-1){0.21}}
\multiput(62.04,64.49)(0.14,-0.2){2}{\line(0,-1){0.2}}
\multiput(61.74,64.89)(0.15,-0.2){2}{\line(0,-1){0.2}}
\multiput(61.43,65.28)(0.1,-0.13){3}{\line(0,-1){0.13}}
\multiput(61.11,65.66)(0.11,-0.13){3}{\line(0,-1){0.13}}
\multiput(60.78,66.03)(0.11,-0.12){3}{\line(0,-1){0.12}}
\multiput(60.44,66.39)(0.11,-0.12){3}{\line(0,-1){0.12}}
\multiput(60.09,66.74)(0.12,-0.12){3}{\line(1,0){0.12}}
\multiput(59.73,67.09)(0.12,-0.11){3}{\line(1,0){0.12}}
\multiput(59.35,67.42)(0.12,-0.11){3}{\line(1,0){0.12}}
\multiput(58.97,67.74)(0.13,-0.11){3}{\line(1,0){0.13}}
\multiput(58.59,68.05)(0.13,-0.1){3}{\line(1,0){0.13}}
\multiput(58.19,68.35)(0.2,-0.15){2}{\line(1,0){0.2}}
\multiput(57.78,68.64)(0.2,-0.14){2}{\line(1,0){0.2}}
\multiput(57.37,68.91)(0.21,-0.14){2}{\line(1,0){0.21}}
\multiput(56.95,69.18)(0.21,-0.13){2}{\line(1,0){0.21}}
\multiput(56.52,69.43)(0.21,-0.13){2}{\line(1,0){0.21}}
\multiput(56.08,69.67)(0.22,-0.12){2}{\line(1,0){0.22}}
\multiput(55.64,69.9)(0.22,-0.11){2}{\line(1,0){0.22}}
\multiput(55.19,70.12)(0.22,-0.11){2}{\line(1,0){0.22}}
\multiput(54.74,70.32)(0.23,-0.1){2}{\line(1,0){0.23}}
\multiput(54.28,70.51)(0.23,-0.1){2}{\line(1,0){0.23}}
\multiput(53.81,70.69)(0.46,-0.18){1}{\line(1,0){0.46}}
\multiput(53.34,70.85)(0.47,-0.16){1}{\line(1,0){0.47}}
\multiput(52.87,71)(0.47,-0.15){1}{\line(1,0){0.47}}
\multiput(52.39,71.14)(0.48,-0.14){1}{\line(1,0){0.48}}
\multiput(51.91,71.26)(0.48,-0.12){1}{\line(1,0){0.48}}
\multiput(51.42,71.37)(0.49,-0.11){1}{\line(1,0){0.49}}
\multiput(50.93,71.47)(0.49,-0.1){1}{\line(1,0){0.49}}
\multiput(50.44,71.56)(0.49,-0.08){1}{\line(1,0){0.49}}
\multiput(49.95,71.62)(0.49,-0.07){1}{\line(1,0){0.49}}
\multiput(49.46,71.68)(0.49,-0.06){1}{\line(1,0){0.49}}
\multiput(48.96,71.72)(0.5,-0.04){1}{\line(1,0){0.5}}
\multiput(48.46,71.75)(0.5,-0.03){1}{\line(1,0){0.5}}
\multiput(47.97,71.76)(0.5,-0.01){1}{\line(1,0){0.5}}
\put(47.47,71.76){\line(1,0){0.5}}
\multiput(46.97,71.75)(0.5,0.01){1}{\line(1,0){0.5}}
\multiput(46.47,71.72)(0.5,0.03){1}{\line(1,0){0.5}}
\multiput(45.98,71.68)(0.5,0.04){1}{\line(1,0){0.5}}
\multiput(45.48,71.62)(0.49,0.06){1}{\line(1,0){0.49}}
\multiput(44.99,71.56)(0.49,0.07){1}{\line(1,0){0.49}}
\multiput(44.5,71.47)(0.49,0.08){1}{\line(1,0){0.49}}
\multiput(44.01,71.37)(0.49,0.1){1}{\line(1,0){0.49}}
\multiput(43.53,71.26)(0.49,0.11){1}{\line(1,0){0.49}}
\multiput(43.04,71.14)(0.48,0.12){1}{\line(1,0){0.48}}
\multiput(42.57,71)(0.48,0.14){1}{\line(1,0){0.48}}
\multiput(42.09,70.85)(0.47,0.15){1}{\line(1,0){0.47}}
\multiput(41.62,70.69)(0.47,0.16){1}{\line(1,0){0.47}}
\multiput(41.16,70.51)(0.46,0.18){1}{\line(1,0){0.46}}
\multiput(40.7,70.32)(0.23,0.1){2}{\line(1,0){0.23}}
\multiput(40.24,70.12)(0.23,0.1){2}{\line(1,0){0.23}}
\multiput(39.79,69.9)(0.22,0.11){2}{\line(1,0){0.22}}
\multiput(39.35,69.67)(0.22,0.11){2}{\line(1,0){0.22}}
\multiput(38.92,69.43)(0.22,0.12){2}{\line(1,0){0.22}}
\multiput(38.49,69.18)(0.21,0.13){2}{\line(1,0){0.21}}
\multiput(38.07,68.91)(0.21,0.13){2}{\line(1,0){0.21}}
\multiput(37.65,68.64)(0.21,0.14){2}{\line(1,0){0.21}}
\multiput(37.25,68.35)(0.2,0.14){2}{\line(1,0){0.2}}
\multiput(36.85,68.05)(0.2,0.15){2}{\line(1,0){0.2}}
\multiput(36.46,67.74)(0.13,0.1){3}{\line(1,0){0.13}}
\multiput(36.08,67.42)(0.13,0.11){3}{\line(1,0){0.13}}
\multiput(35.71,67.09)(0.12,0.11){3}{\line(1,0){0.12}}
\multiput(35.35,66.74)(0.12,0.11){3}{\line(1,0){0.12}}
\multiput(35,66.39)(0.12,0.12){3}{\line(1,0){0.12}}
\multiput(34.65,66.03)(0.11,0.12){3}{\line(0,1){0.12}}
\multiput(34.32,65.66)(0.11,0.12){3}{\line(0,1){0.12}}
\multiput(34,65.28)(0.11,0.13){3}{\line(0,1){0.13}}
\multiput(33.69,64.89)(0.1,0.13){3}{\line(0,1){0.13}}
\multiput(33.39,64.49)(0.15,0.2){2}{\line(0,1){0.2}}
\multiput(33.1,64.09)(0.14,0.2){2}{\line(0,1){0.2}}
\multiput(32.83,63.67)(0.14,0.21){2}{\line(0,1){0.21}}
\multiput(32.56,63.25)(0.13,0.21){2}{\line(0,1){0.21}}
\multiput(32.31,62.82)(0.13,0.21){2}{\line(0,1){0.21}}
\multiput(32.07,62.39)(0.12,0.22){2}{\line(0,1){0.22}}
\multiput(31.84,61.94)(0.11,0.22){2}{\line(0,1){0.22}}
\multiput(31.62,61.5)(0.11,0.22){2}{\line(0,1){0.22}}
\multiput(31.42,61.04)(0.1,0.23){2}{\line(0,1){0.23}}
\multiput(31.23,60.58)(0.1,0.23){2}{\line(0,1){0.23}}
\multiput(31.05,60.12)(0.18,0.46){1}{\line(0,1){0.46}}
\multiput(30.89,59.65)(0.16,0.47){1}{\line(0,1){0.47}}
\multiput(30.74,59.17)(0.15,0.47){1}{\line(0,1){0.47}}
\multiput(30.6,58.69)(0.14,0.48){1}{\line(0,1){0.48}}
\multiput(30.47,58.21)(0.12,0.48){1}{\line(0,1){0.48}}
\multiput(30.36,57.73)(0.11,0.49){1}{\line(0,1){0.49}}
\multiput(30.27,57.24)(0.1,0.49){1}{\line(0,1){0.49}}
\multiput(30.18,56.75)(0.08,0.49){1}{\line(0,1){0.49}}
\multiput(30.11,56.26)(0.07,0.49){1}{\line(0,1){0.49}}
\multiput(30.06,55.76)(0.06,0.49){1}{\line(0,1){0.49}}
\multiput(30.02,55.27)(0.04,0.5){1}{\line(0,1){0.5}}
\multiput(29.99,54.77)(0.03,0.5){1}{\line(0,1){0.5}}
\multiput(29.97,54.27)(0.01,0.5){1}{\line(0,1){0.5}}
\put(29.97,53.77){\line(0,1){0.5}}
\multiput(29.97,53.77)(0.01,-0.5){1}{\line(0,-1){0.5}}
\multiput(29.99,53.28)(0.03,-0.5){1}{\line(0,-1){0.5}}
\multiput(30.02,52.78)(0.04,-0.5){1}{\line(0,-1){0.5}}
\multiput(30.06,52.28)(0.06,-0.49){1}{\line(0,-1){0.49}}
\multiput(30.11,51.79)(0.07,-0.49){1}{\line(0,-1){0.49}}
\multiput(30.18,51.3)(0.08,-0.49){1}{\line(0,-1){0.49}}
\multiput(30.27,50.8)(0.1,-0.49){1}{\line(0,-1){0.49}}
\multiput(30.36,50.32)(0.11,-0.49){1}{\line(0,-1){0.49}}
\multiput(30.47,49.83)(0.12,-0.48){1}{\line(0,-1){0.48}}
\multiput(30.6,49.35)(0.14,-0.48){1}{\line(0,-1){0.48}}
\multiput(30.74,48.87)(0.15,-0.47){1}{\line(0,-1){0.47}}
\multiput(30.89,48.4)(0.16,-0.47){1}{\line(0,-1){0.47}}
\multiput(31.05,47.93)(0.18,-0.46){1}{\line(0,-1){0.46}}
\multiput(31.23,47.46)(0.1,-0.23){2}{\line(0,-1){0.23}}
\multiput(31.42,47)(0.1,-0.23){2}{\line(0,-1){0.23}}
\multiput(31.62,46.55)(0.11,-0.22){2}{\line(0,-1){0.22}}
\multiput(31.84,46.1)(0.11,-0.22){2}{\line(0,-1){0.22}}
\multiput(32.07,45.66)(0.12,-0.22){2}{\line(0,-1){0.22}}
\multiput(32.31,45.22)(0.13,-0.21){2}{\line(0,-1){0.21}}
\multiput(32.56,44.79)(0.13,-0.21){2}{\line(0,-1){0.21}}
\multiput(32.83,44.37)(0.14,-0.21){2}{\line(0,-1){0.21}}
\multiput(33.1,43.96)(0.14,-0.2){2}{\line(0,-1){0.2}}
\multiput(33.39,43.55)(0.15,-0.2){2}{\line(0,-1){0.2}}
\multiput(33.69,43.15)(0.1,-0.13){3}{\line(0,-1){0.13}}
\multiput(34,42.76)(0.11,-0.13){3}{\line(0,-1){0.13}}
\multiput(34.32,42.38)(0.11,-0.12){3}{\line(0,-1){0.12}}
\multiput(34.65,42.01)(0.11,-0.12){3}{\line(0,-1){0.12}}
\multiput(35,41.65)(0.12,-0.12){3}{\line(1,0){0.12}}
\multiput(35.35,41.3)(0.12,-0.11){3}{\line(1,0){0.12}}
\multiput(35.71,40.96)(0.12,-0.11){3}{\line(1,0){0.12}}
\multiput(36.08,40.63)(0.13,-0.11){3}{\line(1,0){0.13}}
\multiput(36.46,40.31)(0.13,-0.1){3}{\line(1,0){0.13}}
\multiput(36.85,39.99)(0.2,-0.15){2}{\line(1,0){0.2}}
\multiput(37.25,39.7)(0.2,-0.14){2}{\line(1,0){0.2}}
\multiput(37.65,39.41)(0.21,-0.14){2}{\line(1,0){0.21}}
\multiput(38.07,39.13)(0.21,-0.13){2}{\line(1,0){0.21}}
\multiput(38.49,38.87)(0.21,-0.13){2}{\line(1,0){0.21}}
\multiput(38.92,38.61)(0.22,-0.12){2}{\line(1,0){0.22}}
\multiput(39.35,38.37)(0.22,-0.11){2}{\line(1,0){0.22}}
\multiput(39.79,38.14)(0.22,-0.11){2}{\line(1,0){0.22}}
\multiput(40.24,37.93)(0.23,-0.1){2}{\line(1,0){0.23}}
\multiput(40.7,37.72)(0.23,-0.1){2}{\line(1,0){0.23}}
\multiput(41.16,37.53)(0.46,-0.18){1}{\line(1,0){0.46}}
\multiput(41.62,37.36)(0.47,-0.16){1}{\line(1,0){0.47}}
\multiput(42.09,37.19)(0.47,-0.15){1}{\line(1,0){0.47}}
\multiput(42.57,37.04)(0.48,-0.14){1}{\line(1,0){0.48}}
\multiput(43.04,36.9)(0.48,-0.12){1}{\line(1,0){0.48}}
\multiput(43.53,36.78)(0.49,-0.11){1}{\line(1,0){0.49}}
\multiput(44.01,36.67)(0.49,-0.1){1}{\line(1,0){0.49}}
\multiput(44.5,36.57)(0.49,-0.08){1}{\line(1,0){0.49}}
\multiput(44.99,36.49)(0.49,-0.07){1}{\line(1,0){0.49}}
\multiput(45.48,36.42)(0.49,-0.06){1}{\line(1,0){0.49}}
\multiput(45.98,36.36)(0.5,-0.04){1}{\line(1,0){0.5}}
\multiput(46.47,36.32)(0.5,-0.03){1}{\line(1,0){0.5}}
\multiput(46.97,36.29)(0.5,-0.01){1}{\line(1,0){0.5}}
\put(47.47,36.28){\line(1,0){0.5}}
\multiput(47.97,36.28)(0.5,0.01){1}{\line(1,0){0.5}}
\multiput(48.46,36.29)(0.5,0.03){1}{\line(1,0){0.5}}
\multiput(48.96,36.32)(0.5,0.04){1}{\line(1,0){0.5}}
\multiput(49.46,36.36)(0.49,0.06){1}{\line(1,0){0.49}}
\multiput(49.95,36.42)(0.49,0.07){1}{\line(1,0){0.49}}
\multiput(50.44,36.49)(0.49,0.08){1}{\line(1,0){0.49}}
\multiput(50.93,36.57)(0.49,0.1){1}{\line(1,0){0.49}}
\multiput(51.42,36.67)(0.49,0.11){1}{\line(1,0){0.49}}
\multiput(51.91,36.78)(0.48,0.12){1}{\line(1,0){0.48}}
\multiput(52.39,36.9)(0.48,0.14){1}{\line(1,0){0.48}}
\multiput(52.87,37.04)(0.47,0.15){1}{\line(1,0){0.47}}
\multiput(53.34,37.19)(0.47,0.16){1}{\line(1,0){0.47}}
\multiput(53.81,37.36)(0.46,0.18){1}{\line(1,0){0.46}}
\multiput(54.28,37.53)(0.23,0.1){2}{\line(1,0){0.23}}
\multiput(54.74,37.72)(0.23,0.1){2}{\line(1,0){0.23}}
\multiput(55.19,37.93)(0.22,0.11){2}{\line(1,0){0.22}}
\multiput(55.64,38.14)(0.22,0.11){2}{\line(1,0){0.22}}
\multiput(56.08,38.37)(0.22,0.12){2}{\line(1,0){0.22}}
\multiput(56.52,38.61)(0.21,0.13){2}{\line(1,0){0.21}}
\multiput(56.95,38.87)(0.21,0.13){2}{\line(1,0){0.21}}
\multiput(57.37,39.13)(0.21,0.14){2}{\line(1,0){0.21}}
\multiput(57.78,39.41)(0.2,0.14){2}{\line(1,0){0.2}}
\multiput(58.19,39.7)(0.2,0.15){2}{\line(1,0){0.2}}
\multiput(58.59,39.99)(0.13,0.1){3}{\line(1,0){0.13}}
\multiput(58.97,40.31)(0.13,0.11){3}{\line(1,0){0.13}}
\multiput(59.35,40.63)(0.12,0.11){3}{\line(1,0){0.12}}
\multiput(59.73,40.96)(0.12,0.11){3}{\line(1,0){0.12}}
\multiput(60.09,41.3)(0.12,0.12){3}{\line(1,0){0.12}}
\multiput(60.44,41.65)(0.11,0.12){3}{\line(0,1){0.12}}
\multiput(60.78,42.01)(0.11,0.12){3}{\line(0,1){0.12}}
\multiput(61.11,42.38)(0.11,0.13){3}{\line(0,1){0.13}}
\multiput(61.43,42.76)(0.1,0.13){3}{\line(0,1){0.13}}
\multiput(61.74,43.15)(0.15,0.2){2}{\line(0,1){0.2}}
\multiput(62.04,43.55)(0.14,0.2){2}{\line(0,1){0.2}}
\multiput(62.33,43.96)(0.14,0.21){2}{\line(0,1){0.21}}
\multiput(62.61,44.37)(0.13,0.21){2}{\line(0,1){0.21}}
\multiput(62.87,44.79)(0.13,0.21){2}{\line(0,1){0.21}}
\multiput(63.13,45.22)(0.12,0.22){2}{\line(0,1){0.22}}
\multiput(63.37,45.66)(0.11,0.22){2}{\line(0,1){0.22}}
\multiput(63.59,46.1)(0.11,0.22){2}{\line(0,1){0.22}}
\multiput(63.81,46.55)(0.1,0.23){2}{\line(0,1){0.23}}
\multiput(64.01,47)(0.1,0.23){2}{\line(0,1){0.23}}
\multiput(64.2,47.46)(0.18,0.46){1}{\line(0,1){0.46}}
\multiput(64.38,47.93)(0.16,0.47){1}{\line(0,1){0.47}}
\multiput(64.55,48.4)(0.15,0.47){1}{\line(0,1){0.47}}
\multiput(64.7,48.87)(0.14,0.48){1}{\line(0,1){0.48}}
\multiput(64.84,49.35)(0.12,0.48){1}{\line(0,1){0.48}}
\multiput(64.96,49.83)(0.11,0.49){1}{\line(0,1){0.49}}
\multiput(65.07,50.32)(0.1,0.49){1}{\line(0,1){0.49}}
\multiput(65.17,50.8)(0.08,0.49){1}{\line(0,1){0.49}}
\multiput(65.25,51.3)(0.07,0.49){1}{\line(0,1){0.49}}
\multiput(65.32,51.79)(0.06,0.49){1}{\line(0,1){0.49}}
\multiput(65.38,52.28)(0.04,0.5){1}{\line(0,1){0.5}}
\multiput(65.42,52.78)(0.03,0.5){1}{\line(0,1){0.5}}
\multiput(65.45,53.28)(0.01,0.5){1}{\line(0,1){0.5}}

\linethickness{0.3mm}
\put(127.74,54.75){\line(0,1){0.5}}
\multiput(127.73,55.75)(0.01,-0.5){1}{\line(0,-1){0.5}}
\multiput(127.7,56.24)(0.03,-0.5){1}{\line(0,-1){0.5}}
\multiput(127.66,56.74)(0.04,-0.5){1}{\line(0,-1){0.5}}
\multiput(127.6,57.23)(0.06,-0.49){1}{\line(0,-1){0.49}}
\multiput(127.53,57.73)(0.07,-0.49){1}{\line(0,-1){0.49}}
\multiput(127.45,58.22)(0.08,-0.49){1}{\line(0,-1){0.49}}
\multiput(127.35,58.71)(0.1,-0.49){1}{\line(0,-1){0.49}}
\multiput(127.24,59.19)(0.11,-0.49){1}{\line(0,-1){0.49}}
\multiput(127.12,59.67)(0.12,-0.48){1}{\line(0,-1){0.48}}
\multiput(126.98,60.15)(0.14,-0.48){1}{\line(0,-1){0.48}}
\multiput(126.83,60.63)(0.15,-0.47){1}{\line(0,-1){0.47}}
\multiput(126.66,61.09)(0.16,-0.47){1}{\line(0,-1){0.47}}
\multiput(126.49,61.56)(0.18,-0.46){1}{\line(0,-1){0.46}}
\multiput(126.3,62.02)(0.1,-0.23){2}{\line(0,-1){0.23}}
\multiput(126.09,62.47)(0.1,-0.23){2}{\line(0,-1){0.23}}
\multiput(125.88,62.92)(0.11,-0.22){2}{\line(0,-1){0.22}}
\multiput(125.65,63.36)(0.11,-0.22){2}{\line(0,-1){0.22}}
\multiput(125.41,63.8)(0.12,-0.22){2}{\line(0,-1){0.22}}
\multiput(125.16,64.23)(0.13,-0.21){2}{\line(0,-1){0.21}}
\multiput(124.89,64.65)(0.13,-0.21){2}{\line(0,-1){0.21}}
\multiput(124.61,65.06)(0.14,-0.21){2}{\line(0,-1){0.21}}
\multiput(124.33,65.47)(0.14,-0.2){2}{\line(0,-1){0.2}}
\multiput(124.03,65.87)(0.15,-0.2){2}{\line(0,-1){0.2}}
\multiput(123.72,66.26)(0.1,-0.13){3}{\line(0,-1){0.13}}
\multiput(123.4,66.64)(0.11,-0.13){3}{\line(0,-1){0.13}}
\multiput(123.06,67.01)(0.11,-0.12){3}{\line(0,-1){0.12}}
\multiput(122.72,67.37)(0.11,-0.12){3}{\line(0,-1){0.12}}
\multiput(122.37,67.72)(0.12,-0.12){3}{\line(1,0){0.12}}
\multiput(122.01,68.06)(0.12,-0.11){3}{\line(1,0){0.12}}
\multiput(121.64,68.4)(0.12,-0.11){3}{\line(1,0){0.12}}
\multiput(121.26,68.72)(0.13,-0.11){3}{\line(1,0){0.13}}
\multiput(120.87,69.03)(0.13,-0.1){3}{\line(1,0){0.13}}
\multiput(120.47,69.33)(0.2,-0.15){2}{\line(1,0){0.2}}
\multiput(120.06,69.61)(0.2,-0.14){2}{\line(1,0){0.2}}
\multiput(119.65,69.89)(0.21,-0.14){2}{\line(1,0){0.21}}
\multiput(119.23,70.16)(0.21,-0.13){2}{\line(1,0){0.21}}
\multiput(118.8,70.41)(0.21,-0.13){2}{\line(1,0){0.21}}
\multiput(118.36,70.65)(0.22,-0.12){2}{\line(1,0){0.22}}
\multiput(117.92,70.88)(0.22,-0.11){2}{\line(1,0){0.22}}
\multiput(117.47,71.09)(0.22,-0.11){2}{\line(1,0){0.22}}
\multiput(117.02,71.3)(0.23,-0.1){2}{\line(1,0){0.23}}
\multiput(116.56,71.49)(0.23,-0.1){2}{\line(1,0){0.23}}
\multiput(116.09,71.66)(0.46,-0.18){1}{\line(1,0){0.46}}
\multiput(115.63,71.83)(0.47,-0.16){1}{\line(1,0){0.47}}
\multiput(115.15,71.98)(0.47,-0.15){1}{\line(1,0){0.47}}
\multiput(114.67,72.12)(0.48,-0.14){1}{\line(1,0){0.48}}
\multiput(114.19,72.24)(0.48,-0.12){1}{\line(1,0){0.48}}
\multiput(113.71,72.35)(0.49,-0.11){1}{\line(1,0){0.49}}
\multiput(113.22,72.45)(0.49,-0.1){1}{\line(1,0){0.49}}
\multiput(112.73,72.53)(0.49,-0.08){1}{\line(1,0){0.49}}
\multiput(112.23,72.6)(0.49,-0.07){1}{\line(1,0){0.49}}
\multiput(111.74,72.66)(0.49,-0.06){1}{\line(1,0){0.49}}
\multiput(111.24,72.7)(0.5,-0.04){1}{\line(1,0){0.5}}
\multiput(110.75,72.73)(0.5,-0.03){1}{\line(1,0){0.5}}
\multiput(110.25,72.74)(0.5,-0.01){1}{\line(1,0){0.5}}
\put(109.75,72.74){\line(1,0){0.5}}
\multiput(109.25,72.73)(0.5,0.01){1}{\line(1,0){0.5}}
\multiput(108.76,72.7)(0.5,0.03){1}{\line(1,0){0.5}}
\multiput(108.26,72.66)(0.5,0.04){1}{\line(1,0){0.5}}
\multiput(107.77,72.6)(0.49,0.06){1}{\line(1,0){0.49}}
\multiput(107.27,72.53)(0.49,0.07){1}{\line(1,0){0.49}}
\multiput(106.78,72.45)(0.49,0.08){1}{\line(1,0){0.49}}
\multiput(106.29,72.35)(0.49,0.1){1}{\line(1,0){0.49}}
\multiput(105.81,72.24)(0.49,0.11){1}{\line(1,0){0.49}}
\multiput(105.33,72.12)(0.48,0.12){1}{\line(1,0){0.48}}
\multiput(104.85,71.98)(0.48,0.14){1}{\line(1,0){0.48}}
\multiput(104.37,71.83)(0.47,0.15){1}{\line(1,0){0.47}}
\multiput(103.91,71.66)(0.47,0.16){1}{\line(1,0){0.47}}
\multiput(103.44,71.49)(0.46,0.18){1}{\line(1,0){0.46}}
\multiput(102.98,71.3)(0.23,0.1){2}{\line(1,0){0.23}}
\multiput(102.53,71.09)(0.23,0.1){2}{\line(1,0){0.23}}
\multiput(102.08,70.88)(0.22,0.11){2}{\line(1,0){0.22}}
\multiput(101.64,70.65)(0.22,0.11){2}{\line(1,0){0.22}}
\multiput(101.2,70.41)(0.22,0.12){2}{\line(1,0){0.22}}
\multiput(100.77,70.16)(0.21,0.13){2}{\line(1,0){0.21}}
\multiput(100.35,69.89)(0.21,0.13){2}{\line(1,0){0.21}}
\multiput(99.94,69.61)(0.21,0.14){2}{\line(1,0){0.21}}
\multiput(99.53,69.33)(0.2,0.14){2}{\line(1,0){0.2}}
\multiput(99.13,69.03)(0.2,0.15){2}{\line(1,0){0.2}}
\multiput(98.74,68.72)(0.13,0.1){3}{\line(1,0){0.13}}
\multiput(98.36,68.4)(0.13,0.11){3}{\line(1,0){0.13}}
\multiput(97.99,68.06)(0.12,0.11){3}{\line(1,0){0.12}}
\multiput(97.63,67.72)(0.12,0.11){3}{\line(1,0){0.12}}
\multiput(97.28,67.37)(0.12,0.12){3}{\line(1,0){0.12}}
\multiput(96.94,67.01)(0.11,0.12){3}{\line(0,1){0.12}}
\multiput(96.6,66.64)(0.11,0.12){3}{\line(0,1){0.12}}
\multiput(96.28,66.26)(0.11,0.13){3}{\line(0,1){0.13}}
\multiput(95.97,65.87)(0.1,0.13){3}{\line(0,1){0.13}}
\multiput(95.67,65.47)(0.15,0.2){2}{\line(0,1){0.2}}
\multiput(95.39,65.06)(0.14,0.2){2}{\line(0,1){0.2}}
\multiput(95.11,64.65)(0.14,0.21){2}{\line(0,1){0.21}}
\multiput(94.84,64.23)(0.13,0.21){2}{\line(0,1){0.21}}
\multiput(94.59,63.8)(0.13,0.21){2}{\line(0,1){0.21}}
\multiput(94.35,63.36)(0.12,0.22){2}{\line(0,1){0.22}}
\multiput(94.12,62.92)(0.11,0.22){2}{\line(0,1){0.22}}
\multiput(93.91,62.47)(0.11,0.22){2}{\line(0,1){0.22}}
\multiput(93.7,62.02)(0.1,0.23){2}{\line(0,1){0.23}}
\multiput(93.51,61.56)(0.1,0.23){2}{\line(0,1){0.23}}
\multiput(93.34,61.09)(0.18,0.46){1}{\line(0,1){0.46}}
\multiput(93.17,60.63)(0.16,0.47){1}{\line(0,1){0.47}}
\multiput(93.02,60.15)(0.15,0.47){1}{\line(0,1){0.47}}
\multiput(92.88,59.67)(0.14,0.48){1}{\line(0,1){0.48}}
\multiput(92.76,59.19)(0.12,0.48){1}{\line(0,1){0.48}}
\multiput(92.65,58.71)(0.11,0.49){1}{\line(0,1){0.49}}
\multiput(92.55,58.22)(0.1,0.49){1}{\line(0,1){0.49}}
\multiput(92.47,57.73)(0.08,0.49){1}{\line(0,1){0.49}}
\multiput(92.4,57.23)(0.07,0.49){1}{\line(0,1){0.49}}
\multiput(92.34,56.74)(0.06,0.49){1}{\line(0,1){0.49}}
\multiput(92.3,56.24)(0.04,0.5){1}{\line(0,1){0.5}}
\multiput(92.27,55.75)(0.03,0.5){1}{\line(0,1){0.5}}
\multiput(92.26,55.25)(0.01,0.5){1}{\line(0,1){0.5}}
\put(92.26,54.75){\line(0,1){0.5}}
\multiput(92.26,54.75)(0.01,-0.5){1}{\line(0,-1){0.5}}
\multiput(92.27,54.25)(0.03,-0.5){1}{\line(0,-1){0.5}}
\multiput(92.3,53.76)(0.04,-0.5){1}{\line(0,-1){0.5}}
\multiput(92.34,53.26)(0.06,-0.49){1}{\line(0,-1){0.49}}
\multiput(92.4,52.77)(0.07,-0.49){1}{\line(0,-1){0.49}}
\multiput(92.47,52.27)(0.08,-0.49){1}{\line(0,-1){0.49}}
\multiput(92.55,51.78)(0.1,-0.49){1}{\line(0,-1){0.49}}
\multiput(92.65,51.29)(0.11,-0.49){1}{\line(0,-1){0.49}}
\multiput(92.76,50.81)(0.12,-0.48){1}{\line(0,-1){0.48}}
\multiput(92.88,50.33)(0.14,-0.48){1}{\line(0,-1){0.48}}
\multiput(93.02,49.85)(0.15,-0.47){1}{\line(0,-1){0.47}}
\multiput(93.17,49.37)(0.16,-0.47){1}{\line(0,-1){0.47}}
\multiput(93.34,48.91)(0.18,-0.46){1}{\line(0,-1){0.46}}
\multiput(93.51,48.44)(0.1,-0.23){2}{\line(0,-1){0.23}}
\multiput(93.7,47.98)(0.1,-0.23){2}{\line(0,-1){0.23}}
\multiput(93.91,47.53)(0.11,-0.22){2}{\line(0,-1){0.22}}
\multiput(94.12,47.08)(0.11,-0.22){2}{\line(0,-1){0.22}}
\multiput(94.35,46.64)(0.12,-0.22){2}{\line(0,-1){0.22}}
\multiput(94.59,46.2)(0.13,-0.21){2}{\line(0,-1){0.21}}
\multiput(94.84,45.77)(0.13,-0.21){2}{\line(0,-1){0.21}}
\multiput(95.11,45.35)(0.14,-0.21){2}{\line(0,-1){0.21}}
\multiput(95.39,44.94)(0.14,-0.2){2}{\line(0,-1){0.2}}
\multiput(95.67,44.53)(0.15,-0.2){2}{\line(0,-1){0.2}}
\multiput(95.97,44.13)(0.1,-0.13){3}{\line(0,-1){0.13}}
\multiput(96.28,43.74)(0.11,-0.13){3}{\line(0,-1){0.13}}
\multiput(96.6,43.36)(0.11,-0.12){3}{\line(0,-1){0.12}}
\multiput(96.94,42.99)(0.11,-0.12){3}{\line(0,-1){0.12}}
\multiput(97.28,42.63)(0.12,-0.12){3}{\line(0,-1){0.12}}
\multiput(97.63,42.28)(0.12,-0.11){3}{\line(1,0){0.12}}
\multiput(97.99,41.94)(0.12,-0.11){3}{\line(1,0){0.12}}
\multiput(98.36,41.6)(0.13,-0.11){3}{\line(1,0){0.13}}
\multiput(98.74,41.28)(0.13,-0.1){3}{\line(1,0){0.13}}
\multiput(99.13,40.97)(0.2,-0.15){2}{\line(1,0){0.2}}
\multiput(99.53,40.67)(0.2,-0.14){2}{\line(1,0){0.2}}
\multiput(99.94,40.39)(0.21,-0.14){2}{\line(1,0){0.21}}
\multiput(100.35,40.11)(0.21,-0.13){2}{\line(1,0){0.21}}
\multiput(100.77,39.84)(0.21,-0.13){2}{\line(1,0){0.21}}
\multiput(101.2,39.59)(0.22,-0.12){2}{\line(1,0){0.22}}
\multiput(101.64,39.35)(0.22,-0.11){2}{\line(1,0){0.22}}
\multiput(102.08,39.12)(0.22,-0.11){2}{\line(1,0){0.22}}
\multiput(102.53,38.91)(0.23,-0.1){2}{\line(1,0){0.23}}
\multiput(102.98,38.7)(0.23,-0.1){2}{\line(1,0){0.23}}
\multiput(103.44,38.51)(0.46,-0.18){1}{\line(1,0){0.46}}
\multiput(103.91,38.34)(0.47,-0.16){1}{\line(1,0){0.47}}
\multiput(104.37,38.17)(0.47,-0.15){1}{\line(1,0){0.47}}
\multiput(104.85,38.02)(0.48,-0.14){1}{\line(1,0){0.48}}
\multiput(105.33,37.88)(0.48,-0.12){1}{\line(1,0){0.48}}
\multiput(105.81,37.76)(0.49,-0.11){1}{\line(1,0){0.49}}
\multiput(106.29,37.65)(0.49,-0.1){1}{\line(1,0){0.49}}
\multiput(106.78,37.55)(0.49,-0.08){1}{\line(1,0){0.49}}
\multiput(107.27,37.47)(0.49,-0.07){1}{\line(1,0){0.49}}
\multiput(107.77,37.4)(0.49,-0.06){1}{\line(1,0){0.49}}
\multiput(108.26,37.34)(0.5,-0.04){1}{\line(1,0){0.5}}
\multiput(108.76,37.3)(0.5,-0.03){1}{\line(1,0){0.5}}
\multiput(109.25,37.27)(0.5,-0.01){1}{\line(1,0){0.5}}
\put(109.75,37.26){\line(1,0){0.5}}
\multiput(110.25,37.26)(0.5,0.01){1}{\line(1,0){0.5}}
\multiput(110.75,37.27)(0.5,0.03){1}{\line(1,0){0.5}}
\multiput(111.24,37.3)(0.5,0.04){1}{\line(1,0){0.5}}
\multiput(111.74,37.34)(0.49,0.06){1}{\line(1,0){0.49}}
\multiput(112.23,37.4)(0.49,0.07){1}{\line(1,0){0.49}}
\multiput(112.73,37.47)(0.49,0.08){1}{\line(1,0){0.49}}
\multiput(113.22,37.55)(0.49,0.1){1}{\line(1,0){0.49}}
\multiput(113.71,37.65)(0.49,0.11){1}{\line(1,0){0.49}}
\multiput(114.19,37.76)(0.48,0.12){1}{\line(1,0){0.48}}
\multiput(114.67,37.88)(0.48,0.14){1}{\line(1,0){0.48}}
\multiput(115.15,38.02)(0.47,0.15){1}{\line(1,0){0.47}}
\multiput(115.63,38.17)(0.47,0.16){1}{\line(1,0){0.47}}
\multiput(116.09,38.34)(0.46,0.18){1}{\line(1,0){0.46}}
\multiput(116.56,38.51)(0.23,0.1){2}{\line(1,0){0.23}}
\multiput(117.02,38.7)(0.23,0.1){2}{\line(1,0){0.23}}
\multiput(117.47,38.91)(0.22,0.11){2}{\line(1,0){0.22}}
\multiput(117.92,39.12)(0.22,0.11){2}{\line(1,0){0.22}}
\multiput(118.36,39.35)(0.22,0.12){2}{\line(1,0){0.22}}
\multiput(118.8,39.59)(0.21,0.13){2}{\line(1,0){0.21}}
\multiput(119.23,39.84)(0.21,0.13){2}{\line(1,0){0.21}}
\multiput(119.65,40.11)(0.21,0.14){2}{\line(1,0){0.21}}
\multiput(120.06,40.39)(0.2,0.14){2}{\line(1,0){0.2}}
\multiput(120.47,40.67)(0.2,0.15){2}{\line(1,0){0.2}}
\multiput(120.87,40.97)(0.13,0.1){3}{\line(1,0){0.13}}
\multiput(121.26,41.28)(0.13,0.11){3}{\line(1,0){0.13}}
\multiput(121.64,41.6)(0.12,0.11){3}{\line(1,0){0.12}}
\multiput(122.01,41.94)(0.12,0.11){3}{\line(1,0){0.12}}
\multiput(122.37,42.28)(0.12,0.12){3}{\line(1,0){0.12}}
\multiput(122.72,42.63)(0.11,0.12){3}{\line(0,1){0.12}}
\multiput(123.06,42.99)(0.11,0.12){3}{\line(0,1){0.12}}
\multiput(123.4,43.36)(0.11,0.13){3}{\line(0,1){0.13}}
\multiput(123.72,43.74)(0.1,0.13){3}{\line(0,1){0.13}}
\multiput(124.03,44.13)(0.15,0.2){2}{\line(0,1){0.2}}
\multiput(124.33,44.53)(0.14,0.2){2}{\line(0,1){0.2}}
\multiput(124.61,44.94)(0.14,0.21){2}{\line(0,1){0.21}}
\multiput(124.89,45.35)(0.13,0.21){2}{\line(0,1){0.21}}
\multiput(125.16,45.77)(0.13,0.21){2}{\line(0,1){0.21}}
\multiput(125.41,46.2)(0.12,0.22){2}{\line(0,1){0.22}}
\multiput(125.65,46.64)(0.11,0.22){2}{\line(0,1){0.22}}
\multiput(125.88,47.08)(0.11,0.22){2}{\line(0,1){0.22}}
\multiput(126.09,47.53)(0.1,0.23){2}{\line(0,1){0.23}}
\multiput(126.3,47.98)(0.1,0.23){2}{\line(0,1){0.23}}
\multiput(126.49,48.44)(0.18,0.46){1}{\line(0,1){0.46}}
\multiput(126.66,48.91)(0.16,0.47){1}{\line(0,1){0.47}}
\multiput(126.83,49.37)(0.15,0.47){1}{\line(0,1){0.47}}
\multiput(126.98,49.85)(0.14,0.48){1}{\line(0,1){0.48}}
\multiput(127.12,50.33)(0.12,0.48){1}{\line(0,1){0.48}}
\multiput(127.24,50.81)(0.11,0.49){1}{\line(0,1){0.49}}
\multiput(127.35,51.29)(0.1,0.49){1}{\line(0,1){0.49}}
\multiput(127.45,51.78)(0.08,0.49){1}{\line(0,1){0.49}}
\multiput(127.53,52.27)(0.07,0.49){1}{\line(0,1){0.49}}
\multiput(127.6,52.77)(0.06,0.49){1}{\line(0,1){0.49}}
\multiput(127.66,53.26)(0.04,0.5){1}{\line(0,1){0.5}}
\multiput(127.7,53.76)(0.03,0.5){1}{\line(0,1){0.5}}
\multiput(127.73,54.25)(0.01,0.5){1}{\line(0,1){0.5}}

\linethickness{0.3mm}
\put(65,55){\line(1,0){27}}
\linethickness{0.3mm}
\multiput(32,72)(0.12,-0.12){40}{\line(1,0){0.12}}
\linethickness{0.3mm}
\multiput(31,36)(0.12,0.12){40}{\line(1,0){0.12}}
\linethickness{0.3mm}
\multiput(117,71)(0.12,0.12){40}{\line(1,0){0.12}}
\linethickness{0.3mm}
\multiput(125,65)(0.12,0.08){40}{\line(1,0){0.12}}
\linethickness{0.3mm}
\put(127,50){\line(1,0){5}}
\linethickness{0.3mm}
\multiput(114,38)(0.12,-0.12){40}{\line(0,-1){0.12}}
\put(30,75){\makebox(0,0)[cc]{1}}

\put(27,32){\makebox(0,0)[cc]{$i$}}

\put(120,80){\makebox(0,0)[cc]{2}}

\put(123,73){\makebox(0,0)[cc]{$\bullet$}}

\put(125,71){\makebox(0,0)[cc]{$\bullet$}}

\put(140,70){\makebox(0,0)[cc]{$(i-1)$}}

\put(135,45){\makebox(0,0)[cc]{$(i+1)$}}

\put(127,37){\makebox(0,0)[cc]{$\bullet$}}

\put(130,40){\makebox(0,0)[cc]{$\bullet$}}

\put(45,55){\makebox(0,0)[cc]{1PI}}

\put(5,55){\makebox(0,0)[cc]{{\Huge $\displaystyle \sum_{i=2}^N$}}}

\put(110,55){\makebox(0,0)[cc]{Full}}

\put(120,28){\makebox(0,0)[cc]{$N$}}

\end{picture}
}

\newcommand{\gold}{\VV_{\rm G}}

\newcommand{\goldc}{\VV^c_{\rm G}}

\usepackage{bm}
\usepackage[table]{xcolor}

\newcommand{\scalar}{\VV_{\rm S}}

\newcommand{\fermion}{\VV_{\rm F}}

\newcommand{\wts}{\wt\Sigma}

\newcommand{\wtsp}{\wt\Sigma^c}

\newcommand{\four}{(4)}

\begin{document}

\begin{flushright}
\end{flushright}

\vskip 12pt

\baselineskip 24pt

\begin{center}
{\Large \bf  Supersymmetry Restoration in Superstring Perturbation Theory}

\end{center}

\vskip .6cm
\medskip

\vspace*{4.0ex}

\baselineskip=18pt

\centerline{\large \rm Ashoke Sen}

\vspace*{4.0ex}

\centerline{\large \it Harish-Chandra Research Institute}
\centerline{\large \it  Chhatnag Road, Jhusi,
Allahabad 211019, India}

\vspace*{1.0ex}
\centerline{\small E-mail:  sen@mri.ernet.in}

\vspace*{5.0ex}

\centerline{\bf Abstract} \bigskip

Superstring perturbation theory based on the 1PI effective theory 
approach has been useful for
addressing the problem of mass renormalization and vacuum shift.
We derive Ward identities associated with space-time supersymmetry transformation
in this approach. This leads to a proof of the equality of renormalized masses of bosons
and fermions and identities relating fermionic amplitudes to bosonic amplitudes
after taking into account the effect of mass renormalization. This also relates unbroken
supersymmetry to a given order in perturbation theory to absence of tadpoles of
massless scalars to higher order. The results are valid at the perturbative vacuum as
well as in the shifted vacuum when the latter describes the correct ground state of
the theory. We apply this to SO(32) heterotic string theory on Calabi-Yau 3-folds where
a one loop Fayet-Iliopoulos term apparently breaks supersymmetry at one loop,
but analysis of the low
energy effective field theory indicates that there is a nearby vacuum where 
supersymmetry is restored. We explicitly prove that the perturbative amplitudes
of this theory around the shifted vacuum indeed satisfy the Ward identities
associated with unbroken supersymmetry. We also test the general arguments by 
explicitly verifying the equality of bosonic and fermionic masses at one loop
order in the shifted vacuum, and the appearance of two loop dilaton tadpole in the
perturbative vacuum where supersymmetry is expected to be broken.

\vfill \eject

\baselineskip=18pt










\tableofcontents

\sectiono{Introduction and Summary}

In conventional string perturbation theory we set the external states to be `on-shell'
from the beginning by setting the squared momentum $k^2$ to be equal to $-m^2$
where $m$ is the {\it tree level} mass of the state. Therefore in
this approach we cannot directly compute the renormalized mass of a state at loop
level. For this reason this approach is most efficient for dealing with states whose
masses are not renormalized in perturbation theory -- massless gauge particles and
BPS states. If we attempt to apply this approach to states whose masses {\it are 
renormalized} in perturbation theory, we encounter 
certain infrared divergences in the amplitude that can be taken as a signal of
mass renormalization, but this does not lead to a systematic procedure for computing the
renormalized mass or the S-matrix involving these states.

A related problem in this approach is that the background is
fixed to be a solution to the classical equations of motion from the beginning. 
Therefore we cannot deal with the situation where the classical 
vacuum is destabilized
at the loop level, but there is a nearby stable vacuum. We can detect the
instability in perturbation theory -- again in the form of certain infrared divergences
associated with tadpoles of massless fields, but there is 
no systematic procedure for computing
physical amplitudes by quantizing the theory around the nearby vacuum that does
not solve the classical equations of motion.

Ordinary quantum field theories have a systematic procedure for dealing 
with both
these issues. 
Therefore one might be tempted to resolve these problems by invoking a
field theory for strings. While this is possible for bosonic string 
theory\cite{wittensft,9206084},  despite many 
attempts\cite{wittenssft,9503099,0109100,0406212,0409018,
1312.2948,1312.7197,1407.8485,1412.5281,1403.0940} there is
not yet a fully consistent string field theory for heterotic of type II strings that 
works
at loop level. 
Furthermore by working with string field theory we lose one of the
main advantages of string theory -- while in field theory a given amplitude is
obtained as a sum of many Feynman diagrams, in string theory at each loop order
there is a single term to compute -- given by the integral of certain quantity over the
moduli space of punctured Riemann surfaces of a given genus. For example, amplitudes
computed from bosonic string
field theory   expresses the result as a sum over many terms, with each term describing
the integral over only a small subspace of the moduli space of Riemann surfaces.
When we add up the contribution from all the Feynman diagrams of string field
theory, we recover the integral over the full moduli space\cite{GMW,9206084}.

In \cite{1411.7478,1501.00988} 
(which in turn is based on results of \cite{1311.1257,1401.7014,
1404.6254,1408.0571,1504.00609}) we suggested
a way of overcoming this problem by using the formalism of 
one particle irreducible (1PI) effective theory
that combines some of the good features of quantum
field theories with some of the good features of string theory. In conventional quantum
field theory the 1PI effective action can be used to deal with the problems of
mass renormalization and vacuum shift. This is defined as the generating function
of off-shell 1PI amplitudes, and is free  from all infrared divergences associated
with mass renormalization or massless tadpoles.\footnote{Infrared divergences associated 
with internal lines
in the loop going on-shell are tamed by the usual $i\eps$ 
prescription\cite{berera,1307.5124}.}
Given the 1PI action we are  supposed to first find its local extremum, and
then find plane wave solutions to the {\it classical 
linearized equations of motion} derived
from the action. If these occur at momenta $k$ then the values of $-k^2$ give the
renormalized squared masses of the theory. Once we have determined them, the
{\it tree level S-matrix} computed from the 1PI action gives us the full renormalized
S-matrix of the original quantum field theory.

In quantum field theory most of the complexities of computing loop amplitudes lie
in the computation of the 1PI effective action as it involves sum over many Feynman 
diagrams. It turns out that in string theory it is possible to directly write down the expression
for the 1PI amplitudes without going through the route of first having to formulate
a field theory of strings. A 1PI amplitude 
takes the form of an integral of certain 
correlation function of the underlying world-sheet superconformal field theory 
(SCFT)
over a subspace of the moduli space of Riemann surfaces called the 1PI 
subspace. By construction these 1PI subspaces do not include separating type
degenerations of the moduli space -- the sources of the usual infrared divergences
of ordinary string perturbation theory associated with mass renormalization and
massless tadpoles. Once the 1PI theory is available we can follow the usual route
of quantum field theory, i.e. first find the correct vacuum by solving the classical
equations of motion of this 1PI theory, then determine the renormalized mass by
examining the solutions to the linearized equations of motion around the shifted
vacuum and then compute the tree level S-matrix of the 1PI theory to determine the 
renormalized S-matrix of the full string theory.

The construction of 1PI action requires extra data\cite{nelson} 
 compared to what
is used in conventional superstring perturbation 
theory. These data involve a
choice of local coordinate system around every puncture and locations of picture
changing operators (PCO's) on the Riemann surface.\footnote{It is expected that the
choice of PCO's can be traded in for the choice of local superconformal coordinate system
around the punctures in the formalism 
of \cite{9609220,9703183,9706033,dp,1209.2199,1209.2459,1209.5461,
1304.2832,1304.7798,1306.3621,1307.1749,1403.5494,
1404.6257,1501.02499,1501.02675,1502.03673}, 
but the details of this have not been
worked out fully.} In conventional perturbation theory, where we use BRST invariant
and conformally invariant vertex operators as external states, these data do not
affect the result, but since the construction of the 1PI theory requires 1PI
amplitudes with off-shell external states, the result depends on these additional data.
However it was shown in \cite{1411.7478,1501.00988} following earlier work
of \cite{9301097}
that different 1PI theories associated with two different choices
of local coordinates around the punctures and/or different PCO locations are related
to each other by field redefinition. Since field redefinition does not affect the
renormalized masses or tree level S-matrix elements, we see that the physical 
quantities computed 
from the 1PI effective theory do not suffer from any ambiguity.

It was also found in \cite{1411.7478,1501.00988}
that the 1PI effective theory constructed this way possesses an infinite 
dimensional gauge invariance. This includes the general coordinate invariance, local
supersymmetry etc.\ but also includes gauge invariances associated with
massive fields. Furthermore since the 1PI action includes the effect of loop corrections,
the gauge transformation laws are full quantum corrected gauge transformation laws
of the theory. In specific background some of these gauge symmetries leave the
background invariant. These then represent `global symmetries' of the theory, e.g 
space-time translation invariance, global supersymmetry etc.

Since the full string theory amplitudes are computed from tree amplitudes of the 1PI
theory, once we have identified a gauge symmetry of the 1PI theory, there is no
further scope for any anomaly -- the effect of anomalies would have already been
captured in the 1PI action. Therefore we should be able to use the gauge symmetries 
(and their global counterpart) to derive appropriate Ward identities for the physical
amplitudes. This will be one of the main goals of the present paper. Even though 
our main focus will be on Ward identities associated with local and global 
supersymmetries, much of our analysis carries through for other symmetries as well. 

Supersymmetry Ward identities in string theory have been studied extensively 
 -- most recently in \cite{1209.5461,1304.2832}. 
The difference between the results of \cite{1209.5461,1304.2832} and the
ones discussed here is that while the analysis of \cite{1209.5461,1304.2832} 
had to be restricted to
amplitudes in the perturbative vacuum, for external states which do not suffer
any mass renormalization, our analysis is valid for general external states and also
in vacuum which may not be the perturbative vacuum but a nearby vacuum obtained
by condensation of some scalar fields.  Indeed one of our goals will be to show how
supersymmetry Ward identities can be used to prove the degeneracy between {\it
renormalized} masses of bosons and fermions in the quantum theory. Another goal will be
to show how in some SO(32) heterotic string compactification 
where supersymmetry is broken at the perturbative vacuum 
at one loop order, it is restored in a nearby vacuum by 
condensation of a scalar field.
In particular we shall verify the bose-fermi degeneracy at one loop and vanishing of
tadpoles at two loops in the shifted vacuum, even though at the perturbative
vacuum we do not have these properties.

It is worth re-emphasizing that 
even though the construction of the 1PI action is very similar to that of
string field theory, at least in this paper the motivation is quite opposite to
that of string field theory. String field theory provides us with a triangulation
of the moduli space with different regions of the moduli space coming from
different Feynman diagrams, and the main motivation for string field theory 
stems from
the hope that one can use it to study non-perturbative aspects of string
theory. Instead here our main motivation is to use 1PI action to study
perturbative amplitudes of string theory. With this in mind we try to bring
the expression for the S-matrix elements and Ward identities
computed from the 1PI action to 
a form that is close to the usual Polyakov prescription. We find that
while the final result is very similar to what we get according to the usual
Polyakov prescription, it comes with an in built subtraction procedure 
that removes all the usual infrared
divergences associated with tadpoles and mass renormalization.

Given that much of the analysis is technical, we shall try to summarize our
main results here. 
\begin{enumerate}
\item \label{pone}
Our first main result is to arrive at a prescription for
computing S-matrix elements in string theory involving states that undergo
mass renormalization. We find that the prescription reduces to the usual
Polyakov prescription with two important differences. 
The usual Polyakov prescription suffers from possible infrared divergences
from separating type degeneration where momentum conservation
forces the momentum flowing through the long tube connecting
the  two Riemann surfaces 
to be either zero or equal to one of the momenta carried
by external states. These are the divergences associated with tadpoles
and mass renormalization\cite{1209.5461,1304.2832}.
Our approach leads to the result that in a finite
neighborhood of any such
separating type degeneration, determined by the choice of local coordinate
system used to define the 1PI amplitudes, we have a definite subtraction
that removes the possible divergences associated with tadpoles and
mass renormalization. 
The subtraction involving
tadpole like degenerations 
is 
compensated by having  to sum over arbitrary number 
of insertions of
certain external state -- not necessarily BRST invariant --
which is determined by a recursive procedure.
Both the number of insertions and the required number of recursions
are of course bounded when we work to some fixed 
 order in the string coupling $g_s$.
These recursion relations are given in \refb{esoln}, \refb{esol1} with 
$|\psi_k\rangle$ denoting the external state to be inserted if we want
results up to order $g_s^k$.
Similarly the subtraction associated with degenerations 
corresponding to mass renormalization is compensated by having to
replace the usual
BRST invariant external states by a new set of states -- not in general
BRST invariant -- which 
are determined by a recursive procedure. These recursion relations can be
found in \refb{esa3}-\refb{esa5} with $|\phi_n\rangle$ denoting the state
that replaces the external state if we want result up to order $g_s^n$.
The subtraction procedure for both kinds of divergences 
has been explained in \S\ref{susual}.

\item This procedure is also valid in situations in which there is more than one
possible choice of vacuum at some given order in $g_s$. The existence of
multiple vacua is reflected the existence of multiple solutions to the
recursion relation \refb{esoln}, \refb{esol1} and our result for S-matrix elements
holds in any of the vacua. The results in different vacua differ by the 
choice of the state 
$|\psi_k\rangle$ that
we have to insert into the amplitude arbitrary number of times.

\item \label{ptwo}
Our second main result is a rederivation of the Ward  identities associated
with global and local (super-)symmetry when the external states undergo
mass renormalization and/or the vacuum is shifted from the usual perturbative
vacuum. We find that the Ward identities take forms which are identical to those
given in \cite{1209.5461,1304.2832} except for the kind of 
modifications already mentioned in point \ref{pone}. 
We also derive the equality of {\it renormalized} bosonic and fermionic
masses to all orders in vacua with unbroken supersymmetry. This result
could not be proven with the usual form of the Ward identities that are valid
only in the absence of mass renormalization.

\item \label{pthree} 
Refs.~\cite{1209.5461,1304.2832} 
described how unbroken supersymmetry to a given order in 
perturbation theory implies vanishing of tadpoles to one higher order. We 
rederive this result, but our analysis also includes the cases where the
vacuum under consideration is not the perturbative vacuum but related to
it by a shift of order $g_s$ of the string field. This applies in particular to the
case of SO(32) heterotic string theory on Calabi-Yau 3-folds where often a
Fayet-Iliopoulos term breaks supersymmetry in the perturbative vacuum 
at one loop
order\cite{DSW,ADS,DIS} 
but supersymmetry can be restored in a new vacuum obtained by shifting
the string field.

\item We apply our general method to obtain some explicit results in  
SO(32) heterotic string theory on Calabi-Yau manifolds. 
In particular we explicitly compute the one loop result for masses of a
set of scalars and fermions related by supersymmetry and show that in the
shifted vacuum they are equal as predicted by supersymmetry even though
this equality is absent at the perturbative vacuum. This agrees with the
general results quoted in point \ref{ptwo} above. 
We also show by explicit calculation that the two loop diaton tadpole is
non-vanishing at the perturbative vacuum. This is in agreement with general
arguments\cite{DSW,ADS,DIS,greenseiberg} and explicit 
results\cite{AtickS,1304.2832,1404.5346}.
We also show that the dilaton tadpole 
at the shifted vacuum vanishes to this order (i.e. order $g_s^4$).
\end{enumerate}

The rest of the paper is organized as follows. In \S\ref{sreview} we briefly review the
results of \cite{1411.7478,1501.00988} 
that we shall use in our analysis. In \S\ref{smatrix} we discuss the
construction of the S-matrix elements from the 1PI theory and compare this with the
usual prescription for computing S-matrix in string theory. We show that while the
final results are very similar, there are subtle differences which precisely remove the
divergences associated with massless tadpoles and mass renormalization that we
encounter in the usual perturbation theory. In \S\ref{sglobal} we study the Ward identities
associated with global and local (super-)symmetries. In particular we show how as a
consequence of global supersymmetry we have equality of renormalized masses of 
bosons and fermions and relation between different S-matrix elements involving external
bosons and fermions. Local (super-)symmetry on the other hand can be used to show
how pure gauge states decouple from the S-matrix. In \S\ref{stadpole} we discuss the
relationship between existence of global supersymmetry and tadpoles of massless
fields and show that global supersymmetry to a given order implies vanishing of tadpoles
to one higher order in perturbation theory. This 
result is in the same spirit as in \cite{1209.5461,1304.2832} 
but we use 1PI effective theory instead of
on-shell amplitudes to derive the results. In \S\ref{ehetrev}-\S\ref{sshifted} 
we apply these general
methods to a specific class of theories -- SO(32) heterotic string theory 
compactified on Calabi-Yau 3-folds. 
\S\ref{ehetrev} contains a review of SO(32) heterotic string theory on Calabi-Yau
3-folds.
In \S\ref{srestore} we show that in 1PI effective theory at one loop order
we have multiple vacua, and while supersymmetry is broken at the 
perturbative vacuum,
it is restored at the shifted vacuum. In \S\ref{sbose} we verify one of the consequences of
the supersymmetry restoration by checking that the degeneracy between scalar and
fermion mass for a particular multiplet, that was broken at the perturbative 
vacuum, is restored at the shifted vacuum. In \S\ref{sdilaton} 
we use the method described in \S\ref{stadpole} to compute the
two loop dilaton tadpole at the perturbative vacuum, and find a non-zero value confirming
that supersymmetry is indeed broken there. 
On the other hand we show in \S\ref{sshifted} that to the same order, the dilaton tadpole
at the shifted vacuum vanishes.
In  appendix
\ref{sglossary} we give a glossary of the various symbols that
have been used frequently in the rest of the paper.
Appendices \ref{enewappendix}-\ref{srep} contain various technical
details which have been left out from the main text.  In appendix \ref{slow} we
check the consistency of the results of \S\ref{ehetrev}-\S\ref{sshifted} with
the predictions of supersymmetric low energy effective action.

Throughout this paper we shall work in $\alpha'=1$ unit.

We shall conclude this introduction with one final comment. In the definition of the
1PI amplitudes given in \cite{1408.0571,1411.7478,1501.00988} we have used local coordinate system 
at punctures
/ PCO locations in a way that is gluing compatible at separating type degenerations.
For discussing various aspects of mass renormalization and vacuum shift this is sufficient. Indeed it was shown in \cite{1311.1257,1401.7014,
1411.7478,1501.00988} that once
the choice of local coordinates satisfies this condition, the renormalized masses
and S-matrix are independent of any other details of the choice of local
coordinates and PCO locations.  
However it may happen that for analyzing other aspects of perturbation theory,
{\it e.g.} in dealing with the standard infra-red divergences in loop amplitudes, 
we may
need gluing compatibility at non-separating type degenerations as well. This can be
easily accommodated in our formalism by restricting the possible choice of local coordinate
system and PCO locations in an appropriate manner. 

\sectiono{Review} \label{sreview}

We begin by reviewing our notation and 
some of the main results of \cite{1411.7478,1501.00988}.
Since the analysis in the heterotic and type II string theories are similar,
we shall carry out our discussion mostly in the context of heterotic string
theory and point out briefly where the analysis in type II string theories 
differs.

\subsection{World-sheet theory} \label{s2.1}

The world-sheet theory for any heterotic string compactification
at string tree level contains a
matter superconformal field theory with
central charge (26,15), and a ghost system of  total central charge $(-26,-15)$
containing anti-commuting $b$, $c$, $\bar b$, $\bar c$
ghosts and commuting $\beta, \gamma$ ghosts.
The $(\beta,\gamma)$ system can be bosonized as\cite{FMS}
\be \label{eboserule}
\gamma = \eta\, e^{\phi}, \quad \beta= \p\xi \, e^{-\phi}, \quad \delta(\gamma)
= e^{-\phi}, \quad \delta(\beta) = e^\phi\, ,
\ee
where $\xi, \eta$ are  fermions and $\phi$ is a scalar with background charge.
We assign (ghost number, picture number, GSO) quantum numbers to various fields
as follows:
\ben
&& c, \bar c: (1,0,+), \quad b, \bar b: (-1, 0,+), \quad \gamma: (1,0,-), \quad \beta:(-1,0,-),
\nonumber \\
&& \xi: (-1,1,+), \quad \eta: (1,-1,+), \quad
e^{q\phi}: (0, q, (-1)^q)\, .
\een
The operator products of these fields take the form
\be \label{eghope}
c(z) b(w) =(z-w)^{-1}+\cdots, \quad
\xi(z)\eta(w) = (z-w)^{-1}+\cdots, \quad e^{q_1\phi(z)} e^{q_2\phi(w)} =
(z-w)^{-q_1q_2} e^{(q_1+q_2)\phi(w)}+ \cdots \, ,
\ee
where $\cdots$ denote less singular terms.
We denote by
$\bar T_m(\bar z)$ the anti-holomorphic part of the matter stress tensor, by
$T_{m}(z)$ the holomorphic part of the matter stress tensor,
by $T_{\beta,\gamma}(z)$
the stress tensor of the $(\beta,\gamma)$ system and by 
$T_F(z)$ the 
world-sheet supersymmetry current in the matter sector. 
In terms of these
the BRST charge is given by
\be \label{ebrs1}
Q_B = \ointop dz j_B(z) + \ointop d\bar z \bar j_B(\bar z)\, ,
\ee
where
\be\label{ebrs2}
\bar j_B(\bar z) = \bar c(\bar z) \bar T_m(\bar z)
+\bar b(\bar z) \bar c(\bar z) \bar\p \bar c(\bar z)\, ,
\ee
\be \label{ebrstcurrent}
j_B(z) =c(z) (T_{m}(z) + T_{\beta,\gamma}(z) )+ \gamma (z) T_F(z) 
+ b(z) c(z) \p c(z) 
-{1\over 4} \gamma(z)^2 b(z)\, ,
\ee
and $\ointop$ is normalized so that $\ointop dz/z=1$, $\ointop d\bar z/\bar z=1$.
The picture changing operator (PCO) $\XX$
is given by\cite{FMS,Verlinde:1987sd}
\be \label{epicture}
\XX(z) = \{Q_B, \xi(z)\} = c \partial \xi + 
e^\phi T_F - {1\over 4} \p\eta e^{2\phi} b
- {1\over 4} \p\left(\eta e^{2\phi} b\right)\, .
\ee
This is a BRST invariant dimension zero primary operator 
and carries picture number $1$. Finally to get the signs of various correlation
functions we need to describe our normalization condition for the $SL(2,C)$
invariant vacuum $|0\rangle$. We choose
this to be
\be \label{evacnormprime}
\langle 0| c_{-1} \bar c_{-1} c_0 \bar c_0 c_1 \bar c_1 \xi_0 
e^{-2\phi(z)}|0\rangle =V\, ,
\ee
where $V$ is the volume of space-time (also equal to $(2\pi)^D \delta^{(D)}(0)$
where the argument of the delta function represents zero value of $D$-component
momentum vector). In subsequent discussions we shall set the space-time volume
$V$ to unity. Also for most of our analysis we shall work in the small Hilbert
space\cite{FMS} containing states 
annihilated by $\eta_0$, and include the $\xi_0$ factor in the
definition of the inner product to write
\be \label{evacnorm}
\langle 0| c_{-1} \bar c_{-1} c_0 \bar c_0 c_1 \bar c_1
e^{-2\phi(z)}|0\rangle =1\, .
\ee

For type II string theories the world-sheet theory of matter sector has central charge $(15,15)$.
The ghost system now also includes left-moving $(\bar\beta,\bar\gamma)$ system so that the
total
central charges of the ghost system now is $(-15,-15)$. The left-moving BRST current
$\bar j_B(\bar z)$ now contains extra terms as in \refb{ebrstcurrent} and we have left-handed
PCO $\bar\XX(\bar z)$ given by an expression identical to \refb{epicture} with all right-handed
fields replaced by their left-handed counterpart.

We denote by $\HH_T$ the Hilbert space of GSO even states 
of the matter-ghost CFT with arbitrary ghost and
picture numbers, with coefficients taking values in the grassmann algebra, satisfying the
constraints
\be \label{econd}
|s\rangle \in \HH_T \quad \hbox{iff} \quad b_0^-|s\rangle = 0, \quad L_0^{-}|s\rangle =0\, ,
\quad \eta_0 |s\rangle =0, \quad \bar\eta_0|s\rangle=0 \, \hbox{(in type II)}\, ,
\ee
where $\bar L_n$ and $L_n$ denote total Virasoro generators in the left and right-moving sectors
of the world-sheet theory, and 
\be \label{edefbpm}
b_0^\pm \equiv(b_0\pm\bar b_0), \quad L_0^\pm\equiv(L_0\pm\bar L_0), \quad
c_0^\pm \equiv {1\over 2} (c_0\pm\bar c_0)\, .
\ee
In the heterotic theory $\HH_T$ decomposes into a direct sum of the 
Neveu-Schwarz (NS) sector 
$\HH_{NS}$ and Ramond (R) sector $\HH_R$. In the type II string theories the corresponding
decomposition is $\HH_T=\HH_{NSNS}\oplus \HH_{NSR}\oplus \HH_{RNS}\oplus \HH_{RR}$.
We shall denote by $\{|\vp_r\rangle\}$ and
$\{|\vp^r\rangle\}$ a set of dual basis of $\HH_{T}$
satisfying
\be \label{ebasisbb}
\langle \vp^r | c_0^-|\vp_s\rangle = \delta^r{}_s \qquad \Leftrightarrow \qquad
\langle \vp_s | c_0^-|\vp^r\rangle = \delta^r{}_s  \, ,
\ee
and the completeness relation
\be \label{ecomplete}
|\vp_r\rangle \langle \vp^r| = |\vp^r\rangle \langle \vp_r|=b_0^-\, .
\ee
In heterotic string theory the basis states $|\vp_r\rangle$ in $\HH_{NS}$ are
grassmann even for even ghost number and grassmann odd for odd ghost
number. In $\HH_R$ the situation is opposite. Similar results hold for type II
string theories as well. During the intermediate stages of calculation we shall
also make use of GSO odd operators. The grassmann parity of GSO odd 
operators are opposite of that of GSO even operators for given ghost and
picture numbers.

In the heterotic string theory we define
\be \label{edefgg}
\GG|s\rangle =\begin{cases} {|s\rangle \quad \hbox{if $|s\rangle\in \HH_{NS}$}\cr
\XX_0\, |s\rangle \quad \hbox{if $|s\rangle\in \HH_R$}}
\end{cases}\, ,
\ee
while in type II string theories we define
\be \label{edefggii}
\GG|s\rangle =\begin{cases} {|s\rangle \quad \hbox{if $|s\rangle\in \HH_{NSNS}$}\cr
\XX_0\, |s\rangle \quad \hbox{if $|s\rangle\in \HH_{NSR}$}\cr 
\bar\XX_0\, |s\rangle \quad \hbox{if $|s\rangle\in \HH_{RNS}$}\cr 
\XX_0\bar\XX_0\, |s\rangle \quad \hbox{if $|s\rangle\in \HH_{RR}$}\cr 
}\, ,
\end{cases}
\ee
where
\be \label{edefggr}
\XX_0 \equiv \ointop {dz\over z}\, \XX(z), \quad \bar\XX_0 
\equiv \ointop {d\bar z\over \bar z}\, 
\bar\XX(\bar z).
\ee
Note that
\be \label{eqbg}
[Q_B, \GG]=0\, .
\ee

\subsection{1PI effective string field theory}  \label{s1pi}

Let $\wh\HH_T$ and $\wt\HH_T$ denote the subspaces of $\HH_T$ 
with the following restriction on the picture numbers:
\ben \label{erestrict}
& \hbox{heterotic} \qquad \qquad & \qquad \qquad
\qquad \qquad \hbox{type II} \nonumber \\
\wh\HH_T:\qquad \qquad  & -1, -1/2 \qquad \qquad & (-1,-1), (-1.-1/2), 
(-1/2, -1), (-1/2, -1/2)
\nonumber \\
\wt\HH_T:\qquad \qquad  & -1, -3/2 \qquad \qquad & (-1,-1), (-1.-3/2), 
(-3/2, -1), (-3/2, -3/2) \, . \nonumber \\
\een
This means that 
$\wh\HH_T$ denotes the subspace of $\HH_T$ containing NS sector
states of picture number $-1$ and R sector states of picture number $-1/2$.
In type II string theories $\wh\HH_T$ will contain states of picture numbers
$(-1,-1)$, $(-1,-1/2)$, $(-1/2,-1)$ and $(-1/2,-1/2)$. 
Similar interpretation applies to $\wt\HH_T$.
In \cite{1411.7478,1501.00988} we introduced a multilinear function 
$\{A_1\cdots A_N\}$  of $|A_1\rangle,\cdots |A_N\rangle\in \wh\HH_T$
taking values in the grassmann algebra, and
another multilinear function 
$[A_2\cdots A_N]$ of $|A_2\rangle,\cdots |A_N\rangle\in \wh\HH_T$
taking values in $\wt\HH_T$. 
Physically $\{A_1\cdots A_N\}$ denotes contribution to the off-shell
amplitude with external states $|A_1\rangle,\cdots |A_N\rangle$ from the `1PI region of the
moduli space'.
$[\cdots]$ is related to
$\{\cdots \}$ via
\be \label{edefsquare}
\langle A_1| c_0^- |[A_2\cdots A_N]\rangle = \{A_1\cdots A_N\}
\ee
for all $|A_1\rangle\in\wh\HH_T$. These multilinear functions satisfy
\be \label{ecurlysym}
\{A_1 A_2\cdots A_{i-1}A_{i+1} A_iA_{i+2} \cdots A_N\}
=(-1)^{\gamma_i \gamma_{i+1}} \{A_1A_2\cdots A_N\}\, ,
\ee
\be \label{esymmetry}
[A_1\cdots A_{i-1}A_{i+1} A_iA_{i+2} \cdots A_N]
=(-1)^{\gamma_i \gamma_{i+1}} [A_1\cdots A_N] \, ,
\ee
where $\gamma_i$ is the grassmannality of $|A_i\rangle$.
They also satisfy
\ben \label{eimpid}
&&  \sum_{i=1}^N (-1)^{\gamma_1+\cdots \gamma_{i-1}}\{A_1\cdots A_{i-1} (Q_B A_i)
A_{i+1} \cdots A_N\} \nonumber \\
&=& -  
{1\over 2} \sum_{\ell,k\ge 0\atop \ell+k=N} \sum_{\{i_a;a=1,\cdots \ell\}, \{j_b;b=1,\cdots k\}\atop
\{i_a\}\cup \{j_b\} = \{1,\cdots N\}
}\sigma(\{i_a\}, \{j_b\})
\{A_{i_1} \cdots A_{i_\ell}\GG[A_{j_1} \cdots A_{j_k}]\} 
\, 
\een
and
\ben \label{emain}
&& Q_B[A_1\cdots A_N] + \sum_{i=1}^N (-1)^{\gamma_1+\cdots 
\gamma_{i-1}}[A_1\cdots A_{i-1} (Q_B A_i)
A_{i+1} \cdots A_N] \nonumber \\
&=& -  \sum_{\ell,k\ge 0\atop \ell+k=N} 
\sum_{\{i_a;a=1,\cdots \ell\}, \{j_b;b=1,\cdots k\}\atop
\{i_a\}\cup \{j_b\} = \{1,\cdots N\}
}\sigma(\{i_a\}, \{j_b\})\, 
[A_{i_1} \cdots A_{i_\ell} \GG\, [A_{j_1} \cdots A_{j_k}]] 
\een
where $\sigma(\{i_a\}, \{j_b\})$ is the sign that one picks up while rearranging
$b_0^-,A_1,\cdots A_N$ to\break \noindent
$A_{i_1},\cdots A_{i_\ell}, b_0^-, A_{j_1},\cdots A_{j_k}$.
Finally we also have a relation
\be \label{eneweq2a}
\{A_1\cdots A_k \GG[\wt A_1\cdots \wt A_\ell]\}
= (-1)^{\gamma+\tilde\gamma+\gamma\tilde\gamma}
\{\wt A_1\cdots \wt A_\ell \GG[A_1\cdots A_k]\}\, ,
\ee
where $\gamma$ and $\tilde\gamma$ are the total grassmannalities of
$A_1,\cdots A_k$ and $\wt A_1,\cdots \wt A_\ell$ respectively.

Using these functions we constructed the equations of motion of 
gauge invariant 1PI effective string field theory. The string field $|\Psi\rangle$
is taken to be a state in $\wh\HH_T$ of ghost number 2.
In the heterotic string theory $|\Psi\rangle$
can be decomposed as $|\Psi_{NS}\rangle+
|\Psi_R\rangle$ where $|\Psi_{NS}\rangle$ is a state of picture number $-1$ and
ghost number 2 in $\HH_{NS}$ with coefficients 
multiplying the basis states given by even
elements of the grassmann 
algebra, and $|\Psi_R\rangle$ is a state in $\HH_R$ of picture number $-1/2$ and
ghost number 2  with coefficients multiplying the basis states given by odd
elements of the grassmann algebra. In the type II string theory the string field
is decomposed as $|\Psi_{NSNS}\rangle+
|\Psi_{NSR}\rangle+|\Psi_{RNS}\rangle+
|\Psi_{RR}\rangle$ with similar restriction on picture number and ghost
number. It follows from the discussion below \refb{ecomplete} that
in all theories all components of $|\Psi\rangle$ 
are grassmann even. The equations of motion of the 1PI
effective string field theory takes
the form
\be \label{eeom}
|\EE\rangle=0, \qquad |\EE\rangle\equiv 
Q_B|\Psi\rangle + \sum_{n=1}^\infty {1\over (n-1)!} \GG[\Psi^{n-1}]\, .
\ee
This can be shown to be invariant under the gauge transformation
\be  \label{egauge}
|\delta\Psi\rangle = Q_B|\Lambda\rangle + \sum_{n=0}^\infty {1\over n!} 
\GG[\Psi^n \Lambda]\, ,
\ee
where $|\Lambda\rangle$ is an arbitrary grassmann odd
state in $\wh\HH_T$ carrying  ghost number 1.

Since $[~]$ -- the $n=1$ term on the right hand side of \refb{eeom} --
gets non-zero contribution from Riemann surfaces of genus $\ge 1$, 
$|\Psi\rangle=0$ is not a solution to the equations of motion \refb{eeom}. In
\cite{1411.7478} 
we described a systematic procedure for finding the vacuum solution 
$|\Psi_{\rm vac}\rangle$ -- a solution to \refb{eeom} in the NS sector
carrying zero momentum. This solution
is constructed iteratively as a power series in the string coupling $g_s$.
If $|\Psi_k\rangle$ denotes the solution to order $g_s^k$
then it is obtained iteratively by solving\footnote{Throughout this paper
we shall assume that the vacuum solution has expansion in powers of
$g_s$. This includes the case of perturbative vacuum where the solution
will have expansion in powers of $g_s^2$ -- we simply will get
$|\Psi_{2k+1}\rangle=|\Psi_{2k}\rangle$ for all integer $k$. There may
also be cases where the vacuum solution has an expansion in powers of
$g_s^\alpha$ for some $\alpha$ in the range $0<\alpha<1$. Our analysis 
can be extended to this case as well by replacing $g_s$ by $g_s^\alpha$
everywhere. \label{fgs}}
\be \label{esoln}
|\Psi_{k+1}\rangle = -{b_0^+\over L_0^+} \sum_{n=1}^\infty {1\over (n-1)!} (1-{\bf P})
\GG [\Psi_k^{n-1}]
+ | \psi_{k+1}\rangle\, ,
\ee
where ${\bf P}$ the projection operator into zero momentum
$L_0^+=0$ states and
$| \psi_{k+1}\rangle$ satisfies
\be \label{esol1}
{\bf P}| \psi_{k+1}\rangle = | \psi_{k+1}\rangle, \qquad Q_B| \psi_{k+1}\rangle
= - \sum_{n=1}^\infty {1\over (n-1)!} {\bf P}
\GG [\Psi_k^{n-1}]+\OO(g_s^{k+2})\, .
\ee
Possible obstruction to solving these equations arise from the failure to satisfy
\refb{esol1}. In \cite{1411.7478} 
we showed that the solution to \refb{esol1} exists iff
\be \label{econdaa}
\EE_{k+1}(\phi)\equiv 
\sum_{n=1}^\infty {1\over (n-1)!}  \langle \phi| c_0^- \GG|[\Psi_{k}{}^{n-1}]\rangle 
=\OO(g_s^{k+2})
\, ,
\ee
for any BRST invariant zero momentum
state $|\phi\rangle\in\wt\HH_T$  of ghost number two and
$L_0^+=0$.\footnote{Since we are 
dealing with NS sector states, there is no distinction between 
$\wt\HH_T$ and $\wh\HH_T$ and $\GG$ in \refb{esoln}-\refb{econdaa}
can be replaced by
identity operators.}
Therefore $\EE_{k+1}(\phi)$ represents an obstruction to extending
the vacuum solution beyond order $g_s^k$.
It was shown that the condition is trivially satisfied if $|\phi\rangle$ is BRST
trivial. Hence the non-trivial constraints come from zero momentum
non-trivial elements of the BRST cohomology -- the vertex operators of zero momentum
massless bosonic states. These obstructions correspond to existence of massless tadpoles
in the theory. Therefore the absence of massless tadpoles to order $g_s^{k+1}$ will
correspond to \refb{econdaa}.

While finding solutions to \refb{esol1} we have the
freedom of adding to $|\psi_{k+1}\rangle$ any state of the form
\be \label{emarg}
\sum_\alpha a_\alpha |\vp_\alpha\rangle
\ee
where $\{|\vp_\alpha\rangle\}$ is a basis of zero momentum, NS sector
BRST invariant states in $\wh \HH_T$
and $a_\alpha$'s are arbitrary coefficients.
Some of these $a_\alpha$'s could get fixed while trying to ensure
\refb{econdaa} at higher order. Those that do not get fixed represent moduli
and can be given arbitrary values.

If $|\Psi_{\rm vac}\rangle$ denotes a vacuum solution of \refb{eeom},  
then the gauge symmetries which 
preserve the solution correspond to global symmetries. Therefore they satisfy
\be \label{eglobal}
Q_B|\Lambda_{\rm global}\rangle + 
\sum_{n=0}^\infty {1\over n!} \GG [\Psi_{\rm vac}^n \Lambda_{\rm global}]=0\, .
\ee 
Such global symmetries
arising in the R-sector of heterotic string theory and RNS and NSR sectors
of type II string theories correspond to global supersymmetries. 
In \cite{1501.00988} 
we described a systematic procedure for solving these equations iteratively.
If $|\Lambda_k\rangle$ denotes the solution to \refb{eglobal} to order $g_s^k$
then we have
\be \label{esolam}
|\Lambda_k\rangle = - \sum_{n=0}^\infty {1\over n!} {b_0^+\over L_0^+} (1 - {\bf P})
\GG [\Psi_{\rm vac}^n \Lambda_{k-1}] + |\lambda_k\rangle\, ,
\ee
where ${\bf P}$ denotes the projection operator into $L_0^+=0$ states
and $|\lambda_k\rangle$ is an $L_0^+=0$  state satisfying
\be \label{etauk}
Q_B |\lambda_k\rangle = - \sum_{n=0}^\infty {1\over n!} {\bf P}
\GG [\Psi_{\rm vac}^n \Lambda_{k-1}] +\OO(g_s^{k+1})\, .
\ee
The possible obstruction to solving \refb{eglobal} arises from \refb{etauk}. The latter equation
can be solved only if 
\be \label{esusy2}
\LL_k(\hat\phi) \equiv 
\langle \hat\phi | c_0^- \sum_{n=0}^\infty {1\over n!} \GG |[\Psi_{\rm vac}^n \Lambda_{k-1}]\rangle
=\OO(g_s^{k+1})\, ,
\ee
for any BRST invariant 
state $|\hat\phi\rangle\in\wt\HH_T$ of ghost number 3 and $L_0^+=0$. 
Therefore $\LL_k(\hat\phi)$ represents an obstruction to finding global
(super-)symmetry transformation parameter beyond order $g_s^{k-1}$.
A non-vanishing  $\LL_k(\hat\phi)$
signals spontaneous breakdown of the global (super-)symmetry
at order $g_s^{k}$, with the state  conjugate to
$|\hat\phi\rangle$ representing the candidate goldstone/goldstino state. 

Since one of the main goals of the paper will be to explore the 
possibility of spontaneous breakdown of global 
supersymmetry, it will be useful
to have 
a list of possible candidate 
states $|\goldc\rangle$ for $\hat\phi$ in the fermionic sector. 
For this we shall restrict our discussion to the heterotic string theory, but
generalization to type II string theories is straightforward.
In this case $|\goldc\rangle$ is an element of the BRST cohomology
carrying ghost number 3, picture number $-3/2$ and zero momentum.
The possible candidates  are of the form
\be \label{ep2}
\goldc \qquad \equiv \qquad -4(\p c + \bar\p \bar c) \bar c  c e^{-3\phi/2} V^f, \quad
-4(\partial c +\bar\partial \bar c) \bar c  
c\bar \partial^2 \bar c \, \p\xi e^{-5\phi/2}\, \hat\Sigma
\, ,
\ee
where $V^f$ is a dimension $(1,5/8)$ operator and $\hat\Sigma$ is a dimension 
$(0,5/8)$ operator -- both in the R-sector of the matter CFT -- carrying 
space-time  chirality consistent with GSO projection rules and satisfying
\be \label{etfv}
T_F(z) V^f(w) =\OO\left((z-w)^{-1/2}\right), \quad T_F(z) \hat\Sigma(w) 
=\OO\left((z-w)^{-1/2}\right)\, .
\ee
For simplicity
we have dropped the spinor indices.
$\hat\Sigma$ in fact represents the matter part of an operator 
in the zeroth order global supersymmetry transformation
parameter: $\Lambda_0=\lambda_0 = c e^{-\phi/2} \hat\Sigma$.
For SO(32) heterotic string theory compactified on Calabi-Yau 3-folds,
possible choices for 
$\hat\Sigma$ are listed in \refb{edefSigma}.
The $-4$ factors in the definition of these vertex operators 
will be useful later.
The subscript
`G' and the superscript `c' in $\goldc$
stands for the fact that these states are conjugate to candidate
goldstino states $\gold$:
\be \label{egoldcan}
\gold \qquad \propto \qquad \bar c c e^{-\phi/2} (V^{f})^c, \quad c \eta e^{\phi/2} \hat\Sigma^c
\ee
where $(V^{f})^c$ and $\hat\Sigma^c$ are conjugate operators of $V^f$ and
$\hat\Sigma$ in the matter sector, satisfying relations similar to
\refb{etfv}.  These are
BRST invariant 
states of ghost number 2 and picture number $-1/2$. A non-vanishing
$\LL_k(\goldc)$ would imply that the right hand side of \refb{etauk} has a 
component along $\gold$, leading 
to a failure to solving this equation and consequently
a breakdown of global supersymmetry.

For evaluation of \refb{esusy2} it is also useful to list the operators
\be \label{ep21}
\GG\goldc=\XX_0 \goldc \qquad = \qquad \bar c c \eta e^{\phi/2} V^f, \quad 
\bar c c \bar \partial^2 \bar c \, e^{-\phi/2}\, \hat\Sigma\, .
\ee
One can show that these are also non-trivial elements of the BRST 
cohomology.

Given a string field configuration $|\Psi_{\rm vac}\rangle$  satisfying
\refb{eeom}, we define
\ben \label{eredefined}
&& \{ A_1\cdots A_k\}'' \equiv \sum_{n=0}^\infty {1\over n!} \, \{\Psi _{\rm vac}^n A_1\cdots A_k\}\, ,
\qquad \hbox{for $k\ge 3$}\, ,
\nonumber \\
&& [A_1\cdots A_k]'' \equiv \sum_{n=0}^\infty {1\over n!} \, [\Psi _{\rm vac}^n A_1\cdots A_k]\, ,
\qquad \hbox{for $k\ge 2$}\, , \nonumber \\
&& \{A_1\}'' \equiv 0, \qquad [~]''\equiv 
0,
\qquad \{A_1A_2\}''\equiv 0, \quad  [A_1]''\equiv 0\, , \nonumber \\
&& \wh Q_B|A\rangle \equiv Q_B|A\rangle + \sum_{k=0}^\infty {1\over k!}
\GG [\Psi_{\rm vac}^k A]\, .
\een
$\wh Q_B$ defined in \refb{eredefined} can be expressed
as
\be \label{edefK}
\wh Q_B = Q_B + \GG\, K\,  ,
\qquad
K\, |A\rangle \equiv \sum_{k=0}^\infty {1\over k!}
[\Psi_{\rm vac}{}^k A]\, .
\ee
$\wh Q_B$ and $K$
act naturally on states in $\wh\HH_T$ defined in \refb{erestrict}.
Using the definition of $K$ given in \refb{edefK}, 
the equations of motion \refb{eeom} satisfied by $|\Psi_{\rm vac}\rangle$,
and the identities \refb{ecurlysym}-\refb{eneweq2a} one can prove the
following useful identity:
\be \label{eqbk}
Q_B K + K Q_B + K\GG K =0\, .
\ee
{}From this and \refb{eqbg} it follows that
\be \label{eqbhsq}
\wh Q_B{}^2 = 0\, .
\ee
Using 
\refb{ecurlysym}-\refb{eeom} one also
finds the identities
\be \label{ecurlysympp}
\{ A_1  A_2\cdots  A_{i-1} A_{i+1}  A_i A_{i+2} \cdots  A_N\}''
=(-1)^{\gamma_i \gamma_{i+1}} \{ A_1 A_2\cdots  A_N\}''\, ,
\ee
\ben \label{eimpidpp}
&&  \sum_{i=1}^N (-1)^{\gamma_1+\cdots \gamma_{i-1}}\{ A_1\cdots  A_{i-1} (\wh Q_B  A_i)
 A_{i+1} \cdots  A_N\}'' \nonumber \\
&=& -  
{1\over 2} \sum_{\ell,k\ge 0\atop \ell+k=N} \sum_{\{i_a;a=1,\cdots \ell\}, \{j_b;b=1,\cdots k\}\atop
\{i_a\}\cup \{j_b\} = \{1,\cdots N\}
}\sigma(\{i_a\}, \{j_b\})
\{ A_{i_1} \cdots  A_{i_\ell} \GG [ A_{j_1} \cdots  A_{j_k}]''\}''
\, 
\een
\be \label{esymmetrypp}
[ A_1\cdots  A_{i-1} A_{i+1}  A_i A_{i+2} \cdots  A_N]''
=(-1)^{\gamma_i \gamma_{i+1}} [ A_1\cdots  A_N]'' \, ,
\ee
and
\ben \label{emainpp}
&& \wh Q_B\GG [ A_1\cdots  A_N]'' + \sum_{i=1}^N (-1)^{\gamma_1+\cdots 
\gamma_{i-1}}\GG [ A_1\cdots  A_{i-1} (\wh Q_B  A_i)
 A_{i+1} \cdots  A_N]'' \nonumber \\
&=& -  \sum_{\ell,k\ge 0\atop \ell+k=N} \sum_{\{i_a;a=1,\cdots \ell\}, \{j_b;b=1,\cdots k\}\atop
\{i_a\}\cup \{j_b\} = \{1,\cdots N\}
}\sigma(\{i_a\}, \{j_b\})\, 
\GG [ A_{i_1} \cdots  A_{i_\ell} \GG\, [ A_{j_1} \cdots  A_{j_k}]'']'' \, .
\een

For future use, we also define 
\be \label{edefqbt}
\wt Q_B = Q_B + K\, \GG \, ,
\ee
where $K$ has been defined in \refb{edefK}. 
$\wt Q_B$ acts naturally on states in $\wt\HH_T$ defined 
in \refb{erestrict}.
It is easy to verify using \refb{eqbg},
\refb{edefK}
and \refb{edefqbt} 
that
\be \label{eqbtr}
\wh Q_B \GG =  \GG \wt Q_B\, .
\ee
Furthermore,
using  \refb{edefK} and \refb{eqbk} one can prove the
nilpotence of $\wt Q_B$ and the identities
\be \label{eabqb}
\langle A| c_0^- \wh Q_B|B\rangle = (-1)^{\gamma_A} 
\langle \wt Q_B A| c_0^- |B\rangle\, , \qquad 
\langle B| c_0^- \wt Q_B|A\rangle = (-1)^{\gamma_B} 
\langle \wh Q_B B| c_0^- |A\rangle \, .
\ee

If we expand the string field as $|\Psi\rangle =|\Psi_{\rm vac}\rangle+|\Phi\rangle$ then
the equations of motion for $|\Phi\rangle$ take the form
\be \label{efull}
\wh Q_B |\Phi\rangle + \sum_{n=2}^\infty {1\over n!} \GG [\Phi^n]'' = 0\, .
\ee 
Therefore the linearized equations of motion for $|\Phi\rangle$ are
\be \label{eqexp}
\wh Q_B |\Phi_{\rm linear}\rangle=0\, .
\ee
The spectrum of physical states around the vacuum 
solution $|\Psi_{\rm vac}\rangle$ is given by examining 
the momentum
$k$ carried by $|\Phi_{\rm linear}
\rangle$ for which \refb{eqexp} has solutions. There are 
families of solutions 
to \refb{eqexp} which exist for all momenta, -- these are pure gauge solutions
of the form $\wh Q_B|\Lambda\rangle$ for some $|\Lambda\rangle$. 
There are additional solutions which appear for definite values of $k^2$ -- these represent the
physical states and the values of $k^2$ at which these solutions appear give the
physical mass$^2$ of the states.

In \cite{1411.7478,1501.00988} 
we described a systematic procedure for finding the solutions to \refb{eqexp}
in a power series expansion in $g_s$. 
If $|\Phi_n\rangle$ denotes a solution to \refb{eqexp} 
to order $g_s^n$ then we determine
$|\Phi_n\rangle$ in the `Siegel gauge'\footnote{Siegel gauge here refers
to the gauge in which all states other than those projected by $P$
are annihilated by $b_0^+$. For describing solutions,
we shall follow this convention throughout
this paper not only for the projection operator $P$ but also for the
projection operator ${\bf P}$ introduced earlier.}
using the recursion relation:
\be \label{esa3}
|\Phi_0\rangle = |\phi_n\rangle, \qquad
|\Phi_{\ell+1}\rangle = -{b_0^+\over L_0^+} (1-P) \GG K |\Phi_\ell
\rangle + |\phi_{n}\rangle + 
\OO\left(g_s^{\ell+2}\right)\, ,  \quad \hbox{for} \quad 0\le \ell\le n-1\, ,
\ee
where $|\phi_{n}\rangle$ satisfies
\be \label{esa4}
P|\phi_{n}\rangle = |\phi_{n}\rangle\, , 
\ee
\be \label{esa5}
Q_B|\phi_{n}\rangle = - P 
\GG K |\Phi_{n-1}\rangle +\OO(g_s^{n+1}) \, .
\ee
Here $P$ is a projection operator defined as follows.
Let us choose a basis of states in the CFT which are momentum
eigenstates and also $L_0^+$ eigenstates. 
We define the mass level $m$ via the relation $L_0^+=(k^2+m^2)/2$
i.e. $m^2/2$ is the contribution to the $L_0^+$ eigenvalue from the 
internal part of the matter CFT and the oscillators of free matter and
ghost fields.
$P$ will denote the projection operator to states of some fixed mass
level $m$.
In this case \refb{esa5} at zeroth order,
i.e\ the equation $Q_B|\phi_0\rangle=0$, 
has
non-trivial 
solution  only when $k^2=-m^2$ and hence
$m$ gives the tree level
mass of the state. Therefore
the value of $k^2$ at which \refb{esa3}-\refb{esa5} have
perturbative solution describing a
physical state is expected to be of order $-m^2+\OO(g_s)$ after inclusion of 
quantum corrections. 
At this momentum
the states onto which we project by $P$ all have 
$L_0^+= (k^2+m^2)/2 \sim g_s$, while all other states will have $|L_0^+|\gsim 1$.  
The projection operator 
$(1-P)$ in \refb{esa3} ensures that
the $(L_0^+)^{-1}$ operator in \refb{esa3} 
never gives any  power of $g_s$ in the
denominator. As a result \refb{esa3} leads 
to a well defined expansion of $|\Phi_n\rangle$ in powers of $g_s$, expressing 
it as a linear function of $|\phi_n\rangle$. 
After solving for $|\Phi_n\rangle$ this way we solve 
\refb{esa4}, \refb{esa5} to determine $|\phi_n\rangle$.\footnote{It may 
seem somewhat strange that we first
determine $|\Phi_{\ell+1}\rangle$ for all $\ell$ 
between 0 and $n-1$ iteratively in terms of
$|\phi_{n}\rangle$ and determine $|\phi_n\rangle$ at the end at one step by solving
a linear equation in the subspace projected by $P$. 
The reason for this is that for the
physical states the allowed value of $k^2$ changes 
at each order. Since a small change
in $k$ is not described by small change in the vertex operator it is best not to 
compute $|\phi_n\rangle$ iteratively but compute it at one step at the very end.} 
Since for given momentum $P$ projects onto a finite dimensional
subspace of $\HH_T$, \refb{esa5} gives a finite set of linear equations.
It will have a set of solutions which exist for all $k$. These are 
of the form $P\wh Q_B|\Lambda\rangle$ 
for some ghost number 1 state $|\Lambda\rangle$
carrying momentum $k$, and are associated with
pure gauge states. There is also another class of solutions which
exist for specific values of $-k^2$. These describe 
physical states, with the value of $-k^2$ at which the solution exists
giving the physical mass$^2$.

For future reference we note that \refb{esa3}-\refb{esa5} can be written as
\be \label{esummary}
|\Phi_{\rm linear}\rangle 
= \sum_{\ell=0}^\infty \left(-{b_0^+ \over L_0^+} (1-P) \GG  
K\right)^\ell |\phi\rangle
= \left(1 + {b_0^+ \over L_0^+} (1-P) \GG K\right)^{-1}|\phi\rangle\, ,
\ee
\be
P|\phi\rangle = |\phi\rangle\, ,
\ee
\be \label{e2.49}
Q_B|\phi\rangle = - P \GG K |\Phi_{linear}\rangle\, ,
\ee
where $|\Phi_{\rm linear}\rangle$ denotes the 
solution to the linearized equations of 
motion to any order.
It is understood that
to calculate $|\Phi_{\rm linear}\rangle$ to any 
given order we have to expand the right
hand side of \refb{esummary}
to that order, substitute this into the right hand side of \refb{e2.49}
to get a linear equation for $|\phi\rangle$ in the subspace projected
by $P$, and then solve this equation to determine $|\phi\rangle$
and the momentum carried by $|\phi\rangle$. 

We shall conclude this section by introducing a common notation
for describing states projected by ${\bf P}$ and $P$.
${\bf P}$ projects on to the zero momentum
states with $L_0^+=0$. 
On the other hand for given 
momentum  $k$
at which some states become on-shell, $P$ projects on to states which have
$L_0^+=\OO(g_s)$. These are the states which would have $L_0^+=0$ 
at string tree level, but since the $k$-values shift due to mass renormalization
$L_0^+$ also shifts by terms of order $g_s$. We shall collectively call all
such states (including zero momentum states with $L_0^+=0$) as 
$L_0^+\simeq 0$ states. In this notation both ${\bf P}$ and $P$ project
on to $L_0^+\simeq 0$ states.

\sectiono{S-matrix from the action} \label{smatrix}

In this section we shall discuss the construction of {\it tree level} amplitudes
associated with the equations of motion \refb{efull}; this is supposed to give the
full quantum amplitudes in string theory.  We shall also compare the 
amplitudes obtained this way to the usual amplitudes in string theory expressed
as integrals of certain CFT correlation functions over the moduli spaces of Riemann
surfaces.

\subsection{Construction of off-shell amplitudes and S-matrix elements} \label{s3.1}

Even though the equations of motion can in principle be used for computing the
tree level S-matrix directly, we shall follow a simpler approach. Following 
\cite{1501.00988}
we shall first introduce an
additional set of fields in terms of which one can write down an action from which
the equations of motion can be derived, and then use this action to compute the
S-matrix elements. These additional
fields are described by a grassmann even state 
$|\wt\Psi\rangle\in\wt\HH_T$
carrying ghost number 2, with $\wt\HH_T$ defined as in \refb{erestrict}. 
We now consider the action
\be \label{eactpsi}
S = g_s^{-2}\left[
-{1\over 2} \langle\wt\Psi |c_0^- Q_B \GG |\wt\Psi\rangle 
+ \langle\wt\Psi |c_0^- Q_B |\Psi\rangle + 
\sum_{n=1}^\infty {1\over n!} \{ \Psi^n\}
\right]\, .
\ee
It is easy to see that the action 
\refb{eactpsi} is invariant under the infinitesimal gauge transformation 
\be \label{egaugepsi}
|\delta\Psi\rangle = Q_B|\Lambda\rangle + \sum_{n=0}^\infty {1\over n!} 
\GG [\Psi^n \Lambda]\, , \quad
|\delta\wt\Psi\rangle = Q_B |\wt\Lambda\rangle + \sum_{n=0}^\infty {1\over n!} 
[\Psi^n \Lambda]\, ,
\ee
where $|\Lambda\rangle\in\wh\HH_T$, $|\wt\Lambda\rangle\in \wt\HH_{T}$, 
and both carry ghost number 1. 
The equations of motion derived from \refb{eactpsi} can be written as
\be \label{e01psi}
Q_B (|\Psi\rangle - \GG|\wt\Psi\rangle) = 0\, ,
\ee
\be \label{e02psi}
Q_B |\wt\Psi\rangle + \sum_{n=1}^\infty {1\over (n-1)!} [\Psi^{n-1}]= 0\, .
\ee
Applying $\GG$ on \refb{e02psi} and using  \refb{e01psi} 
we recover the equation of
motion \refb{eeom} of $|\Psi\rangle$.  

Given a vacuum solution 
$|\Psi_{\rm vac}\rangle$ in the NS sector satisfying \refb{eeom}, 
we can find a solution to 
\refb{e01psi} and \refb{e02psi} by setting 
$|\wt\Psi_{\rm vac}\rangle=|\Psi_{\rm vac}\rangle$ since
$\GG|\Psi_{\rm vac}\rangle=|\Psi_{\rm vac}\rangle$. 
Defining shifted
fields
\be \label{eshiftedphi}
|\Phi\rangle = |\Psi\rangle - |\Psi_{\rm vac}\rangle, \qquad
|\wt\Phi\rangle = |\wt\Psi\rangle - |\wt\Psi_{\rm vac}\rangle = 
|\wt\Psi\rangle - |\Psi_{\rm vac}\rangle
\ee
the action \refb{eactpsi} and the gauge transformation laws 
\refb{egaugepsi} can be written as
\be \label{eact}
S = g_s^{-2}\left[
-{1\over 2} \langle\wt\Phi |c_0^-  Q_B \GG |\wt\Phi\rangle 
+ \langle\wt\Phi |c_0^- Q_B |\Phi\rangle +  {1\over 2} \langle \Phi| c_0^- K|\Phi\rangle
+ 
\sum_{n=3}^\infty {1\over n!} \{ \Phi^n\}''
\right]\, ,
\ee
\be
|\delta\Phi\rangle = \wh Q_B|\Lambda\rangle + \sum_{n=1}^\infty {1\over n!} 
\GG [\Phi^n \Lambda]''\, , \quad
|\delta\wt\Phi\rangle = Q_B |\wt\Lambda\rangle  + K|\Lambda\rangle 
+ \sum_{n=1}^\infty {1\over n!} 
[\Phi^n \Lambda]''\, ,
\ee
with $K$ defined as in \refb{edefK}.
The equations of motion derived from \refb{eact} are
\be \label{e01}
Q_B (|\Phi\rangle - \GG|\wt\Phi\rangle) = 0\, ,
\ee
\be \label{e02}
Q_B |\wt\Phi\rangle + K|\Phi\rangle 
+ \sum_{n=3}^\infty {1\over (n-1)!} [\Phi^{n-1}]'' = 0\, .
\ee
Applying $\GG$ on \refb{e02} and using \refb{edefK}, \refb{e01} we recover the equation of
motion \refb{efull} of $|\Phi\rangle$. 

Even though this 1PI action gives the correct equations of motion, 
it has too many degrees of freedom. For example physical 
states in the R-sector will arise both in picture number $-1/2$ and
picture number $-3/2$ sectors. We avoid this by imposing a 
constraint\footnote{This kind of doubling trick accompanied by a 
constraint has been attempted before in the context of Berkovits' open string
field theory\cite{0412215}.}
\be \label{enewcons}
\GG|\wt\Psi\rangle -|\Psi\rangle = 0 \quad \Rightarrow \quad
\GG|\wt\Phi\rangle - |\Phi\rangle = 0
\ee
on the external states. Since this is consistent with the equations of
motion and we are only doing a tree level computation with the 1PI
effecting action, this is a 
consistent truncation. One might wonder if this imposes some additional
restriction on the external states 
compared to the restrictions imposed by
\refb{eqexp}; we have shown in appendix \ref{enewappendix} that as far
as perturbative amplitudes are concerned, \refb{enewcons} does not
give any additional constraint besides \refb{eqexp}.

While determining the Feynman rules for computing tree amplitudes from
the action \refb{eact}, we shall 
use the weight factor of $e^S$ in the path integral. 
Therefore $n$-point vertices in a Feynman diagram with external state
$|A_1\rangle,\cdots |A_n\rangle$ are given directly 
by $g_s^{-2}\{A_1\cdots A_n\}''$, while the propagator is negative of 
the inverse of
the kinetic operator giving the quadratic part of $S$.
If we had
used $e^{-S}$ or $e^{iS}$ then the Feynman rules for the propagators and
vertices will have additional phase. The effect of this will be to change 
each of the tree amplitudes computed from this action by an overall phase
factor, without affecting the relative phases of different terms contributing to a
given amplitude. This is related to the fact that tree level amplitudes computed
from an action are really sensitive only to the equations of motion and
not the normalization of the action. 

In particular we shall see later (see last paragraph of \S\ref{sreal}) that
in our convention the kinetic operator of a field of mass $m$ is  
$-(k^2+m^2)$ up to a positive constant of proportionality. Therefore $S$ has
the correct sign for being interpreted as the action for Lorentzian
signature, and the
correct weight factor in the path integral for Lorentzian signature 
is $e^{i \, S}$.
The amplitudes computed with this rule can be obtained by 
multiplying the amplitudes computed here by a factor of $i$.

We  gauge fix the theory in the Siegel gauge $b_0^+|\Phi\rangle=0$,
$b_0^+|\wt \Phi\rangle=0$. In this gauge $Q_B=c_0^+ L_0^+$ and
the kinetic term of the action \refb{eact} in the ($|\wt\Phi\rangle$, $|\Phi\rangle$)
space takes the form
\be
g_s^{-2}\left[c_0^- c_0^+ L_0^+ \pmatrix{-\GG & 1\cr 1 & 0} + c_0^-
\pmatrix{0 & 0 \cr
0 & K}\right]
\, .
\ee
Inverting this and multiplying by $-1$ we get the propagator
\be \label{esignprop}
- g_s^2 \pmatrix{\check\Delta & \bar\Delta \cr \wt\Delta & \Delta}
\ee
where
\be \label{esiegel-}
\check\Delta = \left[ - {b_0^+\over L_0^+}  K {b_0^+\over L_0^+} 
+ {b_0^+\over L_0^+}  K {b_0^+\over L_0^+} \GG K {b_0^+\over L_0^+} 
+\cdots\right]
b_0^- \delta_{L_0^-}\, ,
\ee
\be\label{esiegela}
\bar\Delta = \left[{b_0^+\over L_0^+}  - {b_0^+\over L_0^+}  K {b_0^+\over L_0^+} \GG 
+ {b_0^+\over L_0^+}  K {b_0^+\over L_0^+} \GG K {b_0^+\over L_0^+} \GG+\cdots\right]
b_0^- \delta_{L_0^-}\, ,
\ee
\be \label{esiegelb}
\wt\Delta = \left[{b_0^+\over L_0^+} - {b_0^+\over L_0^+} \GG K {b_0^+\over L_0^+} 
+ {b_0^+\over L_0^+} \GG K {b_0^+\over L_0^+} \GG K {b_0^+\over L_0^+} +\cdots\right]
b_0^- \delta_{L_0^-}\, , 
\ee
\be \label{esiegel}
\Delta = \left[{b_0^+\over L_0^+} \GG - {b_0^+\over L_0^+} \GG K {b_0^+\over L_0^+} \GG 
+ {b_0^+\over L_0^+} \GG K {b_0^+\over L_0^+} \GG K {b_0^+\over L_0^+} \GG+\cdots\right]
b_0^- \delta_{L_0^-}\, .
\ee
The minus sign in \refb{esignprop} is a reflection of the fact that we use
$e^S$ as the weight factor in the path integral rather than $e^{-S}$.
$\check\Delta$, $\bar\Delta$, $\wt\Delta$ and $\Delta$ act naturally on states 
in $\wh\HH_T$, $\wt\HH_T$, $\wh\HH_T$ and $\wt\HH_T$
to produce states
in $\wt\HH_T$, $\wt\HH_T$, $\wh\HH_T$ and $\wh\HH_T$
 respectively.
One important property of $\Delta$ that will be useful later is the relation
\be \label{econdprop}
\wh Q_B \Delta c_0^- + \Delta c_0^- \wt Q_B = \delta_{L_0^-}\, \GG\, ,
\ee
acting on states in $\wt\HH_T$. This can be derived using \refb{edefqbt},
\refb{eqbk} and
other well-known (anti-)commutators involving $Q_B$.

Our goal will be to compute off-shell Green's functions with external propagators
truncated, involving external $\Phi$ fields, 
using the propagator \refb{esiegel} and the interaction term given by the last term
in \refb{eact}. Since the interaction term as well the external states involve only the 
$\Phi$ fields, only the $\Delta$ term in the propagator is relevant for our computation.
Therefore from now on in our computation we can forget about the $\wt\Phi$ fields
altogether and work only with $\Phi$ fields with propagator $-g_s^2
\Delta$ and interaction terms
$g_s^{-2}\sum_{n=3}^\infty \{\Phi^n\}''/n!$. 
If we denote by 
$g_s^{-2}\Gamma^{(N)}(
| A_1\rangle, \cdots  | A_N\rangle)$ the truncated off-shell Green's function
with external
states $| A_1\rangle,\cdots | A_N\rangle\in \wh\HH_T$ with ghost number 
2, then $g_s^{-2}
\Gamma^{(N)}$
is computed by drawing
all possible
{\it tree} graphs with the states $| A_1\rangle,\cdots | A_N\rangle$ as external states,
$g_s^{-2}\{\Phi^n\}''/n!$ for $n\ge 3$
as vertices and propagator $-g_s^2\Delta$ given in \refb{esiegel}. 
The normalization factor $g_s^{-2}$  in the definition
of $\Gamma^{(N)}$ is by convention, and
has been included so that the 1PI contribution to
$\Gamma^{(N)}(|A_1\rangle,\cdots |A_N\rangle)$ 
is given by $\{A_1\cdots A_N\}$ without any further normalization factor.
S-matrix elements are obtained from $\Gamma^{(N)}$ by setting the external states on-shell
and multiplying the result by $\prod_{i=1}^N (Z_i)^{-1/2}$ where $Z_i$ is the wave-function
renormalization factor associated with the $i$-th external state.

\subsection{Comparison with the usual formulation} \label{susual}

We shall use the description of the S-matrix elements given in \S\ref{s3.1} for most
of our analysis. Nevertheless 
it is useful to compare this definition of S-matrix 
to the usual description  where at each loop order a given amplitude
has a single term involving integration over the full moduli space.
This is what we shall do in this subsection. A summary of the results
of this subsection has been given in \S\ref{ssummary3}; so readers interested
in only the final results can skip this subsection.

If instead of working with the propagator $\Delta$, the interaction
$\sum_{n\ge 3} \{\Phi^n\}''/n!$ and the definition of on-shell states given in
\refb{eqexp},
we had used the unshifted action \refb{eactpsi} to determine the
propagator and interactions 
and used the definition
of on-shell states given by $Q_B|\Psi\rangle=0$, 
then the propagator of the $|\Psi\rangle$ field will be given 
by\cite{1501.00988}
\be \label{edefdelta0}
\Delta_0 \equiv b_0^+ b_0^- \GG (L_0^+)^{-1}\delta_{L_0^-}\, ,
\ee
and the interactions will be given by the 1PI amplitudes in the unshifted
background.
It then follows from standard
results in string field theory (see {\it e.g.} \cite{GMW,9206084}) 
that the on-shell amplitudes computed this way would agree
with the standard amplitudes of string theory. This would of course 
suffer the same divergences associated
with mass and wave-function renormalization and massless tadpoles
as discussed {\it e.g.} in \cite{1209.5461,1304.2832}. Our goal will
be to bring the expression for the S-matrix given in \S\ref{s3.1} as close 
as possible to the expression
involving integration over the full 
moduli space of Riemann surfaces, and at the same
time point out the crucial differences that removes the divergences in the latter
expression associated with mass renormalization and massless tadpoles.

Our strategy will be to carry out a detailed comparison
of the prescription of \S\ref{s3.1} 
with the one where we use
the unshifted action \refb{eactpsi} and solutions to the 
linearized equations $Q_B|\Psi\rangle=0$.
One of the main differences between these two approaches is that 
term $\sum_{n=3}^\infty \{\Phi^n\}''/n!$ in the action does not contain
any one point or two point vertices while 
$\sum_{n=1}^\infty \{\Psi^n\}/n!$ does contain such terms. 
As we shall discuss, at 
a crude level the absence of the $\{\Phi\}''$ term  is
compensated by inclusion of the shift by $|\Psi_{\rm vac}\rangle$ in 
the definition of the 
vertices $\{\Phi^n\}''/n!$ for $n\ge 3$,
while the absence of the $\{\Phi^2\}''/2!$ term
is compensated for by the use of the
modified propagator \refb{esiegel} that includes loop corrections 
encoded in $K$ and
also in the fact that external states are taken to be annihilated by 
$\wh Q_B$ instead of
$Q_B$. Our goal will be to explore to what extent these compensations 
fail to be exact since that encodes the difference between the two approaches.

For clarity we shall divide our analysis into two parts.
\begin{enumerate}
\item First we shall suppose for the sake of argument that $\{A\}=0$ 
(or equivalently
$[~]=0$) so that
there is no constant term in the equations of motion \refb{eeom}  
and we can set  $|\Psi_{\rm vac}\rangle=0$. In this case the one point
function $\{\Psi\}$ vanishes  even in the conventional approach,
and the interaction terms $\{\Psi^n\}/n!$ and $\{\Phi^n\}''/n!$ 
give the same vertex
for $n\ge 3$. Therefore
the difference between the conventional approach and our approach
lies in the 
two point vertex computed from
$\{\Psi^2\}/2!$ which is present in the conventional 
approach but is absent in our approach. 
We shall examine how use of the modified
propagator $\Delta$ instead of $\Delta_0$ and modified external states
compensate for it.
\item Next we shall relax the assumption that $\{A\}=0$ so that 
$|\Psi_{\rm vac}\rangle$ no longer vanishes. Now the conventional approach will
have also one point function that is absent in our approach. But the interaction
vertices computed from the $\{\Phi^n\}''/n!$ term in our approach now differ
from the interaction vertices computed from
$\{\Psi^n\}/n!$ in the conventional approach.
We shall examine to what extent these two effects compensate each other.
\end{enumerate}

So let us begin with the first step, assuming that $\{A\}=0$, $|\Psi_{\rm vac}
\rangle=0$. In this case 
$\langle B| c_0^- K|A\rangle=\langle B| c_0^- |[A]\rangle=\{BA\}$ 
represents the usual two point  amplitude in the conventional approach
based on \refb{eactpsi}.
Therefore  the
two point function corresponds to the operator $c_0^-K$ and 
has to be inserted between a pair of internal propagators $\Delta_0$. 
Summing over multiple insertions of this type 
would convert the internal propagators to
\be\label{edd0}
-\Delta_0 + \Delta_0 (c_0^- K) \Delta_0 -  \Delta_0 (c_0^- K) \Delta_0 (c_0^- K) \Delta_0 + \cdots\, .
\ee
This is precisely the
propagator $-\Delta$ given in \refb{esiegel}. 

This shows that in our approach,
the effect of using the propagator $-\Delta$ instead of
$-\Delta_0$ precisely compensates for the missing two point vertices 
inserted on the internal lines in the conventional approach.
However since the
definition of $\Gamma^{(N)}$ involves truncating the full external propagator 
$-\Delta$,
while the usual string amplitudes would correspond to first expressing $\Delta$ 
as in \refb{edd0} and then truncating the left-most $-\Delta_0$, the prescription of
\S\ref{s3.1} would appear to
be missing the contributions involving (multiple) 
insertion of the two point function 
$(c_0^-K)$ on the external legs.
The result can still be expressed as an integral over the moduli space, but the
integration runs over only a subspace of the full moduli space.

To understand how these missing contributions arise in our formalism, we have to
recall that in the computation of the S-matrix element the external states are taken
to be annihilated by $\wh Q_B$ and not by $Q_B$. For states which undergo mass
renormalization this shifts the on-shell value of the momentum. However this is not
the only difference -- there are other differences which affect even states which do
not undergo mass renormalization. For this we recall that in \refb{esa3}-\refb{esa5} 
we described a
systematic way of constructing solution to the $\wh Q_B|\Phi_{\rm linear}
\rangle=0$ equation in 
a power series in the string coupling $g_s$. 
Let us examine how the modification of the external state as described there
affects the computation of the S-matrix.  
As described in \refb{esummary}, 
the equation \refb{esa3}
for $|\Phi_{\rm linear}\rangle$ can be summarized as
\be \label{ex.1}
|\Phi_{\rm linear}\rangle 
= \sum_{\ell=0}^\infty (-\Delta_0' c_0^- K)^\ell |\phi\rangle
= (1 + \Delta_0' c_0^- K)^{-1}|\phi\rangle\, .
\ee
where
\be \label{emodified}
\Delta_0' = {b_0^+ b_0^-\over L_0^+} \delta_{L_0^-} (1-P) \GG\, ,
\ee
denotes a modified propagator in which we have removed 
the contributions from a subset
of states using the projection operator $(1-P)$. While expressing 
\refb{esummary} as \refb{ex.1} we have used $\delta_{L_0^-}K=K$,
$b_0^-K=0$.
\refb{ex.1}
denotes repeated application of $-\Delta_0'$
and the two point vertex $c_0^-  K$. 
If we replace $\Delta_0'$ by $\Delta_0$, and 
$|\phi\rangle$ by a $Q_B$ invariant state
in \refb{ex.1}, then \refb{ex.1} would exactly supply the missing 
contribution involving
insertions of $c_0^-K$ on external legs that is needed to make our
prescription agree with the usual amplitude computed in string theory.
However since $\Delta_0'$ is not $\Delta_0$ and $|\phi\rangle$ to any
given order in $g_s$ is obtained by solving eqs.\refb{esa3}-\refb{esa5}
instead of $Q_B|\phi\rangle=0$, our prescription 
is not exactly the same
as that used for computing usual string amplitude.
Due to this projector $(1-P)$ in $\Delta_0'$ our
procedure is free from the infrared divergences associated with
mass renormalization. The price we pay is that the external state that has to be
used for computing the amplitude is $|\phi\rangle=
P|\Phi_{\rm linear}\rangle$ obtained by solving \refb{esummary}-\refb{e2.49}. 
Even though this is a linear
combination of the states at the same mass level as the state that we have at the
string tree level, this is not in general a BRST invariant state.

Now let us consider the general case when $\{A\}$ and $[~]$ are non-zero
and hence  $|\Psi_{\rm vac}\rangle$ is also non-zero. 
Since by Lorentz invariance $\{A\}$ vanishes when $|A\rangle\in\HH_{R}$,
we can take $|\Psi_{\rm vac}\rangle$ to be in the NS sector.
Now the rules for computing the amplitude described in \S\ref{s3.1} differ from the
conventional prescription based on the action \refb{eactpsi} expanded 
around $|\Psi\rangle=0$ on two
counts. First of all in the conventional approach we have to 
include the effect of one point function $\{\Psi\}$ in 
the computation of the amplitudes.
This will require us to include tadpole graphs which are absent in
our approach. Second, in our approach the vertex
uses $\{\Phi^n\}''/n!$ whereas in the unshifted background the vertex uses 
$\{\Psi^n\}/n!$. 
Now using the vertex computed from $\{\Phi^n\}''/n!$ 
is equivalent to using the vertex computed from
$\{\Psi^n\}/n!$ but inserting arbitrary number
of  $\Psi_{\rm vac}$ in the amplitude.
To analyze the effect of these insertions we recall that 
in \refb{esoln}, \refb{esol1} we described
an algorithm 
for computing $|\Psi_{\rm vac}\rangle$ in a power series expansion in the string coupling.
It is easy to convince oneself that if in \refb{esoln} 
we did not have the $|\psi_{k+1}\rangle$ term
and replaced $(1-{\bf P})$ by 1, then  arbitrary number of insertions of $|\Psi_{\rm vac}\rangle$
in the amplitude in our approach
will generate back the missing tadpole contributions that arise in the
conventional approach. However the
presence of the projector $1-{\bf P}$ in our approach
has the effect that it removes the contribution
of the $L_0^+=0$ states from the zero momentum propagator which will arise in the
computation of the tadpole diagrams, thereby rendering the amplitude manifestly
free from infrared divergences associated with propagators of massless states 
carrying zero momentum. However we have to now
supplement this by inserting into the amplitude
arbitrary number of insertions of $L_0^+=0$ states
$|\psi\rangle$, whose order $g_s^{k+1}$ expression $|\psi_{k+1}\rangle$
is obtained 
by solving \refb{esol1}. Since $|\psi\rangle$ has a power series
expansion in $g_s$ starting at order $g_s$ or higher power of $g_s$, in any given
order in $g_s$ we only need a finite number of $|\psi\rangle$ insertions. 

\subsection{Orientation of the moduli space} \label{sorientation}

Once we have fixed our conventions, the amplitudes
are determined including the overall sign and hence comes with a
prescription for how to choose the orientation of the moduli space over
which we integrate. We shall now illustrate how this works for 
the genus zero four point function. Since the picture changing operators
play no role in this analysis we shall carry out the analysis in bosonic string
theory. Identical results will hold in superstring theory. (What we refer
to as matter sector here will stand for matter plus the $\beta$, $\gamma$
ghost system in superstring theory.)

Let us suppose that we have four on-shell vertex operators $\VV_i$ of the
form $\bar c c V_i$ where $V_i$ are matter sector primaries of dimension 
(1,1). Now the 1PI part of the four point amplitude has the form
\be \label{eori1}
{1\over 2\pi i} \int dz \wedge d\bar z \, \left\langle
\ointop_z b(w) dw \ointop_z d\bar w \bar
b(\bar w) \VV_1(z_1) \VV_2(z_2) \VV_3(z_3) \VV_4(z) \right\rangle
\ee
where $\ointop_z$ denotes integration along a contour enclosing $z$ with the
normalization convention $\ointop_z dw/(w-z)=1$,  $\ointop_z d\bar w/(\bar w
-\bar z)=1$. The overall normalization in \refb{eori1} is part of the
definition of the normalization convention of $\{A_1\cdots A_N\}$ --
the complete factor being $(2\pi i)^{-(3g-3+n)}$ for a $g$-loop 
$n$-point contribution. Now we can carry out the contour integrals
yielding the result
\be \label{eori2}
{1\over 2\pi i} \int dz \wedge d\bar z \, \left\langle
 \VV_1(z_1) \VV_2(z_2) \VV_3(z_3) V_4(z) \right\rangle\, .
\ee
Let us suppose that we do not {\it a priori} know what the
orientation of the sphere labelled by $z$ is. In that case we can
write
\be \label{eori3}
dz\wedge d\bar z = -2\, i\, d ({\rm Re} z) \wedge d({\rm Im} z)
= - 2\, i\, \ve\, d^2 z
\ee
where $d^2z$ by definition is an integration measure whose integral over
any subspace of the $z$-plane gives positive result. $\ve$ is a sign
which we want to determine.

A direct determination of $\ve$ requires recalling some part of the
analysis in \cite{1408.0571,1411.7478,1501.00988}. 
There in the proof of various
factorization identities of the integration measure
 we had
used the plumbing fixture relation $zw=q$ with $q=\exp(-s+i\theta)$ and
had taken $ds\wedge d\theta$ to define positive integration measure.
This would correspond to $dq\wedge d\bar q$ describing $i$ times a positive
integration measure. Since $z$ is related to $q$ by an analytic transformation,
we see that $dz\wedge d\bar z$ should also correspond to $i$ times a 
positive integration measure and hence we must have $\ve=-1$. This gives
\be \label{eorifin}
dz \wedge d\bar z = 2\, i\, d^2 z\, .
\ee

We shall now verify this by directly analyzing the amplitudes. 
Even though \refb{eori2} is supposed to describe the 1PI part of the vertex,
the contribution from the 1PR parts, obtained by gluing two three point
vertices by a propagator, should have the same form with the integration
over $z$ covering different parts of the moduli space. Together they will
cover the whole moduli space. 
Now suppose that there is a matter sector vertex operator $V$ of dimension
$(1+h,1+h)$ for small $h$ that appears in the operator product of 
$V_4(z)$ and $V_3(z)$
\be \label{eoriope}
V_4(z) V_3(z_3) = A\, (z-z_3)^{-1+h} (\bar z -\bar z_3)^{-1+h}
V(z_3) +\cdots
\ee
where $A$ is some coefficient. We shall assume that $V$ is normalized as
\be  \label{evnorm}
V(z) V(w) = (z-w)^{-2-2h} (\bar z - \bar w)^{-2-2h}\, .
\ee
In this case the integral \refb{eori2} will receive a contribution from the 
$z\simeq z_3$ region of the form
\be
-{1\over \pi} \, \ve \, A\, \int d^2 z  (z-z_3)^{-1+h} (\bar z -\bar z_3)^{-1+h}
\left\langle
 \VV_1(z_1) \VV_2(z_2) \VV(z_3) \right\rangle
 \ee
 where
 \be
 \VV(z_3) = \bar c  c V(z_3)\, .
 \ee 
 This can be expressed as
 \be \label{eori4}
-{1\over \pi} \, \ve\, A\, \left\langle
 \VV_1(z_1) \VV_2(z_2) \VV(z_3) \right\rangle
 \int_0^1 2\pi r dr r^{-2+2h} 
 \simeq -\ve\, A\,  h^{-1} \, \left\langle
 \VV_1(z_1) \VV_2(z_2) \VV(z_3) \right\rangle \, .
\ee

Let us now compare this with what we expect from the 1PR amplitude in
the effective field theory. The contribution comes from the diagram
where a three point vertex containing $\VV_1$, $\VV_2$ and $\VV$
is connected by a propagator to a three point vertex connecting
$\VV_3$, $\VV_4$ and $\VV$.  If we denote by $\VV^c$ the state
conjugate to $\VV$:
\be \label{enucnorm}
|\VV^c\rangle \equiv c_0\bar c_0 \bar c_1 c_1|V\rangle,
\quad \langle \VV^c|\VV\rangle =1\, ,
\ee
then, using the fact that the zeroth order propagator is given by 
$-b_0^+b_0^- \delta_{L_0^-} (L_0^+)^{-1}$, we get the 
1PR contribution to the four point function to be
\be \label{eori5}
-\langle \VV_1 (z_1)\VV_2(z_2) \VV(w_1) \rangle \langle \VV(w_2)
\VV_3(z_3) \VV_4(z_4)\rangle 
\langle \VV^c | b_0^+ b_0^- \delta_{L_0^-} (L_0^+)^{-1}
|\VV^c\rangle
\ee
where $w_1$ and $w_2$ are arbitrary points on which the result
does not depend to leading order in $h$. Using the result
$b_0^+b_0^-=2\bar b_0 b_0$, $L_0^+ |\VV^c\rangle=2h|\VV^c\rangle$
and eqs.\refb{evnorm}, \refb{enucnorm}, \refb{eoriope},
\refb{evacnorm} we can express \refb{eori5} as
\be \label{eori6}
A\, h^{-1} \, \langle \VV_1 (z_1)\VV_2(z_2) \VV(z_3) \rangle 
\ee
Comparing this with \refb{eori4} we see that we must have 
$\ve=-1$.

We could recover the more standard convention $dz\wedge d\bar z=-2i d^2 z$
by including an additional factor of $(-1)^{3g-3+n}$ in the definition of
$\{A_1\cdots A_n\}$ since the latter involves integration over 
$3g-3+n$ dimensional complex manifold. This would 
replace the normalization factor
$(2\pi i)^{-(3g-3+n)}$ in the definition of $\{A_1\cdots A_n\}$ by
$(-2\pi i)^{-(3g-3+n)}$.
We shall not attempt to do so in this paper.

\subsection{Reality conditions on string fields} \label{sreal}

In computing tree level Feynman amplitudes from a field theory we
do not need to know what the reality condition on the fields are -- the path
integral over various fields can be taken to run over arbitrary 
contours passing through the origin in
the complexified field space. However when
we discuss the choice of vacuum solutions encoded in the
$a_\alpha$'s entering \refb{emarg}, then in order
to see what classical backgrounds are physically possible we need to 
know the reality conditions on the $a_\alpha$'s. 
This is the issue we shall
address now.
For definiteness we shall restrict our analysis
to the heterotic string theory, but the generalization to type II string theories
is straightforward.

We shall choose the basis $|\vp_\alpha\rangle$ to be normalized as
\be \label{eno0}
\langle \vp_\alpha | c_0^- c_0^+ | \vp_\beta\rangle = -{1\over 2}
\delta_{\alpha\beta}\, .
\ee
The reason for choosing the $-$ sign will be apparent soon. Most zero momentum
BRST invariant states satisfying this normalization are of the form
\be \label{eno1}
|\vp_\alpha\rangle = \bar c c e^{-\phi} V_\alpha (0) |0\rangle\, ,
\ee
where $V_\alpha$ is a dimension (1,1/2) superconformal primary in the
matter sector satisfying 
\be \label{eno2}
V_\alpha(z) V_\beta(w) = (z-w)^{-1} (\bar z-\bar w)^{-2} +\cdots \, ,
\ee
where $\cdots$ denotes less singular terms.
Using \refb{eghope},
\refb{evacnorm}, \refb{edefbpm}
and the fact that $e^{-\phi}$ anti-commutes with $V_\alpha$
we can easily verify that \refb{eno2} implies \refb{eno0}

Let us for definiteness suppose that we are analyzing a four point function
of the form given in \refb{eori1}, but in heterotic string theory instead of
in the bosonic string theory.  
Let us take $\VV_4$ to be the operator $\vp_\alpha$ given in \refb{eno1},
and the other three vertex operators to be some fixed operators in the
NS sector.
In that case we need to insert 
two PCO's. We shall take the location of one of them  
to coincide with the location of the vertex $\VV_4$. This converts 
$\VV_4=\vp_\alpha$
to the form 
\be \label{enewvert}
\XX_0 |\vp_\alpha\rangle = \bar c c \wt V_\alpha (0) |0\rangle
-{1\over 4} \bar c \eta e^\phi V_\alpha(0)|0\rangle\, ,
\ee
where 
\be \label{epicchange}
|\wt V_\alpha\rangle = -\ointop dz T_F(z) |V_\alpha\rangle \equiv -G_{-1/2}
|V_\alpha\rangle\, ,
\ee
and $G_n\equiv \ointop dz \, z^{n+1/2}T_F(z)$.
It follows from \refb{eno2} and $\{G_{1/2},G_{-1/2}\}=(L_0)_{matter}/2$
 that
\be \label{enormV}
\wt V_\alpha(z) \wt V_\beta(w) ={1\over 4}
 (z-w)^{-2} (\bar z - \bar w)^{-2} +\cdots \, .
\ee
Let us consider the effect of the first 
term on the right hand side of \refb{enewvert}. Following the steps that
led from \refb{eori1} to \refb{eori2} we see that 
the contour integrals of $b$ and $\bar b$ remove the $\bar c c$ factor from
the vertex operator 
and leaves us with only the operator $\wt V_\alpha$.
It now follows from  \refb{eori2} and \refb{eorifin}
that the effect of 
switching on a background of the form given in \refb{emarg}
is to insert $(1/\pi) \sum_\alpha a_\alpha
\int d^2 z \wt V_{\alpha}(z,\bar z)$ into the CFT correlation functions. 
From the normalization condition
\refb{enormV} it follows that this is an allowed deformation of the CFT
for real $a_\alpha$.
Therefore a field
configuration of the form given in \refb{emarg} should be declared real for 
real $a_\alpha$.

As an aside we note from \refb{eno0}, \refb{evacnorm} and the fact that
the $L_0^+$ eigenvalues of the GSO even states in $\wh H_T$ are of
the form $(k^2+m^2)/2$ with positive constant $m^2$, that the kinetic term
of the NS sector states in the action given in \refb{eactpsi} comes with 
coefficient proportional to $-(k^2+m^2)$ multiplying the
square of the real fields. Therefore $e^S$ has damped kinetic term for large
real values of the fields. This is the conventional definition of the euclidean
path integral; so our action $S$ is really the negative of the conventional
euclidean action. 
Even though we shall not use the result, it may be useful to keep this in
mind for future applications.

\subsection{Summary} \label{ssummary3}

Since the analysis has been 
somewhat technical, we shall now summarize the main results of
this section. Our main aim will be to explain how the prescription for computing
S-matrix elements that
we arrive at differs from the usual Polyakov prescription for computing
amplitudes. 
\begin{enumerate}
\item 
The usual Polyakov prescription expresses an amplitude as
an integral over the moduli space of Riemann surfaces with punctures, 
with the integrand
given by certain correlation functions of vertex operators inserted at
the punctures, ghost fields and
picture changing operators inserted in an appropriate way
as explained in \cite{1504.00609}. 
For describing our modified prescription we first 
need to divide the integral over moduli space into separate
sectors containing 1PI contributions and various 
one particle reducible
(1PR) contributions. The latter contain 
Riemann surfaces obtained by
gluing 1PI Riemann surfaces using plumbing fixture relation
\be
zw = e^{-s+i\theta}, \quad 0\le s<\infty, \quad 0\le\theta<2\pi\, ,
\ee
where $z$ and $w$ are the local coordinates at the punctures that are glued.
The integration over $s$ and $\theta$ -- which describe two of the moduli of the
1PR Riemann surface -- yields the propagator
\be \label{eregprop}
-{1\over 2\pi} \int_0^\infty ds \, \int_0^{2\pi} d\theta \, e^{-s L_0^+}
e^{i\theta L_0^-} b_0^+ b_0^- \GG = -
{b_0^+ b_0^-\over L_0^+} \delta_{L_0^-} \GG\, .
\ee
The minus sign is convention dependent and has been explained 
below \refb{esiegel}.
This by itself of course
does not change the result for the integral over the moduli space, but just involves
organizing the integral as sum over different terms each of which in turn can be
regarded as 1PI amplitudes connected by propagators. We shall now describe how
the prescription we arrive at differs from the standard prescription.
\item \label{point2}
In a 1PR contribution, 
any propagator that is forced to carry zero momentum due to momentum
conservation (and hence is
part of a `tadpole' contribution) 
is replaced by the modified propagator 
\be
-{b_0^+ b_0^-\over L_0^+} \delta_{L_0^-} (1-{\bf P}) \GG
\, ,
\ee
where ${\bf P}$ is the projector into the zero momentum $L_0^+=0$ states. 
This makes these contributions
free from infrared divergences.
\item  If we want results accurate up to order
$g_s^n$, then 
in any amplitude we sum over arbitrary number $k$ of insertions of the state
$|\psi_n\rangle$ satisfying \refb{esol1},  weighted by a factor
of $1/k!$, for $0\le k\le n$.
\item The external momenta are set equal to their renormalized on-shell values
for which eqs.\refb{esa3}-\refb{esa5} have solutions, and not their classical
on-shell values obtained by demanding BRST invariance of the vertex operator.
\item The external states that are inserted into the amplitude
are not what we had at tree level, but are given by the states 
$|\phi_n\rangle$ obtained
by solving \refb{esa3}-\refb{esa5} if we want results accurate up to order
$g_s^n$.
\item \label{point6}
In a 1PR contribution, 
if any propagator is forced to carry momentum equal to the momentum $k_i$
carried by one of the external states (and hence is part of a mass / wave-function
renormalization diagram on the external leg)
we replace it by
\be 
-{b_0^+ b_0^-\over L_0^+} \delta_{L_0^-} (1-P_i) \GG
\, ,
\ee
where $P_i$ is the projector into the states carrying momentum $k_i$ and
mass level equal to the tree level mass of that external state. These are the 
states carrying $L_0^+\simeq 0$ in the sense described at the end of
\S\ref{sreview}.
Therefore the insertion of $(1-P_i)$ removes  
states with $L_0^+\simeq 0$ and makes the 
amplitude free from the IR divergences associated with mass renormalization.
\item Finally we multiply 
the result by $\prod_{i=1}^N (Z_i)^{-1/2}$ where $Z_i$ is the wave-function
renormalization factor associated with the $i$-th external state. These can be
computed from the residues at the poles of the two point function.
\end{enumerate}
Steps 2 and 3 above deal with massless tadpole contributions while the steps
4, 5, 6 and 7 deal with mass and wave-function renormalization issues.
The prescription of steps \ref{point2} and \ref{point6} imply
that whenever a propagator is forced to 
carry momentum that either vanishes or is equal to one of the external
momenta, we remove the
contribution from the $L_0^+\simeq 0$ states from it, where $L_0^+\simeq 0$
states have been defined at the end of \S\ref{sreview}. 

There are two more unrelated but useful results derived in
this section. In \S\ref{sorientation} we derived the sign of the integration
measure over the moduli space that is compatible with the rest of our
conventions. We found for example that if $(z,\bar z)$ denote the 
complex coordinates denoting the location of a vertex operator on the
Riemann surface then we must define the orientation of the moduli
space such that
\be \label{efirstuse}
dz\wedge d\bar z = 2\, i\, d^2 z\, ,
\ee
where the integration measure $d^2z$ is defined such that its integral
over any subspace of the Riemann surface gives a positive result.
Alternatively we can use the conventional orientation where 
$dz\wedge d\bar z = - 2\, i \, d^2 z$, but multiply every genus $g$,
$n$-point 1PI amplitude by an additional factor of $(-1)^{3g-3+n}$.

The second result deals with the reality condition on the string fields.
Let us suppose that in heterotic string theory 
we have a field configuration of the form
\be \label{eno1pre}
\lambda\, \bar c c e^{-\phi} V (0) |0\rangle\, ,
\ee
where $\lambda$ is a complex number and
$V$ is a dimension (1,1/2) superconformal primary in the
matter sector satisfying 
\be \label{eno2pre}
V(z) \, V(w) = (z-w)^{-1} (\bar z-\bar w)^{-2} +\cdots \, .
\ee
Then this field configuration is real if $\lambda$ is real.

\sectiono{Consequences of global and local (super-)symmetry} \label{sglobal}

In this section we shall discuss consequences of local (super-)symmetry
and also unbroken global 
(super-)symmetry. 
To simplify notation we shall drop the spinor index carried by the
supersymmetry generator and the fermionic vertex operators. Our formul\ae\ may be
interpreted as relations involving particular components of the supersymmetry generator
and fermionic vertex operators.

The results of this section may be summarized by saying that they are
identical to those found in \cite{1209.5461,1304.2832} except for the 
following differences:
\begin{enumerate}
\item These Ward identities hold
also for external states which undergo mass renormalization and for backgrounds
which undergo non-trivial vacuum shift.
\item The amplitudes that enter the Ward identities are the modified amplitudes
computed according to the prescription summarized in \S\ref{ssummary3}.
As a result these amplitudes are free from infrared divergences.
\item Using these Ward identities we can prove the equality of renormalized
masses of bosons and fermions paired by supersymmetry. This is possible only
due to the fact that the Ward identities hold for external states which
undergo mass renormalization.
\end{enumerate}

\subsection{Bose-Fermi degeneracy for global supersymmetry} \label{sbosefermi}

Let $|\Psi_{\rm vac}\rangle$ be a vacuum solution to the equations
of motion \refb{eeom}, and let us suppose that we have a global supersymmetry
transformation parameter $|\Lambda_{\rm global}\rangle$ that 
preserves this vacuum solution.
Therefore $|\Lambda_{\rm global}\rangle$ satisfies \refb{eglobal} 
which can also be expressed as
\be \label{es.1}
\wh Q_B|\Lambda_{\rm global}\rangle=0\, .
\ee
Let $|\Phi_{\rm linear}\rangle$ 
be a solution to the linearized equations of motion around the background
 i.e. it satisfies $\wh Q_B|\Phi_{\rm linear}\rangle=0$.
Then it follows from \refb{eredefined}, 
\refb{emainpp} and \refb{es.1} that
\be \label{es.3}
\wh Q_B \GG\,  [\Lambda_{\rm global} \Phi_{\rm linear}]'' = 0\, .
\ee
Therefore $\GG[\Lambda_{\rm global}\Phi_{\rm linear}]''$ also 
satisfies the linearized equations of motion. Since 
$|\Lambda_{\rm global}\rangle$ is fermionic, this 
provides a map between the bosonic and 
fermionic solutions to the linearized equations of motion. Since 
$|\Lambda_{\rm global}\rangle$ carries
zero momentum, these solutions occur at
the same values of momentum.
Furthermore 
if the solution $|\Phi_{\rm linear}\rangle$ exists for all values of 
momenta so does the solution
$\GG[\Lambda_{\rm global}\Phi_{\rm linear}]''$ and if the solution 
$|\Phi_{\rm linear}\rangle$ exists for special values of $k^2$,
the solution $\GG[\Lambda_{\rm global} \Phi_{\rm linear}]''$ also exists 
for the same special values of $k^2$. 
Therefore this procedure pairs pure gauge solutions in the bosonic and fermionic sector
and also physical solutions in the two sectors. In particular since the physical
solutions occur at the same values of the $k^2$, it establishes the equality of
the masses of bosons and fermions (even though each of them
may get
renormalized by perturbative 
corrections of string theory).\footnote{The only exception
to this is the situation where 
$\GG[\Lambda_{\rm global}\Phi_{\rm linear}]''$ vanishes. 
However typically in such
situations one can identify another component of the supersymmetry transformation 
parameter which does the pairing.}

Note that in general $\GG[\Lambda_{\rm global}\Phi_{\rm linear}]''$ is not in the 
Siegel gauge. Therefore if
$|\wh\Phi_{\rm linear}\rangle$ denotes the Siegel gauge physical state 
in the same sector
obtained by solving \refb{esa3}-\refb{esa5}, then we have
\be \label{ens}
\GG[\Lambda_{\rm global}\Phi_{\rm linear}]'' = \beta \, |\wh\Phi_{\rm linear}\rangle 
+ \wh Q_B|\wh\Lambda\rangle
\ee
for some $|\wh\Lambda\rangle$. Here $\beta$ is a normalization factor and we
have assumed that there is a unique state with the required quantum numbers
and physical mass. If there is a degeneracy even after quantum corrections
then $\beta$ will in general be a matrix.

For later use it will be useful to develop a procedure for computing the
constant $\beta$.
For this we recall from \refb{esummary}-\refb{e2.49} 
that $|\wh\Phi_{\rm linear}\rangle$ can be regarded as 
the result of recursively solving the equations
\be \label{e4.4}
|\wh\Phi_{\rm linear}\rangle = -{b_0^+\over L_0^+} (1-P) 
\GG K |\wh\Phi_{\rm linear}\rangle 
+ |\wh\phi\rangle \, , \ee
where $|\wh\phi\rangle$ satisfies
\be \label{e4.5}
P|\wh\phi\rangle = |\wh\phi\rangle\, , 
\quad Q_B|\wh\phi\rangle = - P 
\GG K |\wh\Phi_{\rm linear}\rangle \, .
\ee
$|\wh\Phi_{\rm linear}\rangle$ and $|\wh\phi\rangle$ are states 
in $\wh\HH_T$
of ghost number 2.
Let $|\wt\Phi_{\rm linear}\rangle$ and $|\wt\phi\rangle$ be conjugate states 
in $\wt\HH_T$ of ghost
number 3, satisfying
\be \label{e4.7}
|\wt\Phi_{\rm linear}\rangle = -{b_0^+\over L_0^+} (1-P) K\GG |\wt\Phi_{\rm linear}\rangle 
+ |\wt\phi\rangle \, ,
\ee
\be \label{e4.8}
P|\wt\phi\rangle = |\wt\phi\rangle\, , 
\quad Q_B|\wt\phi\rangle = - P 
K \GG |\wt\Phi_{\rm linear}\rangle \, ,
\ee
\be\label{e4.6}
\langle \wt\phi|c_0^- |\wh\phi\rangle=1\, .
\ee
It follows from \refb{e4.4}-\refb{e4.6} that
\be  \label{einnerphi}
\langle \wt\Phi_{\rm linear}| c_0^- | \wh\Phi_{\rm linear}\rangle = \langle \wt\phi| c_0^- |\wh\phi\rangle
=1\, ,
\ee
and 
\be \label{ess}
\wt Q_B|\wt\Phi_{\rm linear}\rangle = 0\, , 
\ee
where $\wt Q_B$ has been defined in \refb{edefqbt}.
Taking the inner product of $\langle \wt\Phi_{\rm linear}|c_0^-$ with \refb{ens},
and using  \refb{eabqb}, \refb{einnerphi}, \refb{ess} we get
\be \label{e4.10}
\beta = \langle\wt\Phi_{\rm linear}|c_0^- 
\GG[\Lambda_{\rm global}\wh\Phi_{\rm linear}]''\rangle = 
\{ (\GG\wt\Phi_{\rm linear}) \Lambda_{\rm global} \wh\Phi_{\rm linear}\}''\, .
\ee
The right hand side of \refb{e4.10} is the 1PI three point function of external
states $\GG|\wt\Phi_{\rm linear}\rangle$, $|\Lambda_{\rm global}\rangle$ and 
$|\wh\Phi_{\rm linear}\rangle$. Following the
analysis of \S\ref{susual} 
we can interpret this as the usual three point
amplitude with external states $\GG|\wt\phi\rangle$, $|\lambda\rangle$ 
and $|\wh\phi\rangle$ -- the projections of 
$\GG|\wt\Phi_{\rm linear}\rangle$, $|\Lambda_{\rm global}\rangle$ and 
$|\wh\Phi_{\rm linear}\rangle$ 
to the
$L_0^+\simeq 0$ sector, and
arbitrary number of insertions of $|\psi\rangle$ -- the $L_0^+=0$ component
of the vacuum solution, 
with the contribution from $L_0^+\simeq 0$
sectors removed from the propagators. As a result $\beta$ is free from
infrared divergences.

\subsection{Ward identities for local (super-)symmetry}

In this subsection we shall
derive the Ward identities for S-matrix elements computed from the 
1PI effective theory. As remarked at the end of \S\ref{s3.1}, 
the S-matrix elements can be computed from the truncated Green's functions 
$\Gamma^{(N)}$ by
setting the external states
on-shell and multiplying the result by appropriate wave-function renormalization 
factors for each external leg.
We shall first show that the $\Gamma^{(N)}$'s satisfy the identities:
\be \label{egrid}
\sum_{i=1}^N (-1)^{\gamma_1+\cdots \gamma_{i-1}} \Gamma^{(N)}(
| A_1\rangle, \cdots | A_{i-1}\rangle,
\wh Q_B| A_i\rangle, | A_{i+1}\rangle, \cdots | A_N\rangle) =0\, .
\ee

The proof of \refb{egrid} goes as follows. In any Feynman
diagram the external legs are divided
into subsets, with the members of a given subset all part of the same 1PI
vertex. We group the sum over $i$ in the left hand side of
\refb{egrid} by these subsets.
The result can now be organized using \refb{eimpidpp}, with two kinds of
contributions. In the first kind we have terms with $\wh Q_B$ acting on the
external states of the 1PI vertex which 
are not external states of $\Gamma^{(N)}$. These
are the terms that would be present on the left hand side
of \refb{eimpidpp} but are missing from the 
left hand side
of \refb{egrid}.
In the second kind we have contributions coming from the right hand side of
\refb{eimpidpp}. The first kind of terms, taken in pairs in which $\wh Q_B$
acts on the states at the two ends of a given {\it internal propagator}, 
can be simplified using
\refb{econdprop}. 
It is easy to see that after summing over all graphs, these two kinds
of terms cancel, with a pair of terms of the first kind, combined using 
\refb{econdprop}, canceling a term of the second type. 

We shall illustrate this
with an example.
Consider a contribution where two 1PI vertices are joined by a
propagator, with external states $| A_1\rangle,\cdots | A_M\rangle$ 
attached to the first 1PI vertex and $| A_{M+1}\rangle,\cdots | A_N\rangle$
attached to the second 1PI vertex, and $\wh Q_B$ acts in turn on each 
of these external
states as on the left hand side of \refb{egrid}. 
In that case terms of the first kind have the form
\ben \label{eyx1}
&& (-1)^{\gamma_1+\cdots \gamma_M} \{ A_1\cdots  A_M(\wh Q_B \vp_r) \}''
\langle \vp^r| c_0^- \Delta c_0^- | \vp^s\rangle  
\{ \vp_s  A_{M+1}\cdots  A_N\}''  \nonumber \\
&& + (-1)^{\gamma_1+\cdots \gamma_M+\gamma_{\vp_s}} 
\{ A_1\cdots  A_M \vp_r \}''
\langle \vp^r| c_0^- \Delta c_0^- | \vp^s\rangle  
\{ (\wh Q_B\vp_s)  A_{M+1}\cdots  A_N\}'' \, .
\een
There are overall minus signs in the two terms from having to take these
terms from the left to the right hand side of \refb{eimpidpp}, but they cancel 
the minus sign in the propagator \refb{esignprop}. 
Now we
apply \refb{edefsquare} in reverse 
on the last factor in each term, and use the results
\ben\label{euse}
\wh Q_B |\vp_r\rangle \langle \vp^r | c_0^- \Delta c_0^- |\vp^s \rangle \langle \vp_s|
c_0^-
&=& \wh Q_B b_0^- c_0^- \Delta c_0^- b_0^- c_0^-= \wh Q_B \Delta  c_0^-\nonumber \\
|\vp_r\rangle \langle \vp^r | c_0^- \Delta c_0^- |\vp^s \rangle \langle  \wh Q_B\vp_s|
c_0^- &=&  b_0^- c_0^- \Delta c_0^- |\vp^s \rangle \langle  \vp_s|  c_0^- \wt Q_B(-1)^{\gamma_{\vp_s}}
= b_0^- c_0^- \Delta c_0^- b_0^- c_0^- \wt Q_B 
(-1)^{\gamma_{\vp_s}} \nonumber \\
&=& (-1)^{\gamma_{\vp_s}} \Delta c_0^- \wt Q_B\, .
\een
In arriving at \refb{euse} we have used \refb{ecomplete},
\refb{eabqb} and
the fact that $\Delta$ defined in \refb{esiegel} is 
annihilated by $b_0^-$ from the left as well as from 
the right. Substituting these into
\refb{eyx1} we can bring it to the form
\ben \label{e3.9}
&&  (-1)^{\gamma_1+\cdots \gamma_M} \left\{ 
A_1\cdots  A_M \left((\wh Q_B \Delta c_0^-
+ \Delta c_0^-  \wt Q_B)
[ A_{M+1}\cdots  A_N]''\right)\right\}'' \nonumber \\
&=&  (-1)^{\gamma_1+\cdots \gamma_M} \{ A_1\cdots  A_M 
\GG [ A_{M+1}\cdots  A_N]''\}'' 
\een
where in arriving at the second expression we have used \refb{econdprop} and the fact that
\be
\delta_{L_0^-}[ A_{M+1}\cdots  A_N]''=[ A_{M+1}\cdots  A_N]''\, .
\ee
We shall now combine this with contributions that come from a single 1PI vertex
with external states $| A_1\rangle,\cdots | A_N\rangle$, with $\wh Q_B$ acting on
one of the states on the left hand side. 
After summing over the $N$ terms in which
$\wh Q_B$ acts in turn on $| A_1\rangle,\cdots | A_N\rangle$, 
a direct application of \refb{eimpidpp}
reduces this to a sum of terms of the second kind, one of which takes the
form
\be \label{e3.11}
-(-1)^{\gamma_1+\cdots \gamma_M} \{ A_1\cdots  A_M 
\GG[ A_{M+1}\cdots  A_N]''\}'' \, .
\ee
The overall minus sign
comes from the minus sign on the right hand side of \refb{eimpidpp}. 
The factor of 1/2 on the right hand side of \refb{eimpidpp} has been cancelled by
a factor of 2 that arises by combining pairs of terms -- one where $ A_{M+1}\cdots  A_N$
is inside the square bracket and the other where $ A_1\cdots  A_{M}$ is inside
the square bracket. We now see that \refb{e3.11} cancels \refb{e3.9}. Similar cancellations
occur for all terms, leading to \refb{egrid}.

Let us now suppose that we have a set of physical 
external states $|\AAA_1\rangle,\cdots 
|\AAA_N\rangle$ satisfying
\be \label{erenor}
\wh Q_B |\AAA_i\rangle = 0\, , \qquad \hbox{for} \qquad 1\le i\le N\, .
\ee
Let us also suppose that we have a local gauge  transformation parameter
$|\Lambda\rangle$ belonging either to the fermionic sector or to the bosonic sector.
Then $\wh Q_B|\Lambda\rangle$ represents a pure gauge state. It now follows
from \refb{egrid} with $N$ replaced by $N+1$ and the states $| A_1\rangle,\cdots
| A_{N+1}\rangle$ replaced by $|\Lambda\rangle, |\AAA_1\rangle,\cdots |\AAA_N\rangle$ 
that
\be \label{efinlocal}
\Gamma^{(N+1)}(\wh Q_B|\Lambda\rangle, |\AAA_1\rangle,\cdots |\AAA_N\rangle)
= 0\, .
\ee
Since S-matrix elements with external states $\wh Q_B|\Lambda\rangle, |\AAA_1\rangle,
\cdots |\AAA_N\rangle$ are given by multiplying 
$\Gamma^{(N+1)}(\wh Q_B|\Lambda\rangle, |\AAA_1\rangle,\cdots |\AAA_N\rangle)$
by wave-function renormalization factors, vanishing of \refb{efinlocal} 
will also imply the
vanishing of this S-matrix element. This shows that pure gauge states of the form
$\wh Q_B|\Lambda\rangle$ decouple from the
S-matrix of physical states. Note that since we have taken $|\AAA_i\rangle$'s to satisfy
\refb{erenor} which takes into account the effect of string loop corrections in the definition
of $\wh Q_B$, the decoupling of pure gauge states occurs even in the presence
of external states that
suffer mass renormalization.

The $\Gamma^{(N+1)}$ appearing in \refb{efinlocal} corresponds to truncated
Green's functions and hence, regarded as amplitudes with external states
$|\AAA_i\rangle$ and $\wh Q_B |\Lambda\rangle$, 
it will be given by integral over a
subspace of the full moduli space of Riemann surfaces in which the contributions
corresponding to diagrams 1PR in external legs are removed. However
since $|\AAA_i\rangle$ are physical states annihilated by $\wh Q_B$, we can use
the arguments given below \refb{emodified} 
to show that the missing 1PR parts associated
with the external states $|\AAA_i\rangle$ are added back to the amplitude 
if we replace $|\AAA_i\rangle$'s by $P|\AAA_i\rangle$'s, 
$P$ being the projection operator to $L_0^+\simeq 0$ states defined below 
\refb{esa5} -- except that
the contribution due to the $L_0^+\simeq 0$
states are subtracted from the propagators
on the external lines and hence the amplitude does not suffer from any divergence
due to mass / wave-function renormalization. However this does not directly
apply to the external line $\wh Q_B|\Lambda\rangle$. We shall now show how
this can also be rearranged so that we include also the contribution from the
regions of the moduli space that are 1PR in the leg $\wh Q_B|\Lambda\rangle$. For
this we choose
\be 
|\Lambda\rangle = (1+\Delta_0 c_0^- K)^{-1} |\bar\Lambda\rangle 
= \sum_{k=0}^\infty (-\Delta_0 c_0^- K)^k |\bar\Lambda\rangle 
\ee
for some state $|\bar\Lambda\rangle$. Now it follows from \refb{eqbk}
and the expression for $\Delta_0$ given in \refb{edefdelta0} that
\be
[Q_B, \Delta_0 c_0^- K] =(1+\Delta_0 c_0^- K)  \GG K
\ee
acting on states in $\wh \HH$,
and hence 
\be
[Q_B, (1+\Delta_0 c_0^- K)^{-1}] = - (1+\Delta_0 c_0^- K)^{-1} 
[Q_B, \Delta_0 c_0^- K] (1+\Delta_0 c_0^- K)^{-1} =-\GG K 
(1+\Delta_0 c_0^- K)^{-1} \, .
\ee
This gives
\be  \label{eright}
\wh Q_B |\Lambda\rangle 
= (Q_B + \GG K) (1+\Delta_0 c_0^- K)^{-1} |\bar\Lambda\rangle
= (1+\Delta_0 c_0^- K)^{-1} Q_B |\bar\Lambda\rangle\, .
\ee
Therefore in \refb{efinlocal} we can replace $\wh Q_B|\Lambda\rangle$ by the
right hand side of \refb{eright}. Expanding this as
$\sum_{k=0}^\infty (-\Delta_0 c_0^- K)^k  Q_B |\bar\Lambda\rangle$ we see
that we get back the missing 1PR contributions, but the external state
$\wh Q_B |\Lambda\rangle$ is replaced by $Q_B|\bar\Lambda\rangle$.

To summarize, we can interpret \refb{efinlocal}  as a statement of vanishing
of the full off-shell string amplitude involving external states 
$Q_B|\bar \Lambda\rangle$ and the $L_0^+\simeq 0$ components
of the $|\AAA_i\rangle$'s. However in computing
this amplitude we have to subtract the  contribution due to the
$L_0^+\simeq 0$ states from all internal propagators which carry momentum
equal to that of any of the external momenta carried by the states
$|\AAA_i\rangle$.  This removes the potential infrared divergences 
associated with mass renormalization. 
We also need to remove the contribution of $L_0^+=0$ states from internal
propagators carrying zero momentum and sum over arbitrary number of
insertions of $|\psi\rangle$ -- the $L_0^+=0$ component of the vacuum 
solution. This removes the potential infrared divergences associated with
tadpole graphs.
This agrees with the general form of the
Ward identity described in \cite{1209.5461,1304.2832}, 
except for the subtraction of the infrared
divergent contribution mentioned above and the use of $L_0^+\simeq 0$
components of $|\AAA_j\rangle$
instead of BRST invariant states as external states.

\subsection{Ward identities for global (super-)symmetry}

\begin{figure}
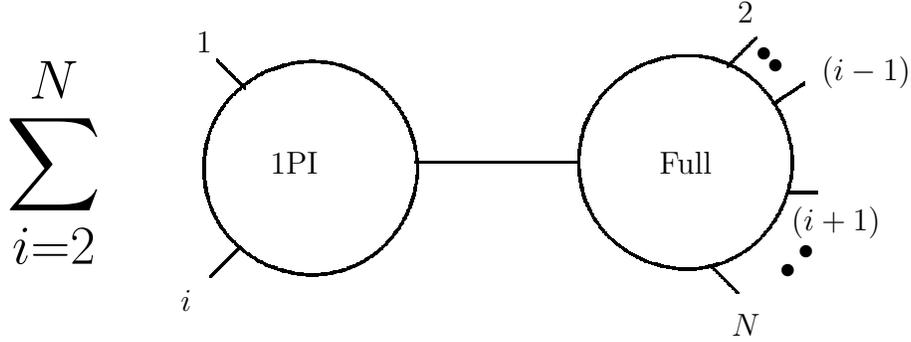


\begin{center}

\figwtgam

\end{center}

\vskip -.9in 

\caption{The contributions to be excluded from the definition of $\wt\Gamma^{(N)}$. 
Here the blob marked 1PI represents the 1PI vertex $\{~\}''$, the blob marked Full represent
the full  truncated Green's function, 
the horizontal line connecting the two blobs
represent the full propagator $\Delta$ and the short lines represent external states.
\label{figwtgam}}
\end{figure}

We shall now explore the consequences of global (super-)symmetry
on the S-matrix. The existence of
such a symmetry is signaled by a gauge transformation parameter 
$|\Lambda_{\rm global}\rangle$
satisfying
\be \label{eglobala}
\wh Q_B|\Lambda_{\rm global}\rangle=0\, .
\ee
Typically $|\Lambda_{\rm global}\rangle$ carries zero momentum. 
Now if we used \refb{efinlocal}
with $|\Lambda\rangle$ replaced by 
$|\Lambda_{\rm global}\rangle$ then the resulting 
identity is trivial. To get something
non-trivial we proceed somewhat differently. We first define a new object
$\wt \Gamma^{(N)}(| A_1\rangle,\cdots | A_N\rangle)$ where the first argument
$| A_1\rangle$ plays a somewhat different role compared to the other arguments.
To define $\wt \Gamma^{(N)}$ we begin with the expression for the truncated Green's 
function $\Gamma^{(N)}$ as sum of 
Feynman diagrams built from 1PI vertices and propagators, 
and
delete from this all terms where by removing a single propagator we can 
separate the external states $|\wt A_1\rangle$ and one more $| A_i\rangle$ from
the rest of the $| A_i\rangle$'s. 
This has been shown in Fig.~\ref{figwtgam}.
If we take the $| A_i\rangle$'s to be states 
carrying fixed momenta $k_i$ then this means that we remove all terms where one
of the internal propagators carry momentum $k_1+k_i$ for any $i$ between 2 and $N$.
We can now derive an identity analogous to \refb{egrid} for $\wt\Gamma^{(N)}$ using 
similar method, but now there are additional terms since the contributions
of the first kind involving $\wh Q_B\Delta c_0^-+\Delta c_0^- \wt Q_B$ 
with $\Delta$ representing a propagator
carrying momentum $k_1+k_i$ for $2\le i\le N$ are absent. As a result some of the
contributions of the second kind are not canceled and we get
\ben \label{etildegam}
\sum_{i=1}^N (-1)^{\gamma_1+\cdots \gamma_{i-1}} \wt\Gamma^{(N)}(
| A_1\rangle, \cdots | A_{i-1}\rangle,
\wh Q_B| A_i\rangle, | A_{i+1}\rangle, \cdots | A_N\rangle) 
\qquad \qquad \qquad \quad \nonumber \\ 
+ \sum_{i=2}^N (-1)^{(\gamma_1+1)(\gamma_2+\cdots \gamma_{i-1})} \Gamma^{(N-1)}(
| A_2\rangle, \cdots | A_{i-1}\rangle,
\GG[ A_1 A_i]'' , | A_{i+1}\rangle, \cdots | A_N\rangle) 
=0\, .
\een
Note that the second term involves $\Gamma^{(N-1)}$ and not $\wt\Gamma^{(N-1)}$.
The proof of this relation follows the same logic as the one used in arriving at
\refb{egrid}.

Let us now suppose that we have a set of physical 
external states $|\AAA_1\rangle,\cdots 
|\AAA_N\rangle$ 
satisfying \refb{erenor} and a global (super-)symmetry transformation
parameter $|\Lambda_{\rm global}\rangle$  satisfying \refb{eglobala}. 
Then a direct application of \refb{etildegam} with $N$ replaced by $N+1$, and the states
$| A_1\rangle,\cdots | A_{N+1}\rangle$ taken as 
$|\Lambda_{\rm global}\rangle, |\AAA_1\rangle,
\cdots |\AAA_N\rangle$ gives
\be \label{eglobalid}
\sum_{i=1}^N (-1)^{(\gamma_\Lambda + 1) 
(\gamma_1+\cdots \gamma_{i-1})}
\Gamma^{(N)} (|\AAA_1\rangle, \cdots |\AAA_{i-1}\rangle, 
\GG[\Lambda_{\rm global} \AAA_i]'', |
\AAA_{i+1}\rangle, \cdots |\AAA_N\rangle)=0\, ,
\ee
where $\gamma_\Lambda$ and $\gamma_i$ 
now stand for the grassmannality of
$|\Lambda_{\rm global}\rangle$ and $|\AAA_i\rangle$ respectively. Noting that according to
\refb{es.3}, $\GG[\Lambda_{\rm global} \AAA_i]''$ 
represents the on-shell state which is the transform of
$|\AAA_i\rangle$ under the infinitesimal global (super-)symmetry generated by 
$|\Lambda_{\rm global}\rangle$, we recognize \refb{eglobalid} 
as the Ward identities
associated with the global (super-)symmetry generated by 
$|\Lambda_{\rm global}\rangle$.

We could again examine to what extent the right hand side of \refb{eglobalid}
can be interpreted as usual string amplitudes involving integration over the
full moduli space. For this we use \refb{ens} to replace 
$\GG[\Lambda_{\rm global} \AAA_i]''$
by $\beta_{(i)}|\Phi^{(i)}\rangle+\wh Q_B|\Lambda^{(i)}\rangle$ where 
$|\Phi^{(i)}\rangle$ is a solution to the linearized equations of motion in the
Siegel gauge obtained by solving \refb{esa3}-\refb{esa5}. 
The amplitude involving $\wh Q_B|\Lambda^{(i)}\rangle$ vanishes
by \refb{efinlocal} and we can express \refb{eglobalid} as
\be \label{eglobalid1}
\sum_{i=1}^N (-1)^{(\gamma_\Lambda + 1) (\gamma_1+\cdots \gamma_{i-1})}
\beta_{(i)} \, 
\Gamma^{(N)} (|\AAA_1\rangle, \cdots |\AAA_{i-1}\rangle, |\wh\Phi^{(i)}\rangle, 
|\AAA_{i+1}\rangle, \cdots |\AAA_N\rangle)=0\, .
\ee
Using the
results summarized in \S\ref{ssummary3} we can represent the
amplitude involving $|\wh\Phi^{(i)}\rangle$ and the $|\AAA_j\rangle$'s 
as the usual string amplitude with external states 
given by the projections of $|\wh\Phi^{(i)}\rangle$ and $|\AAA_j\rangle$'s 
to $L_0^+\simeq 0$ sector, arbitrary number of insertions of the
$L_0^+=0$ component of the vacuum solution and
appropriate subtractions from propagators. 
On the other hand using the result obtained in \S\ref{sbosefermi} we can 
interpret $\beta_{(i)}$ as the usual  three point amplitude of string theory
with external states 
$\GG|\wt\phi^{(i)}\rangle$, $|\lambda\rangle$ and $|\phi^{(i)}\rangle$
with  arbitrary number of insertions of the $L_0^+=0$ component of
the vacuum solution and appropriate subtractions from propagators. 
Here $|\lambda\rangle$ and
$|\phi^{(i)}\rangle$ are the projections of $|\Lambda_{\rm global}\rangle$ and
$|\Phi^{(i)}\rangle$ to $L_0^+\simeq 0$ states and $|\wt\phi^{(i)}\rangle$
is the ghost number 3 and picture number $-3/2$ state conjugate to 
$|\phi^{(i)}\rangle$ satisfying $\langle \wt\phi^{(i)}|c_0^- |\phi^{(i)}\rangle=0$
and \refb{e4.7}, \refb{e4.8} with $|\wt\phi\rangle$ replaced by 
$|\wt\phi^{(i)}\rangle$.
Again this form of the Ward identity agrees with those given in
\cite{1209.5461,1304.2832} except for 
the subtraction schemes for the propagators
and modification of external states.

\sectiono{Supersymmetry and massless tadpoles} \label{stadpole}

In our analysis in the previous section,
we have assumed the existence of a vacuum solution 
$|\Psi_{\rm vac}\rangle$  to all orders in $g_s$, and 
have derived various Ward identities 
associated with local and global (super-)symmetries. 
If instead the vacuum solution (and global (super-)symmetry transformation
parameter)
is assumed to exist to a certain order $g_s^n$, then  the 
Ward identities 
will also be valid to that order.
In this section we address a slightly different problem.
We shall assume that we have a vacuum solution 
$|\Psi_{\rm vac}\rangle$
to a certain order in $g_s$ and also unbroken supersymmetry to certain order in
$g_s$. We shall then examine to what extent unbroken supersymmetry may help us
extend the vacuum solution to higher order.

For simplicity we shall carry out our discussion in the context of heterotic string theory,
but an identical analysis holds for NSNS sector tadpoles in type II string theory. 
Also we shall assume that the vacuum solution has a power series expansion
in $g_s$ containing both even and odd powers of $g_s$. This allows us to include
the cases where 
under quantum corrections fields may get
shifted by terms of order $g_s$ \cite{1404.6254}. If we are 
considering a vacuum described by a 
string field configuration containing only even powers of $g_s$
then the natural expansion parameter is $g_s^2$ and in 
all subsequent formul\ae\ we have to set the coefficients of all odd powers
of $g_s$ to zero. We shall give the result for this case explicitly in
\refb{efin20}. As mentioned in footnote \ref{fgs}, there may also be
cases where the vacuum solution has expansion in powers of $g_s^\alpha$
for some $\alpha$ in the range $0<\alpha<1$. In such cases we have to replace
$g_s$ by $g_s^\alpha$ in the following analysis.

We shall begin by assuming that we have a translationally invariant
vacuum solution $|\Psi_p\rangle$ to the classical equations of
motion to order $g_s^p$ for some integer 
$p$, i.e.\footnote{As usual we can restrict the sum to $n\le p+1$ to get terms accurate
to order $g_s^p$. Similar remark holds for subsequent sums.}
\be \label{eeomapprox}
Q_B|\Psi_p\rangle + \sum_{n=1}^\infty {1\over (n-1)!} \GG[\Psi_p{}^{n-1}]
= \OO(g_s^{p+1})\, .
\ee
Our goal will be to see under what condition we can extend $|\Psi_{\rm vac}\rangle$ to the
next order. If we denote by $| \scalar\rangle$ a zero momentum BRST invariant 
NS sector Lorentz scalar state carrying ghost number 2 and picture number $-1$,
then the obstruction to extending $|\Psi_{\rm vac}\rangle$ to the next order is encoded in the matrix
element
\be \label{edefep}
\EE_{p+1}( \scalar)
\equiv \langle \scalar|c_0^- \left( Q_B|\Psi_p\rangle + \sum_{n=1}^\infty {1\over (n-1)!} \GG[\Psi_p{}^{n-1}]\right)\, .
\ee
By \refb{eeomapprox}, $\EE_{p+1}( \scalar)$ is already of order $g_s^{p+1}$. 
If we can show that
$\EE_{p+1}( \scalar)$ is of order $g_s^{p+2}$ for every zero momentum 
BRST invariant state $|\scalar\rangle$, then
we can extend the solution to the next order\cite{1411.7478}. 
This is known to hold as a consequence
of \refb{eeomapprox} when $|\scalar\rangle$ is a BRST trivial state\cite{1411.7478}, 
so we focus on the cases where $| \scalar\rangle$
represents a non-trivial element of the BRST cohomology. Since Lorentz
invariance is unbroken, it will be sufficient to consider only Lorentz scalar
$| \scalar\rangle$'s.

We shall consider string theories with unbroken supersymmetry at tree level.
It has been shown in appendix \ref{sap} that in such theories,
for every $ \scalar$ we can find a BRST
invariant
Ramond sector vertex operator $\fermion$ of ghost number 2 and picture
number $-1/2$, and another BRST
invariant ghost number 1, picture number $-1/2$ state $|\Lambda_0\rangle$ representing
a leading order global supersymmetry transformation, 
such that
\be \label{epair0}
 |\scalar\rangle =  [\Lambda_0\fermion]_0
\ee
up to  addition of BRST trivial states. The subscript 0 on $[\cdots]$ indicates that
we evaluate $[\cdots]$ only at genus 0.  In the definition of
$|\fermion\rangle$ and $|\Lambda_0\rangle$ we shall not include
multiplication by grassmann odd variable as would be required to
promote them to fermionic
string fields and supersymmetry transformation parameters
respectively.
Therefore $|\fermion\rangle$ is grassmann odd and $|\Lambda_0\rangle$ is
grassmann even.

Let us now 
suppose that 
$\Lambda_0$ can be extended to a global supersymmetry transformation
parameter to
order $g_s^q$ for some $q\le p$, 
i.e.\ there is a state 
$|\Lambda_q\rangle\in \HH_T$ 
carrying zero momentum, ghost number
1 and picture number $-1/2$, satisfying
\be \label{elambapprox}
\wh Q_B|\Lambda_q\rangle \equiv (Q_B + \GG K_p) 
|\Lambda_q\rangle = \OO(g_s^{q+1})\, , \quad |\Lambda_q\rangle
=|\Lambda_0\rangle +\OO(g_s)\, ,
\ee
where
$K_p$ is the operator $K$ defined in \refb{edefK} with $|\Psi_{\rm vac}\rangle$ replaced by
its $p$-th order solution $|\Psi_p\rangle$.
Then we can use \refb{epair0} to write
\be \label{escalarapprox}
| \scalar\rangle = [\Lambda_q\fermion] + \OO(g_s)
= \sum_{m=0}^\infty {1\over m!} [\Lambda_q\fermion\Psi_p{}^m]
+ \OO(g_s)\, ,
\ee
up to addition of BRST trivial terms. 
In the last step we have included a sum over $m$ by exploiting the
fact that since $\Psi_p$ has its expansion beginning at order $g_s$,
all but the $m=0$ term
will be of order $g_s$.
Using \refb{edefep}, \refb{escalarapprox} and \refb{eeomapprox} we get
\be\label{e5.10pre}
\EE_{p+1}( \scalar)
= \sum_{m=0}^\infty {1\over m!} \langle [\Lambda_q\fermion\Psi_p{}^m]| 
c_0^- \left(Q_B|\Psi_p\rangle + \sum_{n=1}^\infty {1\over (n-1)!} 
\GG[\Psi_p{}^{n-1}]\right) + \OO(g_s^{p+2}) \, ,
\ee
Using  
the fact that $|\fermion\rangle$ is grassmann odd 
whereas 
$|\Lambda_q\rangle$ 
and $|\Psi_p\rangle$ are grassmann even
we can express \refb{e5.10pre} as
\ben \label{ekxa}
\EE_{p+1}( \scalar) &=& -\sum_{m=0}^\infty {1\over m!} \left\{\Lambda_q
\fermion\Psi_p{}^m (Q_B\Psi_p)\right\} \nonumber \\ &&
- \sum_{m=0}^\infty \sum_{n=1}^\infty {1\over m! {(n-1)}!} \left\{\Lambda_q
\fermion\Psi_p{}^m
\GG[\Psi_p{}^{n-1}]
\right\} 
+ \OO(g_s^{p+2})  \, . 
\een
Using \refb{eimpid}, \refb{eneweq2a} and BRST invariance of $\fermion$,
we can rewrite this
equation as
\ben \label{ekxb}
\EE_{p+1}( \scalar) 
&=& -\sum_{m=0}^\infty {1\over m!} 
\left\{(Q_B\Lambda_q)\fermion
\Psi_p{}^m \right\}
- \sum_{m=0}^\infty \sum_{n=0}^\infty {1\over m! n!} \left\{ \Lambda_q 
\Psi_p{}^m \GG [\fermion \Psi_p{}^n]\right\} \nonumber \\ &&
+ \sum_{m=0}^\infty \sum_{n=0}^\infty {1\over m! n!} \left\{ \Lambda_q 
\fermion
\Psi_p{}^m \GG [\Psi_p{}^n]\right\} \nonumber \\ &&
- \sum_{m=0}^\infty \sum_{n=1}^\infty {1\over m! {(n-1)}!} \left\{\Lambda_q
\fermion \Psi_p{}^m
\GG[\Psi_p{}^{n-1}]
\right\} + \OO(g_s^{p+2}) \nonumber \\
&=& -\sum_{m=0}^\infty {1\over m!} 
\left\{(Q_B\Lambda_q)
\fermion\Psi_p{}^m \right\}
- \sum_{m=0}^\infty \sum_{n=0}^\infty {1\over m! n!} \left\{ \Lambda_q 
\Psi_p{}^m \GG [\fermion \Psi_p{}^n]\right\}  
\nonumber \\ &&
+ \OO(g_s^{p+2}) \, .
\een
The first three terms in the middle expression of \refb{ekxb}
arise from manipulating the first term
on the right hand side of \refb{ekxa} using \refb{eimpid} and the symmetry property
\refb{eneweq2a}. 
Using the definition of $K_p$ that follows from \refb{edefK}  
we can express \refb{ekxb} as
\be \label{ekxd}
\EE_{p+1}( \scalar) = -\langle K_p\fermion| c_0^-  
\wh Q_B |\Lambda_q\rangle+ \OO(g_s^{p+2})
=\OO(g_s^{q+2}, g_s^{p+2}) \, ,
\ee
where $\OO(g_s^{q+2},
g_s^{p+2})$ denotes that the error is given by the dominant term among
$g_s^{q+2}$ and 
$g_s^{p+2}$. 
The $\OO(g_s^{q+2})$ correction comes from possible
contributions of order $g_s$ to $K_p |\fermion\rangle$ and
order $g_s^{q+1}$ to $\wh Q_B|\Lambda_q\rangle$. 

Since $q\le p$, we have to consider two possibilities. For $q\le p-1$,
the order $g_s^{q+2}$ term dominates.
Since this is of order $g_s^{p+1}$ or larger, \refb{ekxd}
does not allow us to extend the vacuum solution to the next order;
this would require the order $g_s^{p+1}$ contribution
to  $\EE_{p+1}( \scalar) $
to vanish.
On the other hand for $q=p$, this equation gives
$\EE_{p+1}( \scalar)=\OO(g_s^{p+2})$. This now allows us to extend
$|\Psi_p\rangle$ to satisfy the classical
equations of motion to next order, i.e. replace $p$ by $p+1=q+1$ 
in \refb{eeomapprox}. 
Therefore our analysis shows that if supersymmetry is unbroken to order $g_s^q$
then
we can construct the vacuum
solution to order $g_s^{q+1}$. 
This is in the spirit of the results of 
\cite{1209.5461,1304.2832}. Note however that in this analysis we have not
made any assumption about $|\Psi_p\rangle$ being the perturbative vacuum.
In particular by allowing $|\Psi_p\rangle$ to begin its expansion at
order $g_s$ instead of at order $g_s^2$, we have allowed for the possibility
of non-trivial background $|\Psi_p\rangle$ -- this will be illustrated in the
next few sections.

Since for given $q$ we can extend the classical solution to order
$g_s^p$ for $p=q+1$, we shall from now on set $p=q+1$.
\refb{ekxd} can now be used to compute the order $g_s^{q+2}$ 
contribution to 
$\EE_{q+2}(\scalar)$ -- referred henceforth as the tadpole contribution
since in the conventional perturbation theory a non-vanishing 
$\EE_{q+2}(\scalar)$ will show up as a tadpole divergence.
In this case the order $g_s^{q+2}$ contribution to
the tadpole is given by the first term in the middle
expression of \refb{ekxd}. 
Now in $\HH_T$ one can choose a basis
in which the inner product
$\langle A|c_0^-|B\rangle$ pairs physical states with physical states and pure gauge
states with unphysical states where physical states are defined as those which are
$Q_B$ invariant but not $Q_B$ trivial, unphysical states are not $Q_B$ invariant
and pure gauge states are $Q_B$ trivial. This allows us to express $c_0^-$ appearing in 
\refb{ekxd} as
\be \label{elast0}
c_0^-=c_0^- \, |{\rm phys}\rangle \langle {\rm phys}| c_0^- +
c_0^- |{\rm unphys}\rangle \langle 
{\rm unphys}|Q_B c_0^-
+ c_0^- Q_B|{\rm unphys}\rangle \langle {\rm unphys}|c_0^- \, ,
\ee
where we have used the fact that pure gauge states can be written as
$Q_B|{\rm unphys}\rangle$. On the right hand side of \refb{elast0} it is
understood that a term like $|{\rm phys}\rangle \langle {\rm phys}|$ 
represents sum over all physical states in $\HH_T$.
Furthermore, 
 as
a consequence of  \refb{eeomapprox}, \refb{elambapprox}
with $p>q$ and BRST invariance of $|\fermion\rangle$, we have the 
relations\cite{1411.7478,1501.00988}
\ben \label{exqx0}
&& Q_B \sum_{m=0}^\infty {1\over m!} [\Lambda_q \Psi_{q+1}{}^m] 
= \OO(g_s^{q+2}),
\quad Q_B \sum_{m=0}^\infty {1\over m!} [\fermion \Psi_{q+1}{}^m] 
= \OO(g_s^{2}),
\nonumber \\
&& Q_B \wh Q_B |\Lambda_q\rangle = \OO(g_s^{q+2}), \quad
Q_B K_{q+1} |\fermion\rangle = \OO(g_s^{2})\, ,
\een
where the equations in the second line follow from those in the first line and
the definitions of $\wh Q_B$, $K_p$. 
Substituting \refb{elast0} into \refb{ekxd} with $p=q+1$, and using
\refb{exqx0} and \refb{elambapprox}
we see that the contribution to the right hand side of \refb{ekxd} from the
last two terms on the right hand side of \refb{elast0} is of order $g_s^{q+3}$.
Therefore we can now express \refb{ekxd} 
for $p=q+1$ as
\ben \label{efin1}
\EE_{q+2} (\scalar)  &=& -\sum_a \langle K_{q+1}\fermion|c_0^-|\zeta_a\rangle
\langle\zeta^a| c_0^- \wh Q_B|\Lambda_q\rangle \nonumber \\
&=& - \sum_a  \left.\left(\sum_{m=0}^\infty {1\over m!} \langle [\fermion \Psi_{q+1}{}^m]| 
c_0^- |\zeta_a\rangle \right)\right|_{g_s}
\left.\left(\sum_{n=0}^\infty {1\over n!} \langle \zeta^a| c_0^- \GG |  [\Lambda_q 
\Psi_{q+1}{}^n]\rangle\right)\right|_{g_s^{q+1}}  \nonumber \\ &&
+\,  \OO(g_s^{q+3})\, ,
\een
where the sum over $a$ runs over finite number of 
physical states $|\zeta_a\rangle\in
\HH_T$ carrying zero momentum and $|\zeta^a\rangle$ are conjugate physical 
states satisfying $\langle \zeta_a|c_0^-|\zeta^b\rangle=\delta_a{}^b$.
In the last step of \refb{efin1} we have used the fact that $Q_B$ annihilates physical states.
The subscripts $g_s$ and $g_s^{q+1}$ denote that we have to compute the
correlation functions to order $g_s$ and $g_s^{q+1}$ respectively, 
-- indeed by assumption these quantities receive possible contribution 
only at these orders and beyond.
Ghost and picture number conservation shows that $|\zeta_a\rangle$
carries ghost number 2 and picture number $-1/2$ while its conjugate state $|\zeta^a\rangle$
carries ghost
number 3 and picture number $-3/2$.

Physically the amplitude
\be \label{efx1}
\left.\left(\sum_{m=0}^\infty {1\over m!} \langle [\fermion \Psi_{q+1}{}^m]| 
c_0^- |\zeta_a\rangle\right)\right|_{g_s} = \sum_{m=0}^\infty {1\over m!} 
\{ \zeta_a \fermion \Psi_{q+1}{}^m\}|_{g_s}
\ee
represents the 1PI two point function of $\fermion$ 
and $\zeta_a$ in the background
$|\Psi_{q+1}\rangle$ to order $g_s$. Since for two point function the 1PI amplitude is also the truncated 
Green's function, we can identify this as 
$g_s\Gamma^{(2,1)}(\zeta_a,\fermion)=-g_s
\Gamma^{(2,1)}(\fermion, \zeta_a)$ where $g_s^k\, \Gamma^{(n,k)}$
denotes the order $g_s^{k}$ contribution to $\Gamma^{(n)}$. 
Similarly the amplitude
\be \label{efx2}
\left.\left(\sum_{n=0}^\infty  {1\over n!} \langle \zeta^a| c_0^- \GG |  [\Lambda_q 
\Psi_{q+1}{}^n]\rangle\right)\right|_{g_s^{q+1}} =\sum_{n=0}^\infty  {1\over n!} 
\{ (\GG\zeta^a) \, \Lambda_q \Psi_{q+1}{}^n\}|_{g_s^{q+1}}
\ee
can be identified as $g_s^{q+1}\Gamma^{(2,q+1)} 
(\GG\zeta^a, \Lambda_q)$.  Since $\zeta^a$ is a Ramond sector state,
$\GG\zeta^a=\XX_0\zeta^a$.
Therefore we can express \refb{efin1} as
\be \label{efin2}
\EE_{q+2} (\scalar)  
=g_s^{q+2}\sum_a
\Gamma^{(2,1)}(\fermion, \zeta_a) \, \Gamma^{(2,q+1)} 
(\XX_0\zeta^a, \Lambda_q)
+ \OO(g_s^{q+3})\, .
\ee

As remarked at the beginning of the section, if we are considering
perturbative vacuum where the vacuum solution has expansion in 
powers of $g_s^2$, then the coefficients of all the odd powers of $g_s$
vanish. In this case all factors of $g_s$ in the above analysis
can be replaced 
by $g_s^2$, and consequently $\Psi_p$, $\EE_p$, $\Lambda_q$,
$K_p$ and 
$\Gamma^{(n,k)}$ factors will 
have to be replaced respectively by $\Psi_{2p}$, $\EE_{2p}$, 
$\Lambda_{2q}$, $K_{2p}$ and
$\Gamma^{(n,2k)}$. 
\refb{efin2} will now take the form
\be \label{efin20}
\EE_{2q+4} (\scalar)  
=g_s^{2q+4}\sum_a
\Gamma^{(2,2)}(\fermion, \zeta_a) \, \Gamma^{(2,2q+2)} 
(\XX_0\zeta^a, \Lambda_{2q})
+ \OO(g_s^{2q+6})\, .
\ee

In \refb{efin2},
$\Gamma^{(2,1)}(\fermion, \zeta_a)$ receives contribution
only from the $m=1$ term in \refb{efx1}, with $\{~\}$ replaced by
its genus zero contribution and $|\Psi_{q+1}\rangle$ replaced by its
order $g_s$ contribution.
Physically, this is the genus zero 3-point amplitude of $\zeta_a$,
$\fermion$ and $\Psi_1$.
Similarly, using
the fact that $\Lambda_q$ is constructed from $\Lambda_0$ by 
the iterative process 
described in \refb{esolam}, \refb{etauk}, one can follow the method of
\S\ref{susual} to conclude that 
$g_s^{q+1}\Gamma^{(2,q+1)} (\XX_0\zeta^a, \Lambda_q)$ factor in
\refb{efin2}
can be interpreted as the full two point function of $(\XX_0\zeta^a)$ and 
$\lambda_q$ -- the projection of $\Lambda_q$ to $L_0^+=0$ state --
 to order $g_s^{q+1}$, with the subtraction and insertion rules
 described in \S\ref{susual}.

If we had chosen to work at the perturbative vacuum where 
$|\Psi_{\rm vac}\rangle$ had its expansion in powers of $g_s^2$, then the 
factor
$\Gamma^{(2,2)}(\fermion, \zeta_a)$ appearing in \refb{efin20} will
receive two contributions 
from the
right hand
side of \refb{efx1}  -- one from genus zero
with one insertion of $|\Psi_{2q+2}\rangle$ at order $g_s^2$ and a
genus one contribution with no insertion of background $|\Psi_{2q+2}\rangle$.
Following the analysis of \S\ref{susual} we can interpret the 
sum of the two terms as the full 
genus one two point function
of $\fermion$ and $\zeta_a$.
Similarly, the  factor
 $\Gamma^{(2,2q+2)} 
(\XX_0\zeta^a, \Lambda_{2q})$ in \refb{efin20}
can be interpreted as the full two point function of $(\XX_0\zeta^a)$ and 
$\lambda_q$ -- the projection of $\Lambda_q$ to $L_0^+=0$ state --
 to order $g_s{}^{2q+2}$, with the subtraction and insertion rules
 described in \S\ref{susual}.  
 With these interpretations, \refb{efin20}
agrees with the results of \cite{1209.5461,1304.2832} for the perturbative
vacuum,
 except for the
modified procedure for dealing with divergences associated with tadpoles and
two point functions on external lines, as summarized  at the beginning of
\S\ref{smatrix}.

We also note that the states $|\zeta^a\rangle$, which represent elements of
BRST cohomology with ghost number 3 and picture number $-3/2$, 
correspond to the candidate 
states $\goldc$ listed in \refb{ep2}, and the conjugate states $|\zeta_a\rangle$
are the goldstino candidates listed in \refb{egoldcan}. Furthermore 
comparing \refb{efx2} with  \refb{esusy2} we see that 
$g_s^{q+1}\Gamma^{(2,q+1)} (\XX_0\zeta^a, \Lambda_q)$ 
can be identified with
$\LL_{q+1}(\zeta^a)$ -- the obstruction to finding global supersymmetry
generator beyond order $g_s^q$. This is of course expected, since if we
could extend the global supersymmetry generator to order $g_s^{q+1}$ then
$\EE_{p+1}(\scalar)$ would vanish to order $g_s^{q+2}$.

\sectiono{SO(32) 
Heterotic string theory on Calabi-Yau three folds} \label{ehetrev}

So far our analysis has been very general without referring to any specific background.
In the rest of the paper we shall 
apply this general analysis to a specific class of
backgrounds -- SO(32) heterotic string theory on Calabi-Yau manifolds, 
with an unbroken U(1) gauge group at the tree level. 
Keeping this in mind in this section we shall review some basic facts about
this theory.

\subsection{Low energy effective field theory description} \label{slowen}

It is known from the 
analysis of \cite{DSW,ADS,DIS} that
one loop
quantum corrections in this theory will generate a Fayet-Iliopoulos 
D-term. 
We shall not review these arguments in detail, but note that 
one of the effects of the D-term is to
generate a one loop effective 
potential for a complex scalar field $\chi$ of the following form\footnote{In general
there can be more than one chiral multiplet charged under the U(1), in which case
the $\chi^* \chi$ term in \refb{epot} is replaced by $\sum_i q_i \chi_i^* \chi_i$ 
where $q_i$ are constants proportional to the U(1) charge carried by $\chi_i$.
We shall assume that even if such multiple scalar fields are present, only
one of them -- which we shall denote by $\chi$ -- acquires a vacuum expectation value.
It is easy to generalize the analysis to cases where multiple fields of 
this type acquire
vacuum expectation values. \label{feffective}}
\be \label{epot}
C \, g_s^{-2}\,  (\chi^* \chi - K\, g_s^2)^2\, ,
\ee
where $g_s$ is the string coupling and $C$ and $K$ are positive
numerical constants which can be computed in any given theory. At tree
level the potential is $Cg_s^{-2} (\chi^*\chi)^2$ and the vacuum is at $\chi=0$. 
However this perturbative vacuum becomes unstable at one loop since  
the field $\chi$ becomes tachyonic and supersymmetry is broken. On the other
hand it is clear from \refb{epot} that there is a stable supersymmetric 
vacuum at
$|\chi|=K^{1/2} g_s$.
We shall for definiteness take the expectation value of $\chi$ to be real,
i.e.\ only the real part $\chi_R\equiv (\chi+\chi^*)/\sqrt 2$ 
of $\chi$ gets a vacuum expectation value.

In order that the minimum of \refb{epot} describes a supersymmetric
extremum we need to assume that there is no F-term potential for the field
$\chi$ and also that there is no other D-term potential for this field. 
It has been shown in \cite{ADS} that left-right symmetric compactification 
of SO(32) heterotic string theory on a
Calabi-Yau manifold always contains a field $\chi$ satisfying these criteria.
The world-sheet properties of the vertex operator of $\chi$ will be described
in \S\ref{scon}.

\subsection{World-sheet superconformal field 
theory} \label{scon}

In this subsection we shall
review some of the properties of the matter sector of the 
world-sheet superconformal field theory 
(SCFT) describing SO(32) heterotic string theory on a Calabi-Yau 3-fold. 
We shall consider backgrounds with `spin connection
embedded in gauge connection' preserving (2,2) world-sheet supersymmetry, but
in principle our analysis can be generalized to compactification preserving
(0,2) world-sheet supersymmetry as well.
We shall only quote the relevant results; more details 
may be found in \cite{ADS} and the references given there.

The SCFT consists of three parts. One part contains four free scalars
describing the non-compact target space coordinates $X^\mu$ for $0\le \mu\le 3$
and their right-moving superpartners $\psi^\mu$. The second part contains 26
free left-moving fermions $\lambda^a$ transforming in the vector representation of
the unbroken SO(26) gauge group. Finally the third part involves an
interacting theory of compact target space coordinates, their right-moving 
superpartners and 6 left-moving fermions transforming in the vector
representation of the SO(6) subgroup of the original
SO(32) gauge group. Together they describe a (2,2) superconformal
field theory with central charge 9.

We shall work in the $\alpha'=1$ unit in which $X^\mu$ and its fermionic
partner $\psi^\mu$ have the following operator product expansion:
\ben \label{ematterope}
&&
\p X^\mu(z) \p X^\nu(w) = -{\eta^{\mu\nu}\over 2 (z-w)^2}+\cdots, \quad
\psi^\mu (z) \psi^\nu(w) = -{\eta^{\mu\nu}\over 2 (z-w)}+\cdots\, ,
\nonumber \\
&& \bar\p X^\mu(\bar z) \bar\p X^\nu(\bar w) 
= -{\eta^{\mu\nu}\over 2 (\bar z-\bar w)^2}
+\cdots, 
\een
where $\cdots$ denote less singular terms whose knowledge will not not be
needed for our analysis.
The matter energy momentum tensor $T(z)$ and its superpartner $T_F(z)$ have
the following form
\ben \label{edefstress}
&& T(z) = - \partial X^\mu \partial X^\nu \eta_{\mu\nu} + \psi_\mu \p \psi^\mu
+ T_{int} , \quad
T_F(z) = - \psi_\mu \p X^\mu + (T_F)_{int}\, , \nonumber \\
&& \bar T(\bar z) = - \bar\partial X^\mu \bar\partial X^\nu \eta_{\mu\nu}  
+ \bar T_{int} + \bar T_\lambda\, ,
\een
where the subscript $~_{int}$ denotes contributions from the 
(2,2) SCFT and $\bar T_\lambda$ denotes the energy momentum tensor
of the 26 free left moving fermions $\lambda^a$.
Following standard procedure of bosonization of the free fermions
$\psi^\mu$ we can introduce spin fields 
$\Sigma^{(4)}_\alpha, \quad 
\Sigma^{\four}_\dotalpha$ of dimension $(0,1/4)$ with 
$\alpha$, $\dotalpha$ each taking two values.
Since they carry dimension 1/4, we cannot assign definite grassmannality
to these fields. Later we shall construct fields with definite grassmannality using
these spin fields, and write down their operator product expansion with each
other and with other fields.

Let us now turn to the interacting SCFT with (2,2) world-sheet supersymmetry.
As a consequence of the (2,2) supersymmetry there is
a right-moving R-symmetry current $J(z)$. We shall normalize
it as 
\be \label{ejjope}
J(z) J(w) = {3\over (z-w)^2} +\cdots\, .
\ee
With the help of this current we can construct conjugate pair of 
internal spin fields 
$\Sigma^{(6)}$ and $\Sigma^{(6) c}$ of dimensions (0,3/8) 
each, carrying $J$ charges $\pm 3/2$\cite{as,banksdixon}. 
Also as a consequence of (2,2) supersymmetry
$(T_F)_{int}$ can be expressed as
$T_F^++T_F^-$ with operator product expansion
\ben \label{eOPE1}
&& T_F^+(z) T_F^-(w) = {3\over 4(z-w)^3} + {1\over 4(z-w)^2} J(w)
+ {1\over 4(z-w)} T(w) + {1\over 8(z-w)} \p J(w) +\cdots, \nonumber \\
&& T_F^+(z) T_F^+(w)=
\hbox{non-singular},
\quad T_F^-(z) T_F^-(w)=\hbox{non-singular} \, .
\een
The $J$-charges carried by $T_F^\pm$, $\Sigma^{(6)}$ and
$\Sigma^{(6)c}$ are as follows:
\be \label{eJ1}
T_F^+: 1, \quad T_F^-: -1, \quad \Sigma^{(6)}:{3\over 2}, \quad \Sigma^{(6) c}: -{3\over 2}
\, .
\ee
This means for example that $J(z) T_F^\pm(w) = \pm (z-w)^{-1} T_F^\pm(w)
+\cdots$  etc.

Finally since the (2,2) supersymmetric theory is left-right symmetric it also has
a left-moving U(1) current $\bar J$ that is responsible for an unbroken U(1) gauge
symmetry of the string theory at tree level. This has operator product expansion
\be \label{ejjopebar}
\bar J(z) \bar J(w) = {3\over (\bar z-\bar 
w)^2}+\cdots\, .
\ee
The left-moving images of the other operators will not play any special role
in our analysis and so we shall not discuss them.

A useful set of composite operators in this theory are 
the dimension (0,5/8) spin fields 
 $\Sigma_\alpha$, $\Sigma^c_{\dotalpha}$, obtained 
 by combining the spin fields coming
 from the compact directions and 
 the non-compact directions:
 \be \label{edefSigma}
\Sigma_{\alpha} = \Sigma^{(6)} \, \Sigma^{(4)}_\alpha, \quad 
\Sigma^c_{\dotalpha} = \Sigma^{(6) c} \, \Sigma^{\four}_\dotalpha\, .
\ee
At the leading order, the global supersymmetry
transformation parameter $\Lambda_0$ in the $-1/2$ picture can be constructed
in terms of these fields. They are
given by 
\be
c \, e^{-\phi/2}  \Sigma_\alpha\, , \quad 
c\, e^{-\phi/2}  \Sigma^c_\dotalpha\, .
\ee
For describing states in the $-3/2$ picture it will also be useful to introduce
spin fields of `wrong chirality' as follows:
\be \label{edefSigmawrong}
\wtsp_{\alpha} = \Sigma^{(6) c} \, \Sigma^{(4)}_\alpha, \quad 
\wts_{\dotalpha} = \Sigma^{(6)} \, \Sigma^{\four}_\dotalpha\, .
\ee

Since the operators $\Sigma^{(6)}$, $\Sigma^{(6) c}$, $\Sigma^{(4)}_\alpha$,
$\Sigma^{\four}_\alpha$, $e^{-\phi/2}$ etc.\ 
have fractional values of $2\times$~dimensions, it is
hard to assign them definite grassmann parities and hence their correlation
functions will have phase ambiguities. 
For this reason we shall now choose spin fields of   
definite grassmann parity by combining the matter and ghost sector spin
fields. They are taken to be
\be
e^{-(2n+1)\phi/2} \Sigma_\alpha, \quad  e^{-(2n+1)\phi/2} \Sigma^c_\dotalpha,
\quad
e^{-(2n-1)\phi/2} \wtsp_\alpha, \quad e^{-(2n-1)\phi/2} \wts_\dotalpha,
\ee
and declared to be GSO even and
grassmann odd for $n$ even and GSO odd and grassmann even for $n$ odd.
This is consistent with the fact that the Ramond sector vertex operators in the
$-1/2$ picture (e.g. $\bar c c e^{-\phi/2} \Sigma_\alpha \bar J$) 
are taken to be grassmann odd and $e^{\pm \phi}$ are also grassmann odd. 
The operator products of
these operators with $e^{m\phi}$ for integer $m$ are determined from 
\refb{eghope} and the fact that $e^{m\phi}$ has grassmann parity $(-1)^m$.
Their operator products with $\psi^\mu$ and $(T_F)_{int}$
and with each other are given as follows:
\ben \label{emainope}
&& \psi^\mu(z) \, e^{-\phi/2} \Sigma_\alpha(w)={i\over 2} (z-w)^{-1/2}
(\gamma^\mu)_\alpha{}^\dotbeta e^{-\phi/2}\wts_\dotbeta(w)+\cdots, \nonumber \\ &&
\psi^\mu(z) \, e^{-\phi/2} \Sigma^c_\dotalpha(w)={i\over 2} (z-w)^{-1/2}
(\gamma^\mu)_\dotalpha{}^\beta e^{-\phi/2}\wtsp_\beta(w)
+\cdots,  \nonumber \\ &&
e^{-\phi/2} \Sigma_\alpha(z) \, \,  e^{-3\phi/2}\wtsp_\beta(w) 
= \ve_{\alpha\beta} (z-w)^{-2} e^{-2\phi}(w)+\cdots \, , \nonumber \\ &&
e^{-\phi/2} \Sigma^c_\dotalpha(z)  \, \,   e^{-3\phi/2}\wts_\dotbeta(w) 
= \ve_{\dotalpha\dotbeta} (z-w)^{-2} e^{-2\phi}(w) +\cdots \, ,\nonumber \\ &&
(T_F)_{int}(z) \, e^{-\phi/2} \Sigma_\alpha(w)\sim  (z-w)^{-1/2}, \quad
(T_F)_{int}(z) \, e^{-\phi/2} \Sigma^c_\dotalpha(w)\sim (z-w)^{-1/2}, \nonumber \\ &&
(T_F)_{int}(z) \, e^{-3\phi/2} \wtsp_\alpha(w)\sim(z-w)^{-1/2}, \quad
(T_F)_{int}(z) \, e^{-3\phi/2} \wts_\dotalpha(w)\sim (z-w)^{-1/2}
\, , \nonumber \\
\een  
where $\ve=\pmatrix{0 & 1\cr -1 & 0}$ and
$\gamma^\mu$ are four dimensional $\gamma$-matrices, normalized as
\be
\{\gamma^\mu, \gamma^\nu\} = 2 \eta^{\mu\nu} \, {\bf 1}\, .
\ee
The operator product of a pair of 
spin fields in which both contain $\Sigma^{(6)}$
or both contain $\Sigma^{(6) c}$ are less singular and will not be needed
for our analysis.

It follows from the first two equations of \refb{emainope} 
and the second equation of
\refb{ematterope} that we also have
\ben \label{esecondope}
&& \psi^\mu(z) \, e^{-\phi/2} \wtsp_\alpha(w)={i\over 2} (z-w)^{-1/2}
(\gamma^\mu)_\alpha{}^\dotbeta e^{-\phi/2}
\Sigma^c_\dotbeta(w)+\cdots, \nonumber \\ &&
\psi^\mu(z) \, e^{-\phi/2} \wts_\dotalpha(w)={i\over 2} (z-w)^{-1/2}
(\gamma^\mu)_\dotalpha{}^\beta e^{-\phi/2}\Sigma_\beta(w)
+\cdots\, .
\een
{}From these we can derive all other operator products, {\it e.g.} we have the
following useful relations involving GSO even operators:
\ben \label{emainopeagain}
&& e^{-\phi}\psi^\mu(z) \, e^{-\phi/2} \Sigma_\alpha(w)={i\over 2} (z-w)^{-1}
(\gamma^\mu)_\alpha{}^\dotbeta e^{-3\phi/2}\wts_\dotbeta(w)+\cdots, \nonumber \\ &&
e^{-\phi}\psi^\mu(z) \, e^{-\phi/2} \Sigma^c_\dotalpha(w)={i\over 2} (z-w)^{-1}
(\gamma^\mu)_\dotalpha{}^\beta e^{-3\phi/2}\wtsp_\beta(w)+\cdots,  \nonumber \\ &&
e^{\phi}\psi^\mu(z) \, e^{-3\phi/2} \wtsp_\alpha(w)=-{i\over 2}  (z-w)
(\gamma^\mu)_\alpha{}^\dotbeta e^{-\phi/2} \Sigma^c_\dotbeta(w)+\cdots, \nonumber \\ &&
e^{\phi}\psi^\mu(z) \, e^{-3\phi/2} \wts_\dotalpha(w)=-{i\over 2} (z-w)
(\gamma^\mu)_\dotalpha{}^\beta e^{-\phi/2} \Sigma_\beta(w)+\cdots
\, , 
\een
\ben \label{ea16}
&& e^{-\phi/2} \Sigma_\alpha(z)  e^{-\phi/2}\Sigma^c_\dotbeta(w) 
= i (z-w)^{-1} \gamma^\mu_{\alpha\dotbeta} e^{-\phi(w)} \psi^\mu(w) +\cdots\, ,
\nonumber \\
&& e^{-\phi/2} \Sigma^c_\dotalpha(z)  e^{-\phi/2}\Sigma_\beta(w) 
= i (z-w)^{-1} \gamma^\mu_{\dotalpha\beta} e^{-\phi(w)} \psi^\mu(w) +\cdots\, ,
\nonumber \\
&& \gamma^\mu_{\alpha\dotbeta}\equiv (\gamma^\mu)_\alpha{}^\dotgamma
\ve_{\dotgamma\dotbeta}, \quad
\gamma^\mu_{\dotalpha\beta}\equiv (\gamma^\mu)_\dotalpha{}^\gamma
\ve_{\gamma\beta}, \quad (\gamma^\mu)_{\alpha\dotbeta}=
(\gamma^\mu)_{\dotbeta\alpha}\, ,
\een
\ben \label{esstildep}
&& e^{-\phi/2} \Sigma_\alpha(z)  e^{\phi/2} \wtsp_\beta(w) 
= -\ve_{\alpha\beta} (z-w)^{-1} +\cdots \nonumber \\
&& e^{-\phi/2} \Sigma^c_\dotalpha(z)  e^{\phi/2} \wts_\dotbeta(w) 
= -\ve_{\dotalpha\dotbeta} (z-w)^{-1} +\cdots \, .
\een
etc.

For our analysis we shall also need the first subleading terms in the operator
product expansion in \refb{esstildep}. 
This must be of the form of a constant multiplying a dimension (0,1)
current. Since $J$ is a dimension (0,1) current, the next term in
each operator product expansion could contain 
$J$ multiplied by a constant. 
The constant can be determined by analyzing
the sphere three point function
\be \label{esph}
\langle e^{-2\phi} J(y) 
e^{-\phi/2} \Sigma_\alpha(z)  e^{\phi/2} \wtsp_\beta(w) 
\rangle
\ee
and its counterpart with dotted indices. Using \refb{eJ1}, \refb{edefSigma},
\refb{edefSigmawrong} and \refb{esstildep} 
we can express this correlator as
\ben \label{esph2}
&& {3\over 2} \left( {1\over y-z} - {1\over y-w}\right) \langle e^{-2\phi}(y)
e^{-\phi/2} \Sigma_\alpha(z)  e^{\phi/2} \wtsp_\beta(w) 
\rangle \nonumber \\
&=& {3\over 2} {(z-w)\over (y-z) (y-w)} (-\ve_{\alpha\beta}) (y-z)^{-1} (y-w) 
(z-w)^{-1}
\nonumber \\
&=& -{3\over 2} \ve_{\alpha\beta}  (y-z)^{-2} \, ,
\een
where in going from the first to the second line we have evaluated the
correlator in the first line using the general form of the three point function
and the operator product expansion \refb{esstildep}.
On the other hand if the operator product 
$e^{-\phi/2} \Sigma_\alpha(z)  e^{\phi/2} \wtsp_\beta(w)$ had contained
a term of the form $c_{\alpha\beta}\, J(w)$ then in the $z\to w$ limit 
\refb{esph} would have behaved as
\be
c_{\alpha\beta} \langle e^{-2\phi} J(y)  J(w)\rangle =
3\, c_{\alpha\beta} (y-w)^{-2}\, ,
\ee
using \refb{ejjope}.  Comparing this with the $z\to w$ limit of \refb{esph2}
we get $c_{\alpha\beta}=-\ve_{\alpha\beta}/2$. A similar analysis for
the correlator involving dotted indices gives the coefficient 
$\ve_{\dotalpha\dotbeta}/2$ of $J$ -- the additional minus sign arising
from the fact that $\Sigma^c_\dotalpha$ and 
$\Sigma_\alpha$ carry opposite $J$ charges.

What other operators could the operator products \refb{esstildep}
contain on the right hand side? First of all they could contain the
ghost current $\p\phi$. They could also contain bilinears of free fermions
$\psi^\mu\psi^\nu$.
Therefore we can write
\ben \label{esstilde}
&& e^{-\phi/2} \Sigma_\alpha(z)  e^{\phi/2} \wtsp_\beta(w) 
= -\ve_{\alpha\beta} \left\{ (z-w)^{-1} + {1\over 2} J(w) \right\}
+\OO(\p\phi) + \OO(\psi^\mu\psi^\nu)+
\OO(z-w)
 \nonumber \\
&& e^{-\phi/2} \Sigma^c_\dotalpha(z)  e^{\phi/2} \wts_\dotbeta(w) 
= -\ve_{\dotalpha\dotbeta} 
\left\{ (z-w)^{-1} - {1\over 2} J(w)\right\} +
\OO(\p\phi) + \OO(\psi^\mu\psi^\nu)+
 \OO(z-w)\, . \nonumber \\
\een

This summarizes the relevant information on the holomorphic and anti-holomorphic 
operators in the (2,2) SCFT. However for our analysis we shall also need to review
the properties of certain non-(anti)-holomorphic operators in this theory -- namely
those which describe the vertex operators associated with target
space fields belonging to a special class of chiral and
anti-chiral multiplets. We describe the relevant properties below:
\begin{enumerate}
\item Let $\sigma$ be a complex scalar field in the target space, 
describing the scalar component of such a massless
chiral superfield. 
The $-1$ picture vertex operators of the field $\sigma$ and its complex 
conjugate take the form
\be \label{echiral}
\VV_{\sigma} = \bar c \,  c\,  e^{-\phi} \Vm_\sigma, \quad \VV_{\sigma^*} 
= \bar c\,  c \, e^{-\phi} \Vm_{\sigma^*},
\ee
where $\Vm_\sigma$, $\Vm_{\sigma^*}$ are matter sector operators of 
dimension $(1,1/2)$ in the (2,2) SCFT associated with the compact directions. 
We normalize
$\Vm_\sigma$, $\Vm_{\sigma^*}$ so that
\be \label{enorm}
\Vm_\sigma(z) \Vm_{\sigma^*}(w) = (z-w)^{-1} (\bar z - \bar w)^{-2},
\quad e^{-\phi} \Vm_\sigma(z) e^{-\phi}
\Vm_{\sigma^*}(w) = -(z-w)^{-2} (\bar z - \bar w)^{-2} e^{-2\phi(w)}\, ,
\ee
in accordance with \refb{eno0}, \refb{eno2}.
\item The $J$-charges carried by $\Vm_{\sigma}$ and $\Vm_{\sigma^*}$ are 
\be \label{ejcharge}
\Vm_{\sigma}\, : \, 1, \qquad \Vm_{\sigma^*} \, :\, -1\, .
\ee
\item The operator product of $T_F^\pm$ with $\Vm_\sigma$, $\Vm_{\sigma^*}$
take the form
\ben \label{edefwtvm}
&&T_F^-(z) \Vm_\sigma(w) = -(w-z)^{-1} \wt\Vm_\sigma(w) +\cdots, \quad
T_F^+(z) \Vm_{\sigma^*}(w) = -(w-z)^{-1} \wt\Vm_{\sigma^*}(w) +\cdots, 
\nonumber \\ &&
T_F^-(z) \Vm_{\sigma^*}(w)=\hbox{non-singular}, 
\quad T_F^+(z) \Vm_{\sigma}(w)=\hbox{non-singular}\, ,
\een
where $\wt\Vm_{\sigma}$ and $\wt\Vm_{\sigma^*}$ are matter sector vertex operators
of dimension (1,1).
It follows from \refb{eJ1}, \refb{ejcharge} that the $J$-charges carried by
$\wt \Vm_\sigma$ and 
$\wt\Vm_{\sigma^*}$ are 
\be \label{ejch1}
\wt\Vm_{\sigma}\, : \, 0, \qquad \wt\Vm_{\sigma^*} \, :\, 0\, .
\ee
$\wt \Vm_\sigma$ and 
$\wt\Vm_{\sigma^*}$ will be useful 
for constructing zero picture vertex operators.
 \item Using the two dimensional superconformal algebra one can show that
these vertex operators satisfy the identities
\ben \label{etfid}
T_F^+(w) \wt \Vm_\sigma(z) &=&  -{1\over 4} (w-z)^{-2} \, \Vm_\sigma(z) - 
{1\over 4} (w-z)^{-1} \p_z \Vm_\sigma(z) 
+ \hbox{non-singular terms} \nonumber \\
T_F^-(w) \wt \Vm_{\sigma^*}(z) &=& -{1\over 4} (w-z)^{-2} \, \Vm_{\sigma^*}(z) - 
{1\over 4} (w-z)^{-1} \p_z \Vm_{\sigma^*}(z) 
+ \hbox{non-singular terms} \nonumber \\
T_F^-(w) \, \wt \Vm_\sigma(z) &=& \hbox{non-singular terms}, \quad
T_F^+(w) \, \wt \Vm_{\sigma^*}(z) \, = \, \hbox{non-singular terms}\, .
\nonumber \\
\een
\item Given a chiral multiplet field $\sigma$ there are also Ramond sector
vertex operators $V^f_\sigma$, $V^f_{\sigma^*}$
of dimension $(1,3/8)$ in the internal CFT from which we can build the full
vertex operator of the space-time fermions.
In the spirit of \refb{edefSigma}, 
\refb{edefSigmawrong} we combine them with the spin fields
$\Sigma^{(4)}_\alpha$, $\Sigma^{(4)}_\dotalpha$ to define
\be \label{edefcomb}
\Vm^f_{\sigma,\alpha}\equiv \Vm^f_{\sigma} \Sigma^{(4)}_\alpha,
\qquad \Vm^f_{\sigma,\dotalpha}\equiv\Vm^f_{\sigma} \Sigma^{(4)}_\dotalpha,
\qquad 
\Vm^f_{\sigma^*,\alpha}\equiv\Vm^f_{\sigma^*} \Sigma^{(4)}_\alpha,
\qquad \Vm^f_{\sigma^*,\dotalpha}\equiv
\Vm^f_{\sigma^*} \Sigma^{(4)}_\dotalpha
\, .
\ee
The vertex operators of
the fermionic partners of $\sigma^*$ and
$\sigma$ in the $-1/2$ picture are given by, respectively,
\be \label{evmf1}
\bar c c e^{-\phi/2} \Vm^f_{\sigma^*,\dotalpha}, \quad 
\bar c c e^{-\phi/2} \Vm^f_{\sigma,\alpha}\, .
\ee
The basic operator products involving these operators are:
\ben \label{ebasicsigma}
\Vm_{\sigma^*}(w) \, e^{-\phi/2} \wts_\dotalpha(z) &=& (w-z)^{-1/2} 
e^{-\phi/2}\Vm^f_{\sigma^*,\dotalpha}(z) +\cdots\, , 
\nonumber \\
\Vm_{\sigma}(w)  \, e^{-\phi/2} \wtsp_\alpha(z) &=& (w-z)^{-1/2}
e^{-\phi/2} \Vm^f_{\sigma,\alpha}(w) +\cdots\, , 
\nonumber \\
\Vm_{\sigma^*}(w) \, e^{-\phi/2} \Sigma_\alpha(z) &=& -(w-z)^{-1/2} 
e^{-\phi/2}\Vm^f_{\sigma^*,\alpha}(z) +\cdots\, , 
\nonumber \\
\Vm_{\sigma}(w)  \, e^{-\phi/2} \Sigma^c_\dotalpha(z) &=& -(w-z)^{-1/2}
e^{-\phi/2} \Vm^f_{\sigma,\dotalpha}(w) +\cdots\, .
\een
The relative minus signs between the first two terms and the last two terms 
in \refb{ebasicsigma} has been included to ensure that the operator product
of $\psi^\mu$ with $e^{-\phi/2} V^f_{\sigma,\alpha}$ etc. follow
the same pattern as the operator product of $\psi^\mu$ with $e^{-\phi/2}
\Sigma_\alpha$ etc. given in \refb{emainope}, \refb{esecondope}:
\ben \label{enewope}
\psi^\mu(w) \, e^{-\phi/2} V^f_{\sigma,\alpha}(z) &=& 
{i\over 2} (w-z)^{-1/2} \, (\gamma^\mu)_\alpha{}^{\dotbeta} \, 
e^{-\phi/2} V^f_{\sigma,\dotbeta}(z) \nonumber \\
\psi^\mu(w) \, e^{-\phi/2} V^f_{\sigma,\dotalpha}(z) &=& 
{i\over 2} (w-z)^{-1/2} \, (\gamma^\mu)_\dotalpha{}^{\beta} \, 
e^{-\phi/2} V^f_{\sigma,\beta}(z) \nonumber \\
\psi^\mu(w) \, e^{-\phi/2} V^f_{\sigma^*,\dotalpha}(z) &=& 
{i\over 2} (w-z)^{-1/2} \, (\gamma^\mu)_\dotalpha{}^{\beta} \, 
e^{-\phi/2} V^f_{\sigma^*,\beta}(z) \nonumber \\
\psi^\mu(w) \, e^{-\phi/2} V^f_{\sigma^*,\alpha}(z) &=& 
{i\over 2} (w-z)^{-1/2} \, (\gamma^\mu)_\alpha{}^{\dotbeta} \, 
e^{-\phi/2} V^f_{\sigma^*,\dotbeta}(z) \, .
\een
In arriving at \refb{enewope} 
we have used the fact that $V_\sigma$ and $V_{\sigma^*}$
anti-commute with $\psi^\mu$.
We also have
\ben
\Vm_{\sigma^*}(w) \, e^{-\phi/2} \Sigma^c_\dotalpha(z) &=& \OO((w-z)^{1/2})\, ,
\nonumber \\
\Vm_{\sigma}(w) \, e^{-\phi/2} \Sigma_\alpha(z) &=&\OO((w-z)^{1/2})\, ,
\nonumber \\
\Vm_{\sigma^*}(w) \, e^{-\phi/2} \wtsp_\alpha(z) &=& \OO((w-z)^{1/2})\, ,
\nonumber \\
\Vm_{\sigma}(w) \, e^{-\phi/2} \wts_\dotalpha(z) &=&\OO((w-z)^{1/2})\, .
\een

Using \refb{enorm} and \refb{ebasicsigma} we get 
\ben \label{ekk1}
\Vm_{\sigma}(w) \, e^{-\phi/2}\Vm^f_{\sigma^*,\dotalpha}(z)
&=& (w-z)^{-1/2}  (\bar w - \bar z)^{-2} e^{-\phi/2} \wts_\dotalpha(z) 
+\cdots
\nonumber \\
\Vm_{\sigma^*}(w) \, e^{-\phi/2} \Vm^f_{\sigma,\alpha}(z) 
&=& (w-z)^{-1/2}  (\bar w - \bar z)^{-2} e^{-\phi/2} \wtsp_\alpha(z)
+\cdots\, , \nonumber \\
\Vm_{\sigma}(w) \, e^{-\phi/2}\Vm^f_{\sigma^*,\alpha}(z)
&=& -(w-z)^{-1/2}  (\bar w - \bar z)^{-2} e^{-\phi/2} \Sigma^c_\dotalpha(z) 
+\cdots
\nonumber \\
\Vm_{\sigma^*}(w) \, e^{-\phi/2} \Vm^f_{\sigma,\dotalpha}(z) 
&=& -(w-z)^{-1/2}  (\bar w - \bar z)^{-2} e^{-\phi/2} \Sigma_\alpha(z)
+\cdots
\, .
\een
From \refb{emainope}, 
\refb{ebasicsigma}, \refb{ekk1} and standard manipulations in 
conformal field theory we can derive other required operator product
expansions. Some useful relations involving GSO even operators
are given below:
\ben \label{evmf}
e^{-\phi/2} \Sigma_\alpha(z) \, \, e^{-\phi} \Vm_{\sigma^*}(w)  
&=& (z-w)^{-1} e^{-3\phi/2}  \Vm^f_{\sigma^*,\alpha}(w) +\cdots\, , 
\nonumber \\
e^{-\phi/2} \Sigma^c_\dotalpha(z) \, \, e^{-\phi} \Vm_{\sigma}(w)  
&=& (z-w)^{-1} e^{-3\phi/2}  \Vm^f_{\sigma,\dotalpha}(w) +\cdots\, , \nonumber \\
e^{-\phi/2} \Sigma^c_\dotalpha(z) \, \, e^{-\phi} \Vm_{\sigma^*}(w)  
&=& \hbox{non-singular terms} \, , \nonumber \\
e^{-\phi/2} \Sigma_\alpha(z) \, \, e^{-\phi} \Vm_{\sigma}(w)  
&=& \hbox{non-singular terms} \, ,   \nonumber \\ 
e^{-3\phi/2} V^f_{\sigma^*,\beta}(z) \, \, 
e^{-\phi/2} V^f_{\sigma,\alpha}(w) &=& (z-w)^{-2} 
(\bar z -\bar w)^{-2}
\ve_{\beta\alpha}
e^{-2\phi(w)} +\cdots\, , \nonumber \\
e^{-3\phi/2} V^f_{\sigma,\dotbeta}(z) \, \, 
e^{-\phi/2} V^f_{\sigma^*,\dotalpha}(w) 
&=& (z-w)^{-2} (\bar z -\bar w)^{-2} \ve_{\dotbeta\dotalpha}
e^{-2\phi(w)} +\cdots \, , \nonumber \\
e^{-\phi/2} \Sigma_\alpha(z) \, \,  e^{-\phi/2}  \Vm^f_{\sigma,\beta}(w) 
&=& (z-w)^{-1} \ve_{\alpha\beta} e^{-\phi} V_\sigma(w) +\cdots\, , 
\nonumber \\
e^{-\phi/2} \Sigma^c_\dotalpha(z) \, \,  e^{-\phi/2}  \Vm^f_{\sigma^*,\dotbeta}(w) 
&=& (z-w)^{-1} \ve_{\dotalpha\dotbeta} e^{-\phi} V_{\sigma^*}(w) +\cdots\, , 
\nonumber \\
e^{-\phi/2} \Sigma_\alpha(z) \, \,  e^{-\phi/2}  \Vm^f_{\sigma^*,\dotbeta}(w) 
&=& \hbox{non-singular terms}\, , 
\nonumber \\
e^{-\phi/2} \Sigma^c_\dotalpha(z) \, \,  e^{-\phi/2}  \Vm^f_{\sigma,\beta}(w) 
&=& \hbox{non-singular terms}\, , 
\nonumber \\e^{\phi}\psi^\mu(z) \, e^{-3\phi/2} \Vm^f_{\sigma^*,\alpha}(w)
&=& -{i\over 2}  (z-w)\, 
(\gamma^\mu)_\alpha{}^\dotbeta 
e^{-\phi/2}  \Vm^f_{\sigma^*,\dotbeta}(w) 
+\cdots, \nonumber \\ 
e^{\phi}\psi^\mu(z) \, e^{-3\phi/2}  \Vm^f_{\sigma,\dotalpha}(w) 
&=&-{i\over 2} (z-w)
(\gamma^\mu)_\dotalpha{}^\beta e^{-\phi/2} 
\Vm^f_{\sigma,\beta}(w) 
+\cdots \, ,
\nonumber \\
e^{-\phi}\psi^\mu(z) \, \, e^{-\phi/2}  \Vm^f_{\sigma^*,\dotalpha}(w)
&=& {i\over 2} (z-w)^{-1}(\gamma^\mu)_\dotalpha{}^\beta e^{-3\phi/2} \Vm^f_{\sigma^*,\beta}(w) +\cdots \, ,
\nonumber \\
e^{-\phi}\psi^\mu(z) \, \, e^{-\phi/2}  \Vm^f_{\sigma,\alpha}(w)
&=&
{i\over 2} (z-w)^{-1} (\gamma^\mu)_\alpha{}^\dotbeta e^{-3\phi/2} 
\Vm^f_{\sigma,\dotbeta}(w) +\cdots\nonumber \\
e^{\phi}\psi^\mu(z) \, \, e^{-5\phi/2}  \Vm^f_{\sigma^*,\dotalpha}(w)
&=& {i\over 2} (z-w)^{2}(\gamma^\mu)_\dotalpha{}^\beta e^{-3\phi/2} \Vm^f_{\sigma^*,\beta}(w) +\cdots \, ,
\nonumber \\
e^{\phi}\psi^\mu(z) \, \, e^{-5\phi/2}  \Vm^f_{\sigma,\alpha}(w)
&=&
{i\over 2} (z-w)^{2} (\gamma^\mu)_\alpha{}^\dotbeta e^{-3\phi/2} 
\Vm^f_{\sigma,\dotbeta}(w) +\cdots\, . \nonumber \\
\een

\item It follows from \refb{eJ1}, \refb{ejcharge} and \refb{ebasicsigma}
that the $J$-charges carried by these
new internal Ramond sector operators are
\be \label{eJ2}
\Vm^f_{\sigma}: -{1\over 2}, \quad \Vm^f_{\sigma^*}: {1\over 2}\, .
\ee
\item Finally let us turn to the $\bar J$ charge carried by the relevant fields. Here there
is no universal result for all chiral and anti-chiral multiplets since different fields may
transform in different representations of the gauge group. However 
the theory contains special chiral multiplet fields which are 
singlets under the
$SO(26)$ gauge group but carry $\pm 2$ unit of charge 
under the U(1) gauge group. 
The number of such fields depends on the Hodge numbers of the Calabi-Yau
manifold. In order to simplify our analysis we shall assume that there is a unique
chiral multiplet field $\chi$ of this type carrying $\bar J$ charge 2 -- this would arise
{\it e.g.} for a Calabi-Yau manifold with $h_{1,1}=1$, $h_{1,2}=0$.
This field
is the one that appears in \refb{epot} and is
important because it will condense to restore supersymmetry which otherwise
is broken by a Fayet-Iliopoulos  term generated at one loop\cite{ADS}.
The $\bar J$ charges carried by the associated vertex operators are
\be \label{ejbarcharge}
\Vm_{\chi}\, : \, 2, \qquad \Vm_{\chi^*} \, :\, -2, \qquad 
\wt \Vm_{\chi}\, : \, 2, \qquad \wt \Vm_{\chi^*} \, :\, -2\, .
\ee
\end{enumerate}

\sectiono{Supersymmetry restoration}  \label{srestore}

We shall now apply the general analysis carried out earlier in this
paper to  SO(32) heterotic string theory compactified on a Calabi-Yau
manifold. As reviewed in \S\ref{slowen}, typically one loop
quantum corrections will generate a Fayet-Iliopoulos 
D-term\cite{DSW,ADS,DIS}. This causes
supersymmetry to be spontaneously broken at the original perturbative vacuum.
However using effective field theory it is easy to see that there should be a nearby
vacuum where supersymmetry is restored. Our goal will be to demonstrate how this
result can be derived from superstring perturbation theory without invoking low energy
effective field theory. This will also tell us how to compute perturbative superstring
amplitudes around the shifted vacuum going beyond what is possible in low energy
effective field theory.

\subsection{Construction of the vacuum solutions in 
superstring perturbation theory} \label{s6.2}

If we work in the subspace of the field space in which 
$\chi=(\chi_R+i\chi_I)/\sqrt 2$ is real, 
then
at the level of order $g_s^3$ contribution to the equations of motion
there are three solutions in the low energy effective field theory 
described by the potential \refb{epot}. These 
correspond to
$\chi_R=0, \pm g_s \sqrt {2\, K}$. The existence of these three solutions would show
up in the analysis of \S\ref{sreview} through the existence of three possible solutions for
$|\psi_1\rangle$ in \refb{esoln}, \refb{esol1} when we study the contribution
to the equation of motion to order
$g_s^3$. Now from the point of view of low energy effective field theory it is clear
that the solution corresponding to $\chi_R=0$ should break supersymmetry 
while the solutions corresponding to $\chi_R=\pm g_s\sqrt {2 K}$ should restore 
supersymmetry. Our goal will be to prove, from the point of view of superstring
perturbation theory, that this is indeed what happens.

Let 
\be \label{edefVchi}
\VV_{\chi_R}=\bar c c e^{-\phi} \Vm_{\chi_R}
\ee 
denote the vertex operator
of $\chi_R$ in the $-1$ picture where $\Vm_{\chi_R}$ is a dimension 
(1,1/2) operator in the matter 
SCFT associated with compact directions. 
We shall normalize $V_\chi$ as in \refb{enorm}. It then follows from the analysis of
\S\ref{sreal}  that the coefficient of
$\VV_{\chi_R}$ in the expansion of the string field is real.
We now proceed to solve the equations of motion following \refb{esoln}, \refb{esol1}, beginning
with the ansatz
\be \label{epsi1}
|\Psi_1\rangle =|\psi_1\rangle = \beta \, g_s|\VV_{\chi_R}\rangle \, ,
\ee
where $\beta$ is a real constant. At this order $\beta$ remains undetermined
since $Q_B|\Psi_1\rangle=0$ for all $\beta$. 
At the next order
we get
\be \label{epsi2}
|\Psi_2\rangle =  -{b_0^+\over L_0^+} (1 - {\bf P})[~]_1 
- {1\over 2} \beta^2 \, g_s^2\, 
{b_0^+\over L_0^+} (1 - {\bf P})  [\VV_{\chi_R}\VV_{\chi_R}]_0 
+|\psi_2\rangle
\ee
\be \label{ets}
|\psi_2\rangle= \beta\, g_s\,
|\VV_{\chi_R}\rangle + |\tilde\psi_2\rangle\, ,
\ee
where $|\tilde\psi_2\rangle$ is a contribution of order $g_s^2$,
determined from the equation
\be \label{epsi4}
Q_B |\tilde\psi_2\rangle = - {\bf P} \left([~]_1 + {1\over 2} \beta^2 \, g_s^2[\VV_{\chi_R}\VV_{\chi_R}]_0 \right)
\, .
\ee
In \refb{epsi2}, \refb{epsi4}, $[\cdots]_g$ denotes that $[\cdots]$ has to be 
computed to $g$-loop
order, and
we have made use of the fact that $[~]_0=0$, $[A]_0=0$ for all $|A\rangle
\in \wh\HH_T$.
We have shown in appendix \ref{svan} that both terms on
the right hand side of \refb{epsi4} vanish and hence we can take
\be \label{evanpsi2}
|\tilde\psi_2\rangle=0\, .
\ee
According to \refb{econdaa},
in order to extend the solution to order $g_s^3$ we need to ensure that for any
BRST invariant physical state $|p\rangle$ 
of ghost number 2, $L_0^+=0$  and
picture number $-1$, we have
\be \label{ean2}
\langle p| c_0^- | ([~]_1 + [\Psi_1]_1 +{1\over 2}  [\Psi_2{}^2]_0 + {1\over 6}
[\Psi_1{}^3]_0) \rangle= \OO(g_s^4)\, .
\ee
In writing down this equation we have used the fact that $[\Psi]$ receives contribution from 1-loop
and higher and hence to get $[\Psi]$ accurate up to order $g_s^3$ 
it is enough the keep terms
up to order $g_s$ in $|\Psi\rangle$. On the other hand $[\Psi^2]$, $[\Psi^3]$ etc.
receive contribution from tree level, with the expansion for $\Psi$ beginning at order $g_s$;
therefore in $[\Psi^2]$ is is enough to keep terms in $|\Psi\rangle$ accurate to order $g_s^2$ and
in $[\Psi^3]$ it is enough to keep terms in $|\Psi\rangle$ accurate up to order $g_s$. 
Substituting \refb{epsi1}-\refb{ets} and
\refb{evanpsi2} into \refb{ean2} we get
\ben \label{ean3}
&& \Big\langle p\Big| c_0^- \Big| \Big( [~]_1 + \beta\, g_s\, [\VV_{\chi_R}]_1 
+
 {1\over 2} \beta^2 g_s^2 \, [\VV_{\chi_R}\VV_{\chi_R}]_0  
- \beta \, g_s \, [\VV_{\chi_R} {b_0^+} (L_0^+)^{-1} (1 - {\bf P}) [~]_1]_0
 \nonumber \\ &&
- {1\over 2} \beta^3 g_s^3 [\VV_{\chi_R} b_0^+ (L_0^+)^{-1} (1-{\bf P}) [\VV_{\chi_R}\VV_{\chi_R}]_0]_0
+{1\over 6} \beta^3 \, g_s^3 \, [\VV_{\chi_R}\VV_{\chi_R}\VV_{\chi_R}]_0
\Big) \nonumber \\ &=& 
\OO(g_s^4)\, .
\een
As already mentioned below \refb{epsi4}, 
the analysis of appendix \ref{svan} shows that 
the first and the
third terms on the left hand side of \refb{ean3} vanish.
To analyze the rest of the terms we recall that 
$\VV_{\chi_R}$ can be expressed as $(\VV_\chi+\VV_{\chi^*})/\sqrt 2$ where
$\VV_{\chi}$ and $\VV_{\chi^*}$ carry $\bar J$ charges 2 and $-2$
respectively (see \refb{ejbarcharge}). 
Since the second and fourth terms each has a single $\VV_{\chi_R}$, in order to get
a non-zero result $|p\rangle$ must carry $\bar J$ charge 2 or $-2$.
Since the fifth and the sixth terms each has three $\VV_{\chi_R}$'s, to get a 
non-zero
result $|p\rangle$ must have $\bar J$ charge $\pm 2$ or $\pm 6$. 
Now, the $\bar L_0$ eigenvalue of a state is bounded from below by 
$-1+\bar j^2/6$,
$\bar j$ being the $\bar J$ charge of the state. Therefore,
there are no states carrying $\bar j=\pm 6$ and $L_0^+=0$.  By
our assumption stated in the last paragraph of \S\ref{ehetrev}, 
the only zero momentum states that can have $\bar j=\pm 2$
and 
$L_0^+=0$  are $\VV_\chi$ and 
$\VV_{\chi^*}$. Furthermore there is a $Z_2$ conjugation symmetry under
which $\chi_R$ is even and the imaginary part $\chi_I$ of $\chi$ is odd.
This fixes $|p\rangle$ uniquely to be $|\VV_{\chi_R}\rangle$ for getting
non-zero contribution to various terms on the left hand side of \refb{ean3}.
Therefore \refb{ean3} can be written as\footnote{If the theory contains
multiple fields carrying $\bar J$ charges $\pm 2$, then there will be more
candidates for $|p\rangle$ leading to more equations. But there will also
be more parameters labelling the solution since the leading order solution 
can be taken to be an arbitrary linear combination of these states. Therefore we
shall have multiple equations involving multiple variables. This is analogous
to the situation in effective field theory stated in footnote \ref{feffective}.}
\ben \label{ean4}
&& -{1\over 2} \beta^3 \, g_s^3 
 \left\{ \VV_{\chi_R}\VV_{\chi_R} \left(b_0^+ (L_0^+)^{-1} (1-{\bf P}) [\VV_{\chi_R}\VV_{\chi_R}]_0
\right)\right\}_0
+ {1\over 6}  \beta^3 \, g_s^3 \{\VV_{\chi_R}\VV_{\chi_R}
\VV_{\chi_R}\VV_{\chi_R}\}_0
\nonumber \\ 
&& +\beta\, g_s \, \{\VV_{\chi_R} \VV_{\chi_R}\}_1 
- \beta \, g_s \, \left\{\VV_{\chi_R}\VV_{\chi_R} 
\left({b_0^+} (L_0^+)^{-1} (1 - {\bf P}) [~]_1\right)\right\}_0 
= \OO(g_s^4)\, ,
\een
where, as for $[\cdots]$, $\{\cdots \}_g$ denotes that we have to compute the contribution
to $\{\cdots\}$ up to genus $g$.

Eq.\refb{ean4} can be expressed in a more convenient form by 
defining
\ben \label{edefG}
\gb^{(4,0)}(0,\chi_R;0,\chi_R;0,\chi_R;0,\chi_R)
&=& \{\VV_{\chi_R}\VV_{\chi_R}
\VV_{\chi_R}\VV_{\chi_R}\}_0
\nonumber \\ && 
- 3 \left\{ \VV_{\chi_R}\VV_{\chi_R} \left(b_0^+ (L_0^+)^{-1} (1-{\bf P}) [\VV_{\chi_R}\VV_{\chi_R}]_0
\right)\right\}_0\, , \nonumber \\
g_s^2 \, \gb^{(2,2)}(0,\chi_R;0,\chi_R) &=&
\{\VV_{\chi_R} \VV_{\chi_R}\}_1 
- \left\{\VV_{\chi_R}\VV_{\chi_R} 
\left({b_0^+} (L_0^+)^{-1} (1 - {\bf P}) [~]_1\right)\right\}_0 \, .
\nonumber \\
\een
We recognize $\gb^{(4,0)}(0,\chi_R;0,\chi_R;0,\chi_R;0,\chi_R)$ as the
tree level on-shell four point function of four external zero momentum 
$\chi_R$ state in the $\beta=0$ vacuum, 
with the first term giving the 1PI part of the amplitude and
the second term giving the 1PR part, with the sum over s, t and u-channel
graphs giving the factor of 3. Similarly $g_s^2\gb^{(2,2)}(0,\chi_R;0,\chi_R)$
gives the one loop on-shell two point function of two external 
 zero momentum 
$\chi_R$ states in the $\beta=0$ vacuum, with the first term giving the 1PI
contribution and the second term giving the 1PR contribution where
a tree level 3-point vertex is attached by a propagator to a one loop
one point vertex.\footnote{Since ${\bf P} [\VV_{\chi_R}\VV_{\chi_R}]_0$ and 
${\bf P} [~]_1$ vanish according to the analysis of appendix
\ref{svan}, we can drop the $1-{\bf P}$ factors in \refb{edefG}. Therefore
$\gb^{(4,0)}(0,\chi_R;0,\chi_R;0,\chi_R;0,\chi_R)$ and
$\gb^{(2,2)}(0,\chi_R;0,\chi_R)$ can be interpreted as
the usual on-shell amplitudes of string theory in the perturbative vacuum
without any need 
for the subtraction, or insertion of external zero momentum state
$|\psi_2\rangle$ of the kind discussed in \S\ref{smatrix}.}
Both $G^{(4,0)}$ and $G^{(2,2)}$ 
have been normalized so that they do not
carry any factor of $g_s$.
\refb{ean4} can now be expressed as
\be \label{etd}
{1\over 6}\,  \beta^3 \, g_s^3 \gb^{(4,0)}(0,\chi_R;0,\chi_R;0,\chi_R;0,\chi_R)
+ g_s^3 \,  \beta\, \gb^{(2,2)}(0,\chi_R;0,\chi_R)=0\, .
\ee
\refb{etd} has a trivial solution $ \beta=0$ that describes
the original perturbative vacuum, and non-trivial solutions at
\be \label{ebetvalue}
 \beta^2 = - 6\,  \gb^{(2,2)}(0,\chi_R;0,\chi_R) / 
\gb^{(4,0)}(0,\chi_R;0,\chi_R;0,\chi_R;0,\chi_R)\, .
\ee
It is possible to show by
explicit calculation that these solutions correspond to real values of $\beta$.
This in turn shows the existence of three solutions
to the equations of motion to order $g_s^3$, in agreement 
with what we get from the
low energy effective field theory.

In the next subsection we shall show that while supersymmetry is broken at order
$g_s^2$ in the $\beta=0$ vacuum, in the other two vacua at $\beta\sim 1$, supersymmetry
is restored at least to order $g_s^2$.

\subsection{Supersymmetry of the vacuum solution} \label{s6.3}

We shall now examine whether or not
the vacuum solutions of \S\ref{s6.2} 
possess global supersymmetry   to
order $g_s^2$. This corresponds to existence of solution to \refb{eglobal} to order $g_s^2$
for some Ramond sector state $|\Lambda_{\rm global}\rangle$. 
Using \refb{esusy2} for $k=2$, and using the
fact that the expansion of $|\Psi_{\rm vac}\rangle$ begins at order $g_s$ and that of $|\Lambda_{\rm global}\rangle$
begins at order $g_s^0$, we see
that the existence of unbroken global supersymmetry 
$|\Lambda_{\rm global}\rangle$
to order $g_s^2$ will require
us to show that
\be \label{etest1}
\LL_1(\goldc)\equiv \langle \goldc| c_0^- \XX_0 |[\Psi_1\Lambda_0]_0\rangle
= \OO(g_s^2)\, ,
\ee
and
\be \label{elphi}
\LL_2(\goldc)\equiv
\Big\langle \goldc \Big| c_0^- \XX_0 \Big|\Big([\Psi_2\Lambda_1]_0
+ [\Lambda_0] _1
+{1\over 2} [\Psi_1\Psi_1\Lambda_0]_0 \Big)\Big\rangle
=\OO(g_s^{3})\, ,
\ee
for any BRST invariant state $|\goldc\rangle$  of ghost number 3, picture number $-3/2$ and $L_0^+=0$.

$|\Lambda_0\rangle$ is given by the zeroth order global supersymmetry
transformation parameter 
\be \label{edeflambda0}
|\Lambda_0\rangle = |\lambda_0\rangle = |c e^{-\phi/2} \Sigma\rangle\, ,
\ee
where $\Sigma$ is a matter sector spin field -- 
one of the operators $\Sigma_\alpha$ or $\Sigma^c_{\dotalpha}$ given in 
\refb{edefSigma}. The existence of $|\Lambda_1\rangle$ requires 
\refb{etest1}, which can be interpreted as $\{ (\XX_0\goldc) \Psi_1\Lambda_0\}_0
=\OO(g_s^2)$.
$\goldc$ is a physical state of ghost number 3 and
picture number $-3/2$. The general form of these states 
is given in \refb{ep2} and the
result of the action of $\XX_0$ on these states is given in \refb{ep21}. Combining this 
with the form of $\VV_{\chi_R}$ and $\lambda_0$ given respectively in \refb{edefVchi} and \refb{edeflambda0} 
we have
\be \label{erep}
\VV_{\chi_R}: \bar c c e^{-\phi} \Vm_{\chi_R}, \quad \lambda_0: c e^{-\phi/2} \Sigma,\quad
\XX_0\goldc: \bar c c\eta e^{\phi/2} V^f, \quad 
\bar c c\bar \partial^2 \bar c \, e^{-\phi/2}\, \hat\Sigma\, ,
\ee
where  both $\Sigma$ and $\hat\Sigma$ represent some operator from the list $(\Sigma_{\alpha},
\Sigma^c_{\dotalpha})$ given in \refb{edefSigma}. 
The contribution to $\{ (\XX_0\goldc) \VV_{\chi_R} \lambda_0\}_0$ from the first
candidate for $\XX_0\goldc$ vanishes by $\phi$ charge conservation (and also by $\xi-\eta$
charge conservation). The contribution from the second candidate for
$\XX_0\goldc$ involves the following matter part of the correlation function:
\be \label{erep1}
\langle \Vm_{\chi_R} \, \Sigma \, \hat\Sigma\rangle\, .
\ee
Since $V_{\chi_R}=(V_\chi+V_{\chi^*})/\sqrt 2$ where $V_\chi$ and 
$V_{\chi^*}$ carry $\bar J$ charges 2 and $-2$ respectively and $\Sigma$,
$\hat\Sigma$ are neutral under $\bar J$, such a three point function 
vanishes by $\bar J$ charge conservation. Therefore \refb{etest1} holds.

$|\Lambda_1\rangle$ can now be computed using 
\refb{esolam}, \refb{etauk}  and \refb{epsi1}:
\be \label{enk1}
|\Lambda_1\rangle = |\lambda_1\rangle - {b_0^+\over L_0^+} (1 - {\bf P}) \XX_0[\Psi_1\Lambda_0]_0
= |\lambda_0\rangle - \beta \, g_s \, {b_0^+\over L_0^+} (1 - {\bf P}) \XX_0[\VV_{\chi_R}\lambda_0]_0\, ,
\ee
\be \label{elambda1}
|\lambda_1\rangle = |\lambda_0\rangle+|\tilde \lambda_1\rangle, 
\qquad Q_B|\tilde\lambda_1\rangle = - {\bf P}\XX_0 [\Psi_1\Lambda_0]_0
= -\beta g_s {\bf P} \XX_0 [\VV_{\chi_R} \lambda_0]_0\, .
\ee
We shall now argue that the right hand side of the second equation
in \refb{elambda1} vanishes. For this we have to show that the inner 
product of $c_0^-\XX_0[\VV_{\chi_R} \lambda_0]_0$ with all states 
in $\HH_T$ with  $L_0^+=0$ vanishes. Due to ghost and picture number
conservation we can restrict to states carrying
ghost number 3 and picture number $-3/2$.
The analysis described in the paragraph containing 
\refb{erep}, \refb{erep1} already shows that the inner product
$\langle\goldc|c_0^- | \XX_0[\VV_{\chi_R} \lambda_0]_0\rangle$ vanishes
for all candidate $\goldc$. There is only one more class of 
states
carrying ghost number 3, picture number $-3/2$ and $L_0^+=0$ --
they are the unphysical  states $\bar c c \bar\p^2\bar c e^{-3\phi/2} \hat\Sigma$
with $\hat\Sigma$ given by one of the operators from the list
$(\Sigma_\alpha, \Sigma^c_\dotalpha)$. However $\bar J$ charge
conservation prevents this from having a non-zero inner product
with $\XX_0[\VV_{\chi_R} \lambda_0]_0$. Therefore we conclude that the right hand
side of the second equation in \refb{elambda1} vanishes, and hence
we can take
\be \label{elambda2}
|\tilde\lambda_1\rangle = 0\, .
\ee

Substituting \refb{epsi1}-\refb{ets}, \refb{evanpsi2} and 
\refb{enk1}-\refb{elambda2} into \refb{elphi} we get
\ben \label{eL1}
\LL_2(\goldc) &=& \beta \, g_s \, 
\{ (\XX_0\goldc) \VV_{\chi_R} \lambda_0\}_0  +\{(\XX_0\goldc)\lambda_0\}_1
\nonumber \\ &&
- \left\{ (\XX_0 \goldc) \lambda_0 \left(
b_0^+ (L_0^+)^{-1} (1 - {\bf P}) [~]_1\right) \right\}_0
\nonumber \\ &&
-{1\over 2} \beta^2 g_s^2 \left\{ (\XX_0\goldc) \lambda_0 
\left({b_0^+} (L_0^+)^{-1} (1 - {\bf P})
[\VV_{\chi_R} \VV_{\chi_R}]_0\right) \right\}_0 
\nonumber \\ &&
+{1\over 2}\beta^2 g_s^2 \{(\XX_0\goldc) \lambda_0 \VV_{\chi_R} \VV_{\chi_R}\}_0 
\nonumber\\&&
-
\beta^2 g_s^2 \left\{ (\XX_0\goldc) \VV_{\chi_R} \left( b_0^+(L_0^+)^{-1}
(1 - {\bf P}) \XX_0[\VV_{\chi_R}\lambda_0]_0 \right)\right\}_0
\nonumber \\ && +\OO(g_s^3)
\, . 
\een
The first term is the same as the one given in \refb{etest1} and has been 
shown to be zero already. Therefore the 
condition for unbroken supersymmetry to order $g_s^2$ can 
now be expressed as
\ben \label{eL2}
&& \{(\XX_0\goldc)\lambda_0\}_1 
- \left\{ (\XX_0 \goldc) \lambda_0 \left(
b_0^+ (L_0^+)^{-1} (1 - {\bf P}) [~]_1\right) \right\}_0  \nonumber \\ &&
-{1\over 2} \beta^2 g_s^2 \left\{ (\XX_0\goldc) \lambda_0 
\left({b_0^+} (L_0^+)^{-1} (1 - {\bf P}) 
[\VV_{\chi_R} \VV_{\chi_R}]_0\right) \right\}_0 
+{1\over 2}\beta^2 g_s^2 \{(\XX_0\goldc) \lambda_0 \VV_{\chi_R} \VV_{\chi_R}\}_0 
\nonumber \\ &&
-
\beta^2 g_s^2 \left\{ (\XX_0\goldc) \VV_{\chi_R} \left( b_0^+(L_0^+)^{-1}
(1 - {\bf P}) \XX_0[\VV_{\chi_R}\lambda_0]_0 \right)\right\}_0
\nonumber \\ 
&=& \OO(g_s^3)\, ,
\een
for all $\XX_0\goldc$ of the form given in \refb{erep}. 
We have shown
in appendix \ref{sunique} that  for all but one 
$\XX_0\goldc$ listed in
\refb{erep},
each term on the left hand side of \refb{eL2} vanishes identically to the required
order. 
The particular $\goldc$ for which the terms do not vanish is of the form
\be \label{eunique}
\XX_0\goldc =\bar c\, c \, \eta \, e^{\phi/2}\, \wts\, \bar J \, ,
\ee
where $\wt\Sigma$ is a matter sector spin field of `wrong chirality' --
one of the operators in the list \refb{edefSigmawrong} -- satisfying
\be 
\wt\Sigma(z) \Sigma (w) \propto (z-w)^{-5/8} + \cdots\, .
\ee
For example if $\Sigma$ is chosen to be $\Sigma_\alpha$ then
we have to take $\wt\Sigma$ to be $\ve^{\alpha\beta} \wt\Sigma_\beta$. 
The $\goldc$ corresponding to \refb{eunique} is given by
$-4(\p c+\bar\p \bar c) \bar c c e^{-3\phi/2} \wt\Sigma \, \bar J$. 
This is turn is
conjugate to $\bar c c e^{-\phi/2} \Sigma \, \bar J$ -- the zero momentum
gaugino vertex operator in the $-1/2$ picture associated with the U(1) gauge 
group -- up to a proportionality constant. 
This is related to the fact that in the
present situation this gaugino acts as the goldstino in the situation when supersymmetry
is broken.

Let us now analyze the fate of global supersymmetry in the three vacua obtained in
\S\ref{s6.2}.
From \refb{eL2} we see that for $\beta=0$ only the first 
two terms survive. 
The sum of these terms
give 
the one loop two point function of $\XX_0\goldc$ and $\lambda_0$ in the
perturbative vacuum.
We shall see
explicitly in \S\ref{sdilaton} (see \refb{e9.9b}) that this two point function 
is non-zero. Therefore at
the $\beta=0$ vacuum supersymmetry is broken at order $g_s^2$.

Since the left hand side of eq.\refb{eL2} is quadratic in $\beta$, 
there are two other values of $\beta\sim 1$, related by a change of sign, where the the left hand side of \refb{eL2} vanishes to order
$g_s^2$.
In fact since the sum of the first two terms in \refb{eL2} 
is the genus 1 two point function, it is of order $g_s^2$. 
Therefore we can factor out an overall factor of $g_s^2$ from the left
hand side of \refb{eL2} and get a $g_s$ independent 
quadratic equation for $\beta$ whose solutions are at $\beta\sim 1$.
Therefore 
at these values of $\beta$ supersymmetry is restored to order $g_s^2$. How are these
related to the value of $\beta$ obtained in \S\ref{s6.2}?
One can of course try to do a 
direct computation and compare the results. However we can avoid doing this by using the
result of \S\ref{stadpole} that unbroken supersymmetry to order $g_s^2$ implies vanishing
tadpole to order $g_s^3$. Therefore once $\beta$ has been adjusted to make the left hand side
of \refb{eL2} vanish to order $g_s^2$, it also makes the left hand side of \refb{ean4} 
vanish to order $g_s^3$. In other words the non-zero solutions for $\beta$ obtained from 
\refb{ean4} and \refb{eL2} must be the same.

This establishes that while supersymmetry is broken at order $g_s^2$ at the perturbative
vacuum $\beta=0$ it is restored at least to order $g_s^2$ at the shifted vacuum where 
$\beta\sim 1$. 

\newcommand{\Beta}{\beta}

\sectiono{Bose-Fermi degeneracy at the shifted vacuum} \label{sbose}

In this section we shall explicitly 
compute the renormalized mass$^2$ of $\chi_R$ and its
fermionic partner to order $g_s^2$ at the shifted vacuum and show that they
are equal.

\subsection{Scalar mass$^2$ to order $g_s^2$} \label{s4}

In this subsection we shall compute the renormalized mass$^2$ of the scalar field 
$\chi_R$ to order $g_s^2$ at the shifted vacuum. For 
this we shall follow the procedure described in \refb{esa3}-\refb{esa5}. 
To order $g_s^2$, the relevant iterative equations are
\ben \label{ekkk1}
|\Phi_0\rangle &=& |\phi_2\rangle \nonumber \\
|\Phi_1\rangle &=& - {b_0^+\over L_0^+} 
(1-P) K |\Phi_0\rangle + |\phi_2\rangle
\nonumber \\
|\Phi_2\rangle &=& - {b_0^+\over L_0^+} 
(1-P) K |\Phi_1\rangle + |\phi_2\rangle
\nonumber \\Q_B|\phi_2\rangle &=& -P K |\Phi_1\rangle 
= P K {b_0^+\over L_0^+} 
(1-P) K |\phi_2\rangle - P K |\phi_2\rangle\, ,
\een
where $P$ now denotes the projection operator onto mass level zero 
states.
Using the definition of $K$, and keeping terms to order $g_s^2$, 
we can express the last equation as
\ben \label{ekkk2}
Q_B|\phi_2\rangle  &=& P[\Psi_1{b_0^+}(L_0^+)^{-1} 
(1-P) [\Psi_1\phi_2]_0]_0  - P[\phi_2]_1 - P[\Psi_2\phi_2]_0 
-{1\over 2} P[\Psi_1{}^2 \phi_2]_0 \nonumber \\
&=& \beta^2 g_s^2 P[\VV_{\chi_R}{b_0^+}(L_0^+)^{-1} 
(1-{P}) [\VV_{\chi_R} \phi_2]_0]_0- P[\phi_2]_1 \nonumber \\ &&
- \beta g_s P[\VV_{\chi_R} \phi_2]_0 
+ P[(b_0^+ (L_0^+)^{-1} (1-{\bf P})[~]_1) \phi_2]_0
\nonumber \\ &&
+{1\over 2} \beta^2 g_s^2 P[(b_0^+ (L_0^+)^{-1} (1-{\bf P})
[\VV_{\chi_R} \VV_{\chi_R}]_0) \phi_2]_0 -{1\over 2} \beta^2 g_s^2 P[
\VV_{\chi_R} \VV_{\chi_R} \phi_2]_0\, ,
\een
where in the last step we have used the expression for the approximations
 $|\Psi_1\rangle$ and $|\Psi_2\rangle$ to the vacuum
solution given in \refb{epsi1}-\refb{evanpsi2}.

So far our discussion has been  for general state, but for computing the 
mass renormalization of $\chi_R$ we shall consider the following
ansatz for $\phi_2$:\footnote{In general we should have allowed
$\phi_2$ to be an arbitrary linear combination of all mass level zero states,
but in the present situation symmetry consideration prevents the mixing 
of $\VV_{\chi_R}$ with other states at mass level zero.}
\be 
\phi_2 = \VV_{\chi_R}(k), \quad  \VV_{\chi_R}(k)
\equiv \VV_{\chi_R} e^{ik\cdot X}
\ee
with $k^2\simeq 0$. The non-trivial part of \refb{ekkk2} comes from the
inner product of this equation with arbitrary state of momentum $-k$ and
$L_0^+\simeq 0$. Using various charge conservation one can show that
the only contribution comes from the inner product with the state
$\langle \VV_{\chi_R}(-k)| c_0^-$. Using the normalization
\be 
 \langle \VV_{\chi_R}(-k)| c_0 \bar c_0 |\VV_{\chi_R}(k)\rangle
 = -1\, ,
 \ee
 which in turn follows from \refb{evacnorm} and the normalization of $\VV_\chi$ given
 in \refb{enorm},
we can express the inner product of  
\refb{ekkk2} with  $\langle \VV_{\chi_R}(-k)| c_0^-$ as
\ben \label{ekkk3}
-{k^2\over 4}  
&=&  
\beta^2 g_s^2 \{ \VV_{\chi_R}(-k) 
\VV_{\chi_R}{b_0^+}(L_0^+)^{-1} 
(1-{P}) [\VV_{\chi_R}  \VV_{\chi_R}(k) ]_0\}_0-  
\{ \VV_{\chi_R}(-k)   \VV_{\chi_R}(k) \}_1 
\nonumber \\ &&
- \beta g_s  \{ \VV_{\chi_R}(-k)  \VV_{\chi_R}  \VV_{\chi_R}(k) \}_0 
+ \{ \VV_{\chi_R}(-k)  (b_0^+ (L_0^+)^{-1} (1-{\bf P})[~]_1) \VV_{\chi_R}(k) \}_0
\nonumber \\ &&
+{1\over 2} \beta^2 g_s^2 \{ \VV_{\chi_R}(-k)  (b_0^+ (L_0^+)^{-1} (1-{\bf P})
[\VV_{\chi_R} \VV_{\chi_R}]_0) \VV_{\chi_R}(k) \}_0
\nonumber \\ &&
-{1\over 2} \beta^2 g_s^2 \{\VV_{\chi_R}(-k)
\VV_{\chi_R} \VV_{\chi_R} \VV_{\chi_R}(k)\}_0\, .
\een
The $\{ \VV_{\chi_R}(-k)  \VV_{\chi_R}  \VV_{\chi_R}(k) \}_0$ term
vanishes due to the by now familiar $\bar J$
charge conservation rule.
Now the renormalized mass $m_B$ of $\chi_R$ is obtained by demanding
that the above equation has a solution at $k^2=-m_B^2$. Since 
we expect $m_B$ to be of order $g_s$, and hence $k^2$ to be of order
$g_s^2$, and since each term on the right hand side is already of order 
$g_s^2$, we can set $k=0$ while evaluating the right hand side. In this case
the projection operator $P$ reduces to ${\bf P}$ and $\VV_{\chi_R}(\pm k)$
reduces to $\VV_{\chi_R}$. 
Setting $k^2=-m_B^2$
on the left hand side of \refb{ekkk3} we get
\ben\label{ekkk4}
{m_B^2\over 4} &=& {3\over 2}
\beta^2 g_s^2 \{ \VV_{\chi_R} 
\VV_{\chi_R}{b_0^+}(L_0^+)^{-1} 
(1-{\bf P}) [\VV_{\chi_R}  \VV_{\chi_R} ]_0\}_0 
- {1\over 2} \beta^2 g_s^2  \{  \VV_{\chi_R} 
\VV_{\chi_R} \VV_{\chi_R}  \VV_{\chi_R}\}_0 \nonumber \\ &&
- \{ \VV_{\chi_R}  \VV_{\chi_R}\}_1 + 
\{ \VV_{\chi_R}   \VV_{\chi_R} (b_0^+ (L_0^+)^{-1} (1-{\bf P})[~]_1)\}_0\, .
\een
Using \refb{edefG} this can be written as
\be\label{etwopoint}
{m_B^2\over 4} =
-{1\over 2} \Beta^2 g_s^2\gb^{(4,0)}(0,\chi_R;0,\chi_R;0,\chi_R;0,\chi_R)
- g_s^2 \, \gb^{(2,2)}(0,\chi_R;0,\chi_R)\, .
\ee
$\Beta$ has been determined in \refb{ebetvalue}.
Substituting \refb{ebetvalue} into \refb{etwopoint}, we get
\be \label{emr2}
m_B^2 =
8 \, g_s^2 \, \gb^{(2,2)}(0,\chi_R;0,\chi_R)
=
- {4\over 3}  \Beta^2 g_s^2 \gb^{(4,0)}(0,\chi_R;0,\chi_R;0,\chi_R;0,\chi_R)
\, .
\ee
Rest of this subsection will be devoted to the computation of
$\gb^{(4,0)}(0,\chi_R;0,\chi_R;0,\chi_R;0,\chi_R)$.

Before we proceed, we need to introduce some notations.
\begin{enumerate}
\item 
As in \S\ref{srestore} we shall denote 
by $\VV_{\chi}$, $\VV_{\chi^*}$ the unintegrated 
vertex operators for zero momentum
$\chi$, $\chi^*$ in $-1$ picture:
\be \label{edefvvchi}
\VV_{\chi} = \bar c\, c \, e^{-\phi} \Vm_\chi, \quad \VV_{\chi^*} = \bar c \, c\, 
e^{-\phi} \Vm_{\chi^*},
\ee
where $\Vm_\chi$, $\Vm_{\chi^*}$ are matter sector operators of dimension $(1,1/2)$. 
$\Vm_\chi$, $\Vm_{\chi^*}$ are normalized as in \refb{enorm}.
$\VV_{\chi_R}$ is obtained from these via the relation
\be \label{evchir}
\VV_{\chi_R} = {1\over \sqrt 2} (\VV_{\chi}+\VV_{\chi^*})\, .
\ee
{}From \refb{edefvvchi} we can calculate the zero picture unintegrated 
vertex operators:\footnote{Even though for
brevity we use the label $z$ to label the argument of a vertex operator, it is to be
understood that it depends both on $z$ and $\bar z$.}
\ben \label{ezerop}
\wt\VV_{\chi}(z) &\equiv& \lim_{w\to z} \XX(w) \VV_{\chi}(z) = \bar c c \wt \Vm_\chi(z)
- {1\over 4} \bar c \, \eta\, e^\phi \, \Vm_\chi(z) \nonumber \\
\wt\VV_{\chi^*}(z) &\equiv& \lim_{w\to z} \XX(w) \VV_{\chi^*}(z) = \bar c c \wt \Vm_{\chi^*}(z)
- {1\over 4}  \bar c \, \eta\, e^\phi \,\Vm_{\chi^*}(z),
\een
where $\wt \Vm$ has been defined in  \refb{edefwtvm}.
\item Using \refb{ebrs1}-\refb{ebrstcurrent}  and \refb{etfid} for
$\sigma=\chi$ it follows that 
\ben \label{ebrsid}
&& \{Q_B, \VV_{\chi}(z)\}=0, \quad \{Q_B, \wt\VV_{\chi}(z)\}=0, \quad
\nonumber \\ &&
\{Q_B, \wt \Vm_{\chi}(z) \} = \p_z \left(
c\wt \Vm_{\chi}(z)
- {1\over 4} \eta\, e^\phi \, \Vm_{\chi}(z)\right) + \p_{\bar z} (\bar c \wt \Vm_{\chi}(z))\, .
\een
There are also similar identities with $\chi$ replaced by $\chi^*$.
\end{enumerate}

We now return to the computation of 
$\gb^{(4,0)}(0,\chi_R;0,\chi_R;0,\chi_R;0,\chi_R)$. For this
we need to choose the locations of the 
two PCO's. 
If we represent the amplitude as the integral of a four point correlation function of
the $\chi_R$ vertex operators in the complex plane, then the PCO's are located
at two points in this plane.
The final result should be independent of their locations
as long as they are chosen in a gluing compatible manner and are compatible
with the permutation symmetries of the external vertices. Gluing compatibility requires
that as we approach a degeneration limit where two of the vertex operators come close
(which is conformally equivalent to other two vertex operators coming close) one of
the PCO's should be close to the two vertex operators which are coming close 
while the other PCO should be at a finite distance away from them or at infinity
(which is conformally equivalent to its being close to the other two vertex operators).
This can be achieved by taking one of the PCO's to coincide with one of the
$\chi_R$ vertex operators for all values of the moduli, converting this to a zero
picture vertex operator and letting  the other PCO be a 
function of the moduli such that near any degeneration the other PCO remains  
(conformally) away from the zero picture 
vertex operator.  In general we have to make the prescription symmetric under the
permutation of the four punctures by taking averages, but in this case that is not
necessary since the vertex operators at the four punctures are identical.

Once we have fixed the choice of the PCO locations, the computation of
$\gb^{(4,0)}$ involves computing the appropriate world-sheet correlator
and integrating the result over the moduli space of four punctured sphere.
For this we can keep three of the punctures at fixed locations and integrate
over the location of the fourth puncture. The final result is independent of
which of the puncture locations we choose to integrate over, and we exploit
this freedom to integrate over the location of the zero picture vertex
operator.

We shall begin by writing down the general expression for the four point
amplitude where the two PCO's are inserted at arbitrary points
$u$ and $v$ and then specialize to the case where one of the PCO's
approach the location of one of the vertices. 
We insert three vertex operators at fixed positions $z_1,z_2,z_3$ 
and the fourth one at a variable position $z$ and let the PCO locations
$u$ and $v$ depend on $z$ and $\bar z$. 
The general expression,
following the rules described in \cite{1408.0571}, is
\ben \label{egengb}
&& \gb^{(4,0)}(0,\chi_R; 0, \chi_R; 0, \chi_R; 0, \chi_R) \nonumber \\
&=&{1\over 2\pi i} \int  dz\wedge d\bar z \bigg\langle
\VV_{\chi_R}(z_1) \VV_{\chi_R}(z_2) \VV_{\chi_R}(z_3) \nonumber \\
&& \bigg( \XX(u) \XX(v) b_z \bar b_{\bar z} 
+ (\XX(v)\p\xi(u) \p_z u+\XX(u) \p\xi(v) \p_z v) \bar b_{\bar z}
\nonumber \\ && 
- (\XX(v)\p\xi(u) \p_{\bar z} u+\XX(u) \p\xi(v) \p_{\bar z} v) b_{z}
+ \p\xi(u) \p\xi(v) (\p_z u \p_{\bar z} v - \p_z v \p_{\bar z} u) \bigg)
 \VV_{\chi_R}(z) 
\bigg\rangle \, , \nonumber \\
\een
where
\be 
b_z\equiv 
\ointop_z b(w) dw, \qquad
\bar b_{\bar z} = \ointop_z \bar b(\bar w) d\bar w \, .
 \ee
Here $\ointop_z$ denotes an integration contour encircling $z$,
normalized so that $\ointop_z dw (w-z)^{-1}=1$, 
$\ointop_z d\bar w (\bar w-\bar z)^{-1}=1$.
The contours must be chosen {\it so as to keep $u$ and $v$ outside the
contours.} Using \refb{evchir} we can now replace each of the 
$\VV_{\chi_R}$ factor in terms of $\VV_{\chi}$ and $\VV_{\chi^*}$. The
resulting correlator will have 16 terms, but only six of them, containing equal
number of $\chi$'s and $\chi^*$'s, will be non-zero. A typical term is
given by 
\ben \label{edefa40}
\AAA^{(4,0)} &\equiv& {1\over 4} {1\over 2\pi i} \int  dz\wedge d\bar z \bigg\langle
\VV_{\chi^*}(z_1) \VV_{\chi}(z_2) \VV_{\chi^*}(z_3) \nonumber \\
&& \bigg( \XX(u) \XX(v) b_z \bar b_{\bar z} 
+ (\XX(v)\p\xi(u) \p_z u+\XX(u) \p\xi(v) \p_z v) \bar b_{\bar z}
\nonumber \\ && 
- (\XX(v)\p\xi(u) \p_{\bar z} u+\XX(u) \p\xi(v) \p_{\bar z} v) b_{z}
+ \p\xi(u) \p\xi(v) (\p_z u \p_{\bar z} v - \p_z v \p_{\bar z} u) \bigg)
 \VV_{\chi}(z)
\bigg\rangle  \, . \nonumber \\ 
\een
The other terms are related to this by different assignments of $\chi$ and
$\chi^*$ to different punctures. They can be generated from \refb{edefa40}
by first summing
over cyclic permutation of $z_1,z_2, z_3$, and then summing over the
exchange of all the $\chi$'s with $\chi^*$'s in each of these terms. Now
the correlator has a symmetry under the exchange of all the $\chi$'s with
$\chi^*$'s; hence summing over this exchange produces a factor of 2.
Therefore if we denote by $\bar \AAA^{(4,0)}$ the result of averaging 
$\AAA^{(4,0)}$ over the cyclic permutation of  $z_1,z_2, z_3$, then
we can write
\be \label{egamreal}
\gb^{(4,0)}(0,\chi_R; 0, \chi_R; 0, \chi_R; 0, \chi_R)
= 6\, \bar \AAA^{(4,0)}\, .
\ee

We now turn to the evaluation of $\AAA^{(4.0)}$. 
Using the result \refb{eorifin} we can replace 
$dz\wedge d\bar z$ by $2id^2 z$.
Now we take the limit
$v\to z$. In this limit $\p_{\bar z} v=0$, $\p_z v=1$. In the process of taking
the limit we have to pass $v$ through the integration contours involved
in the definitions of $b_z$ and $\bar b_{\bar z}$.
Using the relations
$\ointop_v dw b(w) \XX(v)=\p\xi(v)$,
$\ointop_v d\bar w \bar b(\bar 
w) \XX(v)=0$ one can show that all the terms involving
$\p\xi(v)$ cancel. Therefore we are left with
\ben \label{ee6}
&& \AAA^{(4,0)} \nonumber \\
&=& {1\over 4} \int {d^2 z\over \pii} \bigg\langle
\VV_{\chi^*}(z_1) \VV_{\chi}(z_2) \VV_{\chi^*}(z_3) \nonumber \\ &&
\bigg(\XX(u) \ointop_z b(w) dw \ointop_z \bar b(\bar w) d\bar w + \p\xi(u)\, \p_z  u \, \ointop_z \bar b(\bar w) d\bar w 
- \p\xi(u) \, \p_{\bar z} u \,  \ointop_z b(w) dw  \bigg) \wt\VV_{\chi}(z)
\bigg\rangle \nonumber \\
&=& {1\over 4} \int {d^2 z\over \pii} \bigg\langle
\VV_{\chi^*}(z_1) \VV_{\chi}(z_2) \VV_{\chi^*}(z_3) \nonumber \\ &&
\bigg(\XX(u) \wt \Vm_\chi(z)  + \p\xi(u)\, \p_z  u \bigg(
c\wt \Vm_{\chi}(z)
- {1\over 4} \eta\, e^\phi \, \Vm_{\chi}(z)\bigg)
+ \p\xi(u) \, \p_{\bar z} u \,   
\bar c \wt \Vm_{\chi}(z) \bigg)
\bigg\rangle  \, .
\een

For reasons that will become clear soon,
we now introduce an auxiliary quantity
$\wt \AAA^{(4,0)}$  by taking the $u\to z_2$ limit of
\refb{ee6}. In this limit $\p_z u$ and $\p_{\bar z}u$ vanish, and we get
\be \label{e8.18}
\wt\AAA^{(4,0)} = {1\over 4} \int {d^2 z\over \pii} \bigg\langle
\VV_{\chi^*}(z_1) \wt\VV_{\chi}(z_2) \VV_{\chi^*}(z_3) \wt \Vm_{\chi}(z) \bigg\rangle 
\, .
\ee
Using \refb{edefvvchi} and \refb{ezerop} we now see that in \refb{e8.18},
$\phi$ charge conservation forces us to pick the $\bar c c \wt V_\chi(z_2)$
term from the zero picture vertex operators $\wt \VV_\chi(z_2)$.
Therefore the matter part of the correlation function now involves two factors of
$\wt V_\chi$ and two factors of $V_{\chi^*}$.
Eqs.\refb{ejcharge}, \refb{ejch1} then show that the total $J(z)$ charge carried 
by all the 
vertex operators in the correlation function 
add up to $-2$ and hence the result
vanishes by $J$-charge conservation.  
Therefore $\wt\AAA^{(4,0)}$ vanishes
identically and we can write
\be \label{egamdiff}
\AAA^{(4,0)} = \AAA^{(4,0)}
- \wt\AAA^{(4,0)}\, .
\ee
Our strategy will be to express the right hand side of \refb{egamdiff} 
as a total
derivative in the moduli space. 
This can then be expressed as sum of boundary terms 
which are easier to evaluate.

We  use the relations
\be 
\wt\VV_{\chi}(z_2) = \XX(z_2) \VV_{\chi}(z_2)
\ee
and
\be 
\XX(u) - \XX(z_2) = \{Q_B, \xi(u)-\xi(z_2)\}
\ee
to write
\be
\XX(u) \VV_{\chi}(z_2) - \wt\VV_{\chi}(z_2) 
= \{Q_B, \xi(u)-\xi(z_2)\}\, \VV_{\chi}(z_2) \, .
\ee
Using this we get
\ben
&& \AAA^{(4,0)} - \wt \AAA^{(4,0)}
\nonumber \\
&=& {1\over 4} \int {d^2 z\over \pii} \bigg\langle
\VV_{\chi^*}(z_1) \VV_{\chi}(z_2) \VV_{\chi^*}(z_3) \nonumber \\ &&
\left(\{Q_B, \xi(u)-\xi(z_2)\} \wt \Vm_{\chi}(z)  + \p\xi(u)\, \p_z  u \left(
c\wt \Vm_{\chi}(z)
- {1\over 4} \eta\, e^\phi \, \Vm_{\chi}(z)\right)
+ \p\xi(u) \, \p_{\bar z} u \,   
\bar c \wt \Vm_{\chi}(z) \right)
\bigg\rangle\, . \nonumber \\
\een
We can now deform the BRST contour and use the relations \refb{ebrsid}
to arrive at
\ben \label{egamdiffa}
&& \AAA^{(4,0)} - \wt \AAA^{(4,0)}
\nonumber \\
&=& {1\over 4} \int {d^2 z\over \pii}\,  \p_z \bigg\langle (\xi(u) -\xi(z_2))\, 
\VV_{\chi^*}(z_1) \VV_{\chi}(z_2) \VV_{\chi^*}(z_3)  \left(
c\wt \Vm_{\chi}(z)
- {1\over 4} \eta\, e^\phi \, \Vm_{\chi}(z)\right) \bigg\rangle\nonumber \\
&& + {1\over 4} \int {d^2 z\over \pii} \, \p_{\bar z} \bigg\langle (\xi(u) -\xi(z_2))\, 
\VV_{\chi^*}(z_1) \VV_{\chi}(z_2) \VV_{\chi^*}(z_3)  \bigg(
\bar c(\bar z) \wt \Vm_{\chi}(z) \bigg)\bigg\rangle  \, .
\een
Now on the sphere a non-zero correlation function requires the number of $c$ insertions
minus the number of $b$ insertions to be 3, and similarly 
the number of $\bar c$ insertions
minus the number of $\bar b$ insertions to be 3. Since each of 
$\VV_{\chi^*}(z_1)$, $\VV_{\chi}(z_2)$ and $\VV_{\chi^*}(z_3)$ contains a
factor of $\bar c c$, we see that the only non-vanishing term in \refb{egamdiffa} is the
term involving $\eta$. Using \refb{egamdiff} we can now express \refb{egamdiffa} as
\be \label{ex2}
\AAA^{(4,0)} = -{1\over 16}
\int {d^2 z\over \pii}\,  \p_z \bigg\langle (\xi(u) -\xi(z_2))\, 
\VV_{\chi^*}(z_1) \VV_{\chi}(z_2) \VV_{\chi^*}(z_3)  
\eta\, e^\phi \, \Vm_{\chi}(z)\bigg\rangle\, .
\ee

Since this is the integral of a total derivative, the result can be expressed as boundary
contributions. There are three relevant boundaries corresponding to $z$ coming close
to $z_1$, $z_2$ and $z_3$. In order to get a non-zero contribution from the boundary 
$z\to z_i$, the term inside the total derivative must be of the form
\be \label{ebound}
(z-z_i)^{-\alpha} (\bar z-\bar z_i)^{-\alpha-1}
\ee 
for $\alpha\ge 0$. We shall now examine
the contribution near each of these boundaries.
\begin{enumerate}
\item First let us examine the contribution from the boundary near $z=z_2$.
This is controlled by the operator product of $\eta e^\phi \Vm_\chi(z)$ and 
$(\xi(u)-\xi(z_2))\VV_{\chi}(z_2)=(\xi(u)-\xi(z_2))
\bar c c e^{-\phi} \Vm_\chi(z_2)$. 
Possible negative powers of $(\bar z-\bar z_2)$ can come from the 
operator product of $\Vm_\chi(z)$ and $\Vm_\chi(z_2)$. Now since $\Vm_\chi$ carries 
$\bar J$ charge of $2$, any operator appearing in the product 
$\Vm_\chi(z) \Vm_\chi(z_2)$ must have $\bar J$ charge $4$. Standard CFT results
and \refb{ejjope} now tells us that the left-handed conformal weight $\bar h$  of 
such an operator has a lower bound of $4^2/6=8/3$. Therefore the lowest
power of $(\bar z-\bar z_2)$ that we can get in the operator product of
 $\Vm_\chi(z)$ and $\Vm_\chi(z_2)$ is $(\bar z - \bar z_2)^{2/3}$. Comparison with
 \refb{ebound} now shows that it is not possible to get a non-vanishing boundary 
 contribution from $z$ near $z_2$.
\item Next we turn to the boundary contribution from $z$ near $z_1$. In this case
the matter part of the operator product involves the combination 
$\Vm_\chi(z) \Vm_{\chi^*}(z_1)$ carrying total $\bar J$ charge zero, and hence we
may get sufficiently negative power of $(\bar z - \bar z_1)$ so as to get a non-zero
boundary contribution. To evaluate this we have to carefully study the full operator
product expansion. Now as discussed earlier, in this limit we have to keep $u$ away from
$z$ (and hence also $z_1$).
The relevant operator product that could produce a singular term of the form
\refb{ebound} as $z\to z_1$ is
\be \label{ex1}
\eta\, e^\phi \, \Vm_{\chi}(z) \, \VV_{\chi^*}(z_1)
= -\eta(z)\, \bar c(z_1) \, c(z_1) \, e^\phi(z) \, e^{-\phi(z_1)} \, \Vm_{\chi}(z)\Vm_{\chi^*} (z_1) \, .
\ee
Since $e^{\phi(z)} e^{-\phi(z_1)} = (z-z_1) + \OO\left((z-z_1)^2\right)$, in order to get
a term of the form \refb{ebound}, we must pick those terms in the operator
product $\Vm_{\chi}(z)\Vm_{\chi^*} (z_1)$ whose holomorphic part has at least a 
singularity of order $(z-z_1)^{-1}$. Using the fact that $\Vm_\chi$ and $\Vm_{\chi^*}$ are
operators of conformal weight (1,1/2), and the normalization condition
\refb{enorm}, we see that 
the relevant terms in the operator product $\Vm_{\chi}(z)\Vm_{\chi^*} (z_1)$ are of the form
\be \label{evvope}
\Vm_{\chi}(z)\Vm_{\chi^*} (z_1) = (z-z_1)^{-1} (\bar z - \bar z_1)^{-2} 
+  (z-z_1)^{-1} (\bar z - \bar z_1)^{-1}\, \bar\JJ(\bar z_1) + \hbox{less singular terms}
\ee
where $\bar \JJ$ is some dimension (1,0) left-handed current. Assuming that $\bar J$
is the only left-handed U(1) current in the matter CFT associated with the
compact directions, we see that $\bar\JJ$ must be proportional to 
$\bar J$. In order to find the constant of proportionality we use \refb{ejbarcharge} to
write
\ben \label{ejcorr}
&& \left\langle \bar J(\bar w) \Vm_{\chi}(z)\Vm_{\chi^*} (z_1) \right\rangle
= 2 \left( {1\over \bar w - \bar z} - {1\over \bar w - \bar z_1}\right)
\left \langle \Vm_{\chi}(z)\Vm_{\chi^*} (z_1) \right\rangle \nonumber \\ &=&
2 \left( {1\over \bar w - \bar z} - {1\over \bar w - \bar z_1}\right) (z-z_1)^{-1} (\bar z - \bar z_1)^{-2} 
\nonumber \\
&=& 2 (\bar w - \bar z)^{-1} ( \bar w - \bar z_1)^{-1} (z-z_1)^{-1} (\bar z - \bar z_1)^{-1} \, .
\een
Taking $z\to z_1$ limit on both sides and using \refb{evvope} we get
\be
(z-z_1)^{-1} (\bar z - \bar z_1)^{-1} \langle \bar J(\bar w) \bar\JJ (\bar z_1)\rangle
= 2 ( \bar w - \bar z_1)^{-2} (z-z_1)^{-1} (\bar z - \bar z_1)^{-1} \, .
\ee
Comparing with \refb{ejjopebar} we now get
\be \label{ejnorm}
\bar\JJ = {2\over 3} \bar J\, .
\ee
Using \refb{ex1}, \refb{evvope} and \refb{ejnorm} we see that the relevant part of the
operator product expansion involved in the computation of  the boundary contribution to
\refb{ex2} from $z=z_1$ is given by
\be
\eta\, e^\phi \, \Vm_{\chi}(z) \, \VV_{\chi^*}(z_1)
= -\eta(z_1)\, \bar c(z_1) \, c(z_1) \, \left[
(\bar z - \bar z_1)^{-2} 
+  {2\over 3} (\bar z - \bar z_1)^{-1}\, \bar J(\bar z_1) 
\right]\, .
\ee
Comparing with \refb{ebound} we see that only the second term inside the square
bracket contributes to the boundary term from the $z=z_1$ end. Furthermore 
one can easily show\cite{1408.0571}  that 
the boundary contribution from the $z=z_1$ end of
$\int d^2 z \p_z (1/\bar z)$ is given by $-\pi$. Using this the net contribution
to the right hand side of \refb{ex2} from the $z=z_1$ boundary is given by
\be
-{1\over 16} {1\over \pii} (-\pi) (-1) {2\over 3} \left\langle 
(\xi(u) - \xi(z_2)) \eta(z_1)\, \bar c(z_1) \, c(z_1) \bar J(\bar z_1)
\VV_{\chi}(z_2) \VV_{\chi^*}(z_3)
\right\rangle \, .
\ee
This correlation function can be easily evaluated using \refb{evacnorm} and
the known operator product
expansion between various fields, including \refb{ejcorr}. The net result is
\be \label{ec1}
{1\over 12} \, \,  {z_2 - u\over u - z_1} \, \, {z_1 - z_3\over z_2 - z_3}\, .
\ee
\item The analysis of the boundary contribution in the $z\to z_3$ limit 
can be obtained by exchanging $z_1$ and $z_3$ in the above results. This
gives
\be \label{ec3}
{1\over 12}  \, \, {z_2 - u\over u - z_3}  \, \, {z_3 - z_1\over z_2 - z_1}\, .
\ee
\end{enumerate}
Adding \refb{ec1}, \refb{ec3} and averaging over cyclic permutations
of $z_1$, $z_2$ and $z_3$,
we get a net contribution
\be \label{ec5}
\bar\AAA^{(4,0)} 
= -{1\over 12}.
\ee
Substituting \refb{ec5} into \refb{egamreal} we get
\be \label{egbvalue}
\gb^{(4,0)}(0,\chi_R;0,\chi_R;0,\chi_R;0,\chi_R)
=-{1\over 2}\, .
\ee
Note that the final result is independent of the location $u$ 
of the PCO and also of $z_1$, $z_2$ and $z_3$, as is expected 
from the general arguments of \cite{1408.0571,1411.7478}.

Using \refb{emr2} we now get 
\be
m_B^2 = {2\over 3}\,  \Beta^2 g_s^2\, ,
\ee
i.e.
\be \label{emb}
m_B = \sqrt{2\over 3} \,  \Beta \, g_s\, .
\ee

In order to fully determine $m_B$ we need to determine $\Beta$ via 
\refb{ebetvalue}. This in turn requires determination of $\gb^{(2,2)}(0,\chi_R;
0,\chi_R)$. This was done in  \cite{ADS,DIS} and has been partially 
reviewed  in appendix \ref{srep}. 
Using the result for $\gb^{(2,0)}(0,\chi_R;0,\chi_R)$ given
in appendix \ref{srep} and the result for 
$\gb^{(4,0)}(0,\chi_R;0,\chi_R;0,\chi_R;0,\chi_R)$ given in \refb{egbvalue}
we can determine the actual value of $\beta$ and of $m_B$ and verify that they
are real.
However for testing the equality of the masses of fermions and bosons
we shall not need this result, since the fermion mass will also be determined in terms of 
$\beta$.

\subsection{Fermion mass to order $g_s$} \label{sfermion}

We shall now compute the mass of the fermion that is the superpartner of
$\chi_R$ to order $g_s$ and compare this with the scalar mass $m_B$ given
in \refb{emb} to test supersymmetry restoration at the shifted vacuum. For this
we have to solve the linearized equation of motion \refb{eqexp} 
in the fermionic sector
and determine the
on-shell value of $k^2$. Equating this with $-m_F^2$ we can determine the
mass of the fermionic partner of $\chi_R$.

Since  for fermions the quantity that enters the linearized equation of
motion is the mass and not mass$^2$, it is enough 
to compute correction to first order in $g_s$ for determining the mass
to order $g_s$. Therefore the iterative equations
\refb{esa3}-\refb{esa5} take the form
\be \label{eferit}
|\Phi_0\rangle = |\phi_1\rangle, \quad P|\phi_1\rangle = |\phi_1\rangle \, ,
\ee
\be
|\Phi_1\rangle = - {b_0^+\over L_0^+} 
(1-P) \XX_0 K |\Phi_0\rangle + |\phi_1\rangle\, ,
\ee
\be \label{elin}
Q_B|\phi_1\rangle = - P \XX_0 K|\Phi_0\rangle = -P\XX_0[\Psi_1\phi_1]
= -\beta g_s P\XX_0  [\VV_R\phi_1]\, .
\ee
We 
shall look for solution to \refb{elin} using the ansatz
\be \label{ephiexpfer}
|\phi_1\rangle = |Y_\alpha\rangle
f^\alpha +  |Z_\dotalpha\rangle g^\dotalpha + |\tilde\phi_1\rangle\, ,
\ee
where
\be \label{edefyz}
|Y_\alpha\rangle \equiv \bar c c e^{-\phi/2} V^f_{\chi,\alpha} 
e^{ik.X}(0)|0\rangle,
\quad 
|Z_\dotalpha\rangle \equiv {1\over \sqrt 3} 
\bar c c e^{-\phi/2} \Sigma^c_\dotalpha \bar J
e^{ik\cdot X}(0)|0\rangle\, ,
\ee
$f^\alpha$ and $g^\dotalpha$ are grassmann odd variables, and
$|\tilde\phi_1\rangle$ is an order $g_s$ correction. 
We shall for now proceed by ignoring the effect of
$|\tilde\phi_1\rangle$, but will return to discuss its role at the end of this
section.
$|Y_\alpha\rangle$ and $|Z_\dotalpha\rangle$  represent  
respectively the fermionic partners of $\chi$ and the 
U(1) gauge fields at the zeroth order. 
Introducing the states
\be \label{edefab}
|\wt A_\dotbeta\rangle = (\p c+\bar \p \bar c) \bar c c \p\xi
e^{-5\phi/2} V^f_{\chi^*,\dotbeta} e^{-ik\cdot X}(0)
|0\rangle,
\quad 
|\wt B_\beta\rangle = {1\over \sqrt 3} 
(\p c+\bar \p \bar c) \bar c c \p\xi
e^{-5\phi/2} \Sigma_\beta \bar J
e^{-ik\cdot X}(0)|0\rangle\, ,
\ee
the two linearly independent equations derived from \refb{elin} can be taken
to be
\be \label{e8.39}
\langle \wt A_\dotbeta | c_0^- \left( Q_B|\phi_1\rangle + 
\beta g_s \XX_0  [\VV_R\phi_1]\right) =0, \quad 
\langle \wt B_\beta | c_0^- \left( Q_B|\phi_1\rangle + 
\beta g_s \XX_0  [\VV_R\phi_1]\right) \rangle=0\, .
\ee
Note that we have dropped the projection operator $P$ since $|\wt A_\beta\rangle $ and
$|\wt B_\dotbeta\rangle$ are $P$ invariant states.
Using \refb{emainope}, 
\refb{evmf},
\refb{evacnorm}, \refb{ebrs1}-\refb{ebrstcurrent}) and $\bar J$ charge
conservation, we find
\ben
&& \langle \wt A_\dotbeta| c_0^- Q_B |Y_\alpha\rangle = -{1\over 4} 
k_\mu \gamma^\mu_{\dotbeta\alpha},
\quad 
\langle \wt B_\beta| c_0^- Q_B |Z_\dotalpha\rangle = -{1\over 4} 
k_\mu \gamma^\mu_{\beta\dotalpha}, \nonumber \\
&& \langle \wt A_\dotbeta| c_0^- Q_B |Z_\dotalpha\rangle = 0,
\quad 
\langle \wt B_\beta| c_0^- Q_B |Y_\alpha\rangle = 0,
\een
\be 
\langle \wt A_\dotbeta | c_0^- \XX_0 |[\VV_{\chi_R} Y_\alpha]\rangle=0, \quad 
\langle \wt B_\beta | c_0^- \XX_0 |[\VV_{\chi_R} Z_\dotalpha]\rangle=0\, .
\ee
$\gamma^\mu_{\beta\dotalpha}$ and $\gamma^\mu_{\dotbeta\alpha}$ have been 
defined in \refb{ea16}.
Furthermore, using  Lorentz invariance we can write
\be \label{edefcd}
\langle \wt A_\dotbeta | c_0^- \XX_0| [\VV_{\chi_R} Z_\dotalpha]\rangle
= -C \ve_{\dotbeta\dotalpha}, \quad
\langle \wt B_\beta | c_0^- \XX_0 |[\VV_{\chi_R} Y_\alpha]\rangle
= D\, \ve_{\beta\alpha}\, ,
\ee
where $C$ and $D$ are two constants to be determined. This allows us to
express \refb{e8.39} as
\be \label{e8.43}
-{1\over 4}  k_\mu (\gamma^\mu)_{\dotbeta\alpha} 
f^\alpha - \Beta\, g_s\, 
C\, \ve_{\dotbeta\dotalpha}
g^\dotalpha =0, \quad
\Beta \, g_s \, D\, \ve_{\beta\alpha} f^\alpha 
-{1\over 4}  k_\mu (\gamma^\mu)_{\beta\dotalpha}  g^\dotalpha =0\, ,
\ee
ignoring the contribution from $|\tilde\phi_1\rangle$. Multiplying the
first equation by $k_\nu (\gamma^\nu)_\beta{}^\dotbeta$ and 
using the second equation to eliminate $g^\dotalpha$ we get
\be  \label{eons}
\left\{{k^2\over 16} + \Beta^2 \, g_s^{2}\,
C D\right\}\ve_{\beta\alpha} f^\alpha = 0\, .
\ee
Demanding that $f^\alpha$ is non-zero (so that we have a non-trivial solution
to the linearized equations) and
comparing \refb{eons} with the on-shell condition $k^2+m_F^2=0$, we get
the fermion mass $m_F$:
\be \label{ecd}
m_F = 4 \,\Beta \,  g_s\, \sqrt{C D}\, .
\ee

We shall now compute $C$ and $D$.
Using \refb{epicture}, \refb{edefggr}, \refb{ematterope}, \refb{edefstress},
\refb{emainope} and \refb{edefab} we get
\ben
\XX_0 |\wt A_\dotbeta\rangle &=&  {1\over 4}\, |A_\dotbeta\rangle, \quad
|A_\dotbeta\rangle \equiv 
\bar c c e^{-\phi/2} 
V^f_{\chi^*,\dotbeta} e^{-ik\cdot X}(0)
|0\rangle,
\nonumber \\
\XX_0 |\wt B_\beta\rangle &=&  {1\over 4} \, 
 |B_\beta\rangle, 
\quad |B_\beta\rangle \equiv \, {1\over \sqrt 3} 
\bar c c e^{-\phi/2} \Sigma_\beta \bar J
e^{-ik\cdot X}(0)|0\rangle\, .
\een
Comparison with \refb{edefcd} yields
\be  \label{ef34}
{1\over 4} \, \{A_\dotbeta\VV_{\chi_R} Z_\dotalpha\} = -C \, \ve_{\dotbeta\dotalpha}, \quad
{1\over 4} \, \{B_\beta\VV_{\chi_R} Y_\alpha\} = D \, \ve_{\beta\alpha}\, .
\ee
Now using the fact that $\VV_{\chi_R}=(\VV_\chi + \VV_{\chi^*})/\sqrt 2$ and using
$\bar J$ charge conservation, we get 
\be \label{ef35}
\{A_\dotbeta\VV_{\chi_R} Z_\dotalpha\} = {1\over \sqrt 2} \{A_\dotbeta\VV_{\chi} Z_\dotalpha\} \, ,
\quad
\{B_\beta\VV_{\chi_R} Y_\alpha\}  = 
{1\over \sqrt 2} \{B_\beta\VV_{\chi^*} Y_\alpha\}  = -{1\over \sqrt 2} \{ Y_\alpha\VV_{\chi^*}B_\beta\}  
\ee
where in the last step we have used the fact that the $\{\cdots\}$ product is anti-symmetric
under the exchange of two Ramond sector states of ghost number 2.
Using \refb{ef35}, \refb{evacnorm} 
and the various operator product expansions described in  
\S\ref{scon}  we now get
\ben \label{ef36}
\{A_\dotbeta\VV_{\chi_R} Z_\dotalpha\} &=&  {1\over \sqrt 6} \left\langle 
 \left( \bar c c e^{-\phi/2} V^f_{\chi^*,\dotbeta} e^{-ik\cdot X} \right)(z_1)
\left( \bar c 
c e^{-\phi}V_\chi\right)(z_2)   \left( \bar c c e^{-\phi/2}\Sigma^c_\dotalpha \bar J
e^{ik\cdot X}\right)(z_3)\right\rangle
\nonumber \\
&=& -\sqrt {2\over 3}\ve_{\dotbeta\dotalpha} \, ,
\nonumber \\ 
\{B_\beta\VV_{\chi_R} Y_\alpha\}  &=&  -{1\over \sqrt 6}
\left\langle 
 \left( \bar c c e^{-\phi/2}V^f_{\chi,\alpha} e^{ik\cdot X} \right)(z_1)
\left( \bar c c e^{-\phi} V_{\chi^*}\right)
(z_2)   \left( \bar c c e^{-\phi/2}\Sigma_\beta \bar J
e^{-ik\cdot X}\right)(z_3)\right\rangle \nonumber \\
&=& -\sqrt {2\over 3}\ve_{\alpha\beta} \, .
\een
Comparison with \refb{ef34} and \refb{ef36} gives
\be 
C = {1\over 4} \sqrt {2\over 3}, \quad D =
{1\over 4} \sqrt {2\over 3} \, ,
\ee
and hence 
\be \label{emf}
m_F = \sqrt{2\over 3} \, \Beta\, g_s\, .
\ee
This matches $m_B$ given in \refb{emb} confirming the prediction of 
unbroken supersymmetry
at the shifted vacuum.

Let us now discuss the role of $|\tilde\phi_1\rangle$ appearing in
\refb{ephiexpfer}. As mentioned there, $|\tilde\phi_1\rangle$ is an
order $g_s$ contribution. It could contain a
linear combination of unphysical, physical and pure gauge states
 at the same mass level -- where the classification of the states
 into these three categories refer to the property they would
 have at $k^2=0$.
The pure gauge contribution can be removed by a gauge transformation
while the physical state contribution will have the effect of renormalizing 
the constants $f^\alpha$ and $g^\dotalpha$ appearing in \refb{ephiexpfer}.
Therefore we focus on unphysical state contribution to $|\tilde\phi_1\rangle$. 
This can be {\it e.g.} 
of the form $(\bar\p\bar c+\p c) c e^{-\phi/2} \Sigma^c_\dotalpha
e^{ik\cdot X}(0)|0\rangle$. Now using the fact that  
$|\tilde\phi_1\rangle$ is of
order $g_s$, and that $Q_B|\wt A_\dotbeta\rangle \sim g_s$, $Q_B
|\wt B_\beta\rangle\sim g_s$  (counting $k^\mu$ to be of order $g_s$)
one can show 
that the contribution of $|\tilde\phi_1\rangle$ 
to the two equations given in \refb{e8.39}
is of order $g_s^2$. 
Since the rest of the terms in these equations, given in \refb{e8.43},  
are of
order $g_s$ we see that the effect of $|\tilde\phi_1\rangle$ to these
equations is subleading, and hence does not affect the computation of the
fermion mass to the leading order. However $|\tilde\phi_1\rangle$ plays
a crucial role when we take the inner product of \refb{elin} with an
unphysical state $\langle s|c_0^-$. In this case the left hand side
gets a contribution of order $g_s$ from the 
$\langle s|c_0^- Q_B |\tilde\phi_1\rangle$ term.
$|\tilde\phi_1\rangle$ now has to be
adjusted to cancel the order $g_s$ contribution from the rest of the terms.

\sectiono{Two loop dilaton tadpole in the perturbative vacuum} \label{sdilaton}

We shall now use the result of \S\ref{stadpole}  
to compute the two loop dilaton tadpole
in SO(32) heterotic string theory compactified on a Calabi-Yau 3-fold
in the perturbative vacuum corresponding to $\beta=0$.
More precisely we shall compute 
$\EE_4(\scalar)$ in this theory for various zero momentum, ghost number two
states in the BRST cohomology and show that the result does not vanish. 
This will then show that the $\beta=0$ vacuum becomes inconsistent at
two loops.

Now in the $\beta=0$ vacuum the natural expansion parameter is $g_s^2$
since the classical solution is of order $g_s^2$. Therefore, the relevant
equation is \refb{efin20}. 
Since in  \refb{efin20}
the error term is of order $g_s^{2q+6}$, we see that it is enough
to take $q=0$ in order to compute $\EE_4(\scalar)$. Also 
as discussed in the last paragraph of \S\ref{stadpole},
$|\zeta_a\rangle$ must be one of the candidate goldstino states $|\gold\rangle$
appearing in \refb{egoldcan}, and
$|\zeta^a\rangle$
must be one of the conjugate states $|\goldc\rangle$ 
appearing in \refb{ep2}. Furthermore
we can also use the arguments in appendix \ref{sunique} to argue that the choice
of $\goldc$ in this case is given uniquely
by
\be
\XX_0\zeta^a =\XX_0\goldc ={1\over \sqrt 3} \bar c\, 
c \, \eta \, e^{\phi/2}\, \wt\Sigma\, \bar J \, ,
\ee 
where $\wt\Sigma$ is one of the operators that appear in 
\refb{edefSigmawrong} and
the $1/\sqrt 3$ factor has been included for convenience.
This gives
\ben \label{ezaza}
&& \zeta^a =\goldc = -{4\over \sqrt 3} (\p c+\bar\p \bar c) \bar c\, 
c \, e^{-3\phi/2}\, \wt\Sigma\, \bar J 
\nonumber \\
&& \zeta_a = \gold =-{1\over 4\sqrt 3} 
\bar c \, c \, e^{-\phi/2} \, \Sigma^c \bar J\, ,
\een
where
$\Sigma^c$ is a matter sector spin field from the list \refb{edefSigma}
that is conjugate to
$\wt\Sigma$, i.e. satisfies
\be
e^{-\phi/2} \Sigma^c(z)  e^{-3\phi/2} \wt\Sigma(w) = (z-w)^{-2} e^{-2\phi}(w)
\, .
\ee
Eq.~\refb{efin20} with $q=0$ now gives
\be \label{e9.4}
\EE_4(\scalar) = g_s^4 \, \Gamma_P^{(2,2)}(\fermion, \gold) \,
\Gamma_P^{(2,2)}(\XX_0\goldc, \Lambda_0) \, ,
\ee
where the subscript $P$ of $\Gamma$ denotes that we are referring to
the amplitude in the perturbative vacuum ($\beta=0$). Since
there is no two point function on the sphere, and since
for torus amplitudes there is no distinction between truncated Green's
function and full Green's function,
$g_s^2 \, \Gamma_P^{(2,2)}$ denotes 
the full torus two point function in the perturbative vacuum. 
In the next two subsections we 
shall evaluate the two factors on the right hand side of \refb{e9.4}.

\subsection{Goldstino coupling to the supersymmetry generator} \label{sgold}

In this subsection we shall 
evaluate $\Gamma_P^{(2,2)}(\XX_0\goldc, \Lambda_0)$.
We now put in the explicit Lorentz spinor indices and choose
\be \label{evop}
\Lambda_0 = c e^{-\phi/2} \Sigma_\alpha, \quad 
\XX_0\goldc = {1\over \sqrt 3} \bar c\, 
c \, \eta \, e^{\phi/2}\, \wtsp_\beta\, \bar J\, .
\ee
We shall follow the convention of section 7 of \cite{1408.0571} 
for our computation.
We denote by $u$ the coordinate on the torus with the
identification $u\equiv u+1\equiv u+\tau$, place the vertex operator 
$\XX_0\goldc$ at the origin $u=0$ and the vertex operator $\Lambda_0$
at a point $u=y$ that will eventually be integrated over.  Since both vertex
operators are dimension zero primaries, the choice of local coordinate
system does not affect the amplitude. Nevertheless it will be useful to fix
some local coordinate system for the choice of Beltrami differentials.
We take the local coordinates
around the punctures to be $w_1=u$ and $w_2=u-y$. 
We also require the choice of the location of a PCO 
consistent with the factorization property, namely that when the
two vertex operators approach each other, the PCO should be away
from both vertex operators. In the $u$ coordinate
system we take this to be at a point $v$ that in general 
depends on $y,\bar y,\tau,\bar\tau$. 
Then following the procedure described in section 7 of 
\cite{1408.0571} we can express
the amplitude as\footnote{The $(y, \bar y)$ integral will run over only half of the
torus 
due to the involution symmetry 
$u\to -u$. We shall include a factor of 1/2 in the definition
of the correlation function $\langle\cdots \rangle$ and allow the integral
over $(y,\bar y)$ to run over the full torus.}
\be \label{e9.6}
\Gamma_P^{(2,2)}(\XX_0\goldc, \Lambda_0) = -{1\over 4\pi^2}
\int d\tau\wedge d\bar\tau \wedge dy \wedge d\bar y  \bigg\langle
\{\XX(v)
b_\tau \bar b_{\bar\tau} b_y \bar b_{\bar y} +\cdots\}
\XX_0\goldc(0) \, \Lambda_0(y)\bigg\rangle
\ee
where 
\be \label{e9.7}
b_y = \ointop_y dw b(w), \quad \bar b_{\bar y} = \ointop_y d\bar w 
\bar b(\bar w), \quad b_\tau = {1\over 2\pi i}
\int_a dw b(w), \quad \bar b_{\bar\tau}
= -{1\over 2\pi i}\int_a d\bar w \bar b(\bar w)\, .
\ee
Here $\ointop_y$ denotes a contour around $y$ with the normalization
$\ointop_{y} dw (w-y)^{-1}=1$, $\ointop_{y} d\bar w (\bar w-\bar y)^{-1}
=1$, and
$\int_a$ denotes a contour around the $a$-cycle of the torus, connecting
$u$ to $u+1$, with the normalization
$\int_a dw=1$, $\int_{a} d\bar w=1$. The overall factor of $-1/4\pi^2$ arises
from the $(2\pi i)^{-3g+3-n}$ normalization factor in the $g$-loop, $n$-point
amplitude. The $\cdots$ inside the curly bracket is a sum of four terms, in each
one of the factors $b_\tau$, $\bar b_{\bar\tau}$, $b_y$ and 
$\bar b_{\bar y}$ is replaced by $\p\xi(v) \p_\tau v$, $\p\xi(v) 
\p_{\bar\tau} v$, $\p\xi(v) \p_y v$ and $\p\xi(v) \p_{\bar y} v$
respectively and the $\XX(v)$ factor is dropped.

Now from the expression for $\Lambda_0$ given in \refb{evop}  we see that
$\bar b(\bar w)$ has no pole at  $\bar w=\bar y$. Hence the contribution from
the $\bar b_{\bar y}$ term in \refb{e9.6} vanishes, and the only non-vanishing
contribution comes from the term in which $\bar b_{\bar y}$ is replaced
by $\p\xi(v) \p_{\bar y} v$
and the $\XX(v)$ factor is dropped:
\be \label{e9.6a}
\Gamma_P^{(2,2)}(\XX_0\goldc, \Lambda_0) = -{1\over 4\pi^2}
\int d\tau\wedge d\bar\tau \wedge dy \wedge d\bar y \,  \bigg\langle
b_\tau \, \bar b_{\bar\tau} \, b_y  \, \p\xi(v) \, \p_{\bar y} \, v\, 
\XX_0\goldc(0) \, \Lambda_0(y)\bigg\rangle
\ee
Naively one might expect
that the term proportional to $\p_{\bar y} v$ can also be made to vanish by
keeping $v$ fixed at a position away from 0, since this will be consistent
with the factorization relation which requires that in the $y\to 0$ limit the PCO
location $v$ must be away from $0$. This would make the whole amplitude
\refb{e9.6a} vanish. However it is not in general 
possible to keep $v$ fixed at a
$y,\bar y$ independent position due to the existence of spurious poles, and as
a result the amplitude is not identically zero. 
We shall proceed by taking $v$ to be independent of $y$, $\bar y$, $\tau$ and
$\bar \tau$ for most of the range of these variables except for a small tubular
neighborhood around the spurious pole(s). 
Inside the tubular neighborhood we
take the PCO to be located at another constant value $w$ in the $u$ plane.
In that case the non-zero contribution to the amplitude comes only from the boundary
of this tubular neighborhood at which the PCO location jumps discontinuously
from $v$ to $w$. 
The result is\cite{1408.0571,1504.00609}
\be \label{e9.9a}
\Gamma_P^{(2,2)}(\XX_0\goldc, \Lambda_0) = {1\over 4\pi^2}
\int_S d\tau\wedge d\bar\tau \wedge dy \,  \bigg\langle
b_\tau \, \bar b_{\bar\tau}  \, (\xi(v)-\xi(w)) \, 
\XX_0\goldc(0) \, b_y \, \Lambda_0(y)\bigg\rangle\ee
where $S$ denotes the boundary of the tubular neighborhood enclosing
the spurious pole.  For fixed $\tau,\bar\tau$, $S$ corresponds to 
an anti-clockwise
contour enclosing the spurious pole. Note that we have moved the $b_y$ factor next to
$\Lambda_0$ so that we can use it to remove the $c(y)$ factor from
$\Lambda_0$. Since $\XX_0\goldc$ is grassmann even, this generates only one
minus sign from having to pass $b_y$ through $(\xi(v)-\xi(w))$. 
In going from \refb{e9.6a} to \refb{e9.9a},
two other minus signs have cancelled. One of them comes from the fact that since we
are carrying out integration over $\bar y$ first, we have to rearrange
$dy\wedge d\bar y$ as $-d\bar y\wedge dy$. The second minus sign comes
from the fact that the measure is defined so that $d ({\rm Im} \, y) \wedge 
d ({\rm Re} \, y)$ gives positive volume. As this is opposite of what is
used conventionally, we would have an extra minus sign (otherwise the
$y$ integral would run along a clockwise contour). 
We now see from \refb{e9.9a} that if near the location $y_s$ of the spurious
pole we have
\be \label{eresi}
F(y)\equiv \sqrt 3\, \bigg\langle
b_\tau \, \bar b_{\bar\tau} \, (\xi(v)-\xi(w)) \, 
\XX_0\goldc(0) \, b_y  \, \Lambda_0(y)\bigg\rangle = 
{A_s(\tau,\bar\tau)\over y-y_s} +
\hbox{non-singular terms}\, ,
\ee
then we have 
\be \label{e9.9}
\Gamma_P^{(2,2)}(\XX_0\goldc, \Lambda_0) = -{1\over 2\sqrt 3\pi i}
\int_\FF d\tau\wedge d\bar\tau A_s(\tau,\bar\tau)\, ,
\ee
where $\FF$ denotes the fundamental domain of the moduli space of one
punctured genus
one Riemann surface. If there are multiple spurious poles then we have to
sum over the residues at all the poles. The multiplicative factor of
$\sqrt 3$ has been included in \refb{eresi} to get rid of the $1/\sqrt 3$ factor
in $\XX_0\goldc$ given in \refb{evop}.

To compute $A_s$ we need to evaluate the correlation function, locate
the position of the spurious poles and find the residues at the poles.
To locate the spurious poles we use the general expression for the
correlation functions of $\xi$'s, $\eta$'s and $e^{q\phi}$'s in the `large
Hilbert space'\cite{Verlinde:1987sd,lechtenfeld,morozov}. On genus 1
Riemann surface it takes the simple form
\ben \label{espurious}
&& \Big \langle \prod_{i=1}^{n+1} \xi(x_i) \prod_{j=1}^n \eta(y_j) 
\prod_{k=1}^m e^{q_k \phi(z_k)} 
\Big\rangle'_\delta
\nonumber \\
&=& {\prod_{j=1}^n \vt_\delta
(-y_j + \sum_{i=1}^{n+1}  x_i -\sum_{i=1}^n   y_i + \sum_k q_k \, z_k)\over
\prod_{j=1}^{n+1} \vt_\delta(-x_j +\sum_{i=1}^{n+1}  x_i
-\sum_{i=1}^n   y_i + \sum_k q_k \, z_k)}
{\prod_{i<i'} E(x_i, x_{i'}) \, \prod_{j<j'} E(y_j, y_{j'})\over
\prod_{i,j} E(x_i, y_j) \, \prod_{k<\ell} E(z_k, z_\ell)^{q_k q_\ell} 
}\, , \nonumber \\
&& \qquad \times \, \, \delta_{\sum_{k=1}^m q_k, 0}\, ,
\een 
where the prime indicates that we are referring to the correlation function in
the large Hilbert space. $\delta$ labels spin structure, $\vt$ denotes the
Jacobi theta functions and $E(x,y)$ denotes the prime form which on genus
one surface takes the form
\be \label{exy}
E(x,y)=\vt_1(x-y) / \vt_1'(0)\, .
\ee
Note that if $q_i$'s are not integers the correlation function suffers from
the usual phase ambiguity; these will be fixed later. 
Since \refb{espurious}
gives  the correlation function in the `large Hilbert space',
we have one more
$\xi$ compared to $\eta$. 
To compute a correlation function in the small Hilbert space where
there are equal number of $\xi$'s and $\eta$'s, and $\xi$'s always appear
with a derivative or a difference operator acting on it ({\it e.g.} in
\refb{eresi} we have $\xi(v) -\xi(w)$), we can simply insert a factor of
$\xi(p)$ for some arbitrary point $p$ and interpret this as a correlation
function in the large Hilbert space. The general structure of the 
correlators guarantees that the result is independent of the choice of
the point $p$.

Note that \refb{espurious} has poles when 
$\vt_\delta(-x_j +\sum_{i=1}^{n+1}  x_i
-\sum_{i=1}^n   y_i + \sum_k q_k \, z_k)$ vanishes. As this happens when no
operators are coincident in general, they are referred to as spurious poles.
In the following we shall generalize the notion a bit to include any pole that
depends on the position of the PCO, including those which occur when a
PCO collides with another PCO or a vertex operator.

We shall now use \refb{espurious} to compute \refb{eresi}. For this we insert
a factor of $\xi(w)$ into the correlator in
\refb{eresi} to interpret this as a correlation
function in the large Hilbert space. Since the $\xi(w)\xi(w)$ term vanishes,
we get, after using \refb{evop} and using the $b_y$ factor to remove the
$c$ factor of $\Lambda_0$,
\ben \label{efypre}
F(y)
&=&
\bigg\langle b_\tau \, \bar b_{\bar\tau} \xi(w) \xi(v) \bar c\, 
c \, \eta \, e^{\phi/2}\, \wtsp_\beta\, \bar J(0) \,
e^{-\phi/2} \Sigma_\alpha(y)\bigg\rangle' \nonumber \\
&=& -\bigg\langle b_\tau \, \bar b_{\bar\tau} \xi(w) \xi(v) \bar c(0)\, 
c (0)\, \eta(0) \, e^{-\phi/2} \Sigma_\alpha(y) \,
\, e^{\phi/2}\, \wtsp_\beta\, \bar J(0) 
\bigg\rangle'
\, .
\een
We can now use \refb{espurious} to evaluate this correlation function,
and get
\be\label{efyp}
F(y)
= - {1\over 2} \ve\sum_\delta {\vt_\delta(w+v-{y\over 2})\over \vt_\delta(v-{y\over 2})
\vt_\delta(w-{y\over 2})}{E(w,v)\over E(v,0) E(w,0)} (E(y,0))^{1/4}
\bigg\langle  b_\tau \, \bar b_{\bar\tau} \, \bar c(0)\, 
c(0) \Sigma_\alpha(y) \wtsp_\beta(0)\, \bar J(0) 
\bigg\rangle_\delta\, ,
\ee
where $1/2$ is the usual factor accompanying sum over spin structures 
and $\ve$ is a phase to be determined.
Indeed, as it stands the right hand side of \refb{efyp} is ill defined since
there is
a fractional power of $E(y,0)$ and the correlator contains the product
$ \Sigma_\alpha(y)\wtsp_\beta(0) $ which has fractional power
of $y$ in the operator product expansion. This ambiguity is resolved by
noting that in the $v\to 0$ limit the correlation function 
appearing in \refb{efypre} must 
reduce to $v^{-1}$ times the correlation function with the $\xi(v)$ and
$\eta(0)$ factors dropped, while in \refb{efyp} the product of the
$\vt_\delta$'s and $E$'s reduce to 
$(E(y,0))^{1/4} / \left(v\, \vt_\delta(-y/2)
\right)$ 
in this limit. Therefore
an unambiguous way of writing \refb{efyp} is
\ben\label{efy}
F(y)
&=& -{1\over 2} 
\sum_\delta {\vt_\delta(w+v-{y\over 2})\over \vt_\delta(v-{y\over 2})
\vt_\delta(w-{y\over 2})}{E(w,v)\over E(v,0) E(w,0)} \, \vt_\delta(-y/2)\,
\nonumber \\ &&
\times \bigg\langle  b_\tau \, \bar b_{\bar\tau} \, \xi(w) \, \bar c(0)\, 
c(0) e^{-\phi/2}\Sigma_\alpha(y) e^{\phi/2}
\wtsp_\beta(0)\, \bar J(0) 
\bigg\rangle'_\delta\, .
\een

The spurious poles can be identified as the value of $y$ at which the
$\vt_\delta(v-{y\over 2})$ factor in the denominator 
vanishes.\footnote{The other
pole at the zero of $\vt_\delta(w-{y\over 2})$ appears after we have shifted
the location of the PCO from $v$ to $w$ inside the tubular neighborhood
around the spurious pole and plays no role since by construction it is
outside the tubular neighborhood of the original spurious pole.
}
Since we have spin fields in the correlator, we can relate the contribution
from different spin structures by shifting $y$ by the periods of the torus.
This allows us to focus on only one spin structure -- which we shall take to be
the periodic-periodic (PP) spin structure -- at the cost of extending 
the range of $y$ to over a parallelogram of sides 2 and $2\tau$ and picking
up the contribution from all the poles in this range. This gives
\ben
F(y)
&=& - {1\over2} {\vt_1(w+v-{y\over 2})\over \vt_1(v-{y\over 2})
\vt_1(w-{y\over 2})}{E(w,v)\over E(v,0) E(w,0)} \vt_1(-y/2)
\nonumber \\ && 
\qquad \bigg\langle  b_\tau \, \bar b_{\bar\tau} \, \bar c(0)\, 
c(0) e^{-\phi/2}\Sigma_\alpha(y) e^{\phi/2}
\wtsp_\beta(0)\, \bar J(0) 
\bigg\rangle_{PP}\, ,
\een
with the subscript $PP$ denoting periodic-periodic sector (odd spin
structure). Note that we have dropped the $\xi(w)$ factor and returned
to the correlator evaluated in the small Hilbert space.
Within a parallelogram of sides 2 and $2\tau$,
the spurious pole is at $y=2v$.  The residue at this pole is given by
\be
A_s={1\over \vt_1'(0)} {\vt_1(w)\over \vt_1(w-v)} {E(w,v)\over 
E(v,0) E(w,0)}\vt_1(-v)\bigg\langle  b_\tau \, \bar b_{\bar\tau} \, \bar c(0)\, 
c(0) e^{-\phi/2}\Sigma_\alpha(2v) e^{\phi/2}
\wtsp_\beta(0)\, \bar J(0) 
\bigg\rangle_{PP}\, .
\ee
Using \refb{exy} this can be rewritten as
\be \label{eas1}
A_s=- \bigg\langle  b_\tau \, \bar b_{\bar\tau} \, \bar c(0)\, 
c(0) e^{-\phi/2}\Sigma_\alpha(2v) e^{\phi/2}
\wtsp_\beta(0)\, \bar J(0) 
\bigg\rangle_{PP}\, .
\ee

In order to calculate this, we note the following:
\begin{enumerate}
\item The correlator is a doubly periodic function of $v$ with periods 
1 and $\tau$.
\item The operators product singularities of the correlation 
function occur at $2v=0$ mod 1 or $\tau$. This gives
 $v=0, 1/2, \tau/2$ and
$(1+\tau)/2$. 
\item According to \refb{espurious} 
the $\phi$ correlator has a spurious pole due to the $\vt_1(v)$
factor in the denominator.  This occurs at $v=0$. 
\item Since all the poles occur at positions where two vertex operators 
coincide in some spin structure,
we can determine the residues at the singularities 
using \refb{esstilde}.  For example for $v\to 0$,
the leading term in the
expansion is
\be \label{eambi}
\ve_{\alpha\beta}\, (2v)^{-1} 
\bigg\langle  b_\tau \, \bar b_{\bar\tau} \, \bar c(0)\, 
c(0) \bar J(0) 
\bigg\rangle_{PP}\, .
\ee
The leading singularities of \refb{eas1}
near $1/2$, $\tau/2$ and $(1+\tau)/2$ can be found
by noting that translating $v$ by 1/2, $\tau/2$ and $(1+\tau)/2$ in \refb{eas1}
we can access the correlator in the other spin structures. Therefore the 
behavior near one of these poles will be given by
\be \label{eevenspin}
\ve_{\alpha\beta}\, (2v- s_\delta)^{-1} 
\bigg\langle  b_\tau \, \bar b_{\bar\tau} \, \bar c(0)\, 
c(0) \bar J(0) 
\bigg\rangle_{\delta}\, ,
\ee
for some even spin structure $\delta$. $s_\delta$ takes values 1,
$\tau$ and $1+\tau$ for different spin structures.
\item Since in the PP sector there are zero modes of the free fermions
$\psi^\mu$ as well as of the bosonic ghost fields
$\beta,\gamma$, correlation function of the form \refb{eambi} is
somewhat ill defined. For this reason it is simplest to analyze the correlator
\refb{eas1}
for non-zero $v$ first and then take the limit.
Using \refb{espurious} we see that the correlator appearing in
\refb{eas1} in the ghost sector goes as $ E(2v,0)^{1/4} / \vt_1(v)
\sim v^{-3/4}$ for small $v$. The operator product expansion of
the matter sector spin fields goes as $v^{-5/4}$. 
However the spin field correlator
in the free fermion sector in spin structure $\delta$
gives an explicit factor of $\vt_\delta(v)^2$ in the 
numerator\cite{atickas}, which for $PP$ sector translates to $\vt_1(v)^2$. 
Since this goes as $v^2$ for small $v$, we see that the correlator does not
have any singularity for $v\to 0$.
\item  In the case of even spin structure, there are no zero modes
and the correlation function \refb{eevenspin} is unambiguous.
In fact this vanishes due to the vanishing of the one point function 
of $\bar J$
in the matter sector (see the discussion at the end of appendix \ref{svan}).
\end{enumerate}
Therefore the net result is that the correlator \refb{eas1} has no poles in the $v$
plane. Since this is a doubly periodic function of $v$ it must be a constant.
This in turn shows that we can evaluate it by calculating it at any point
in the $v$ plane. We shall choose to evaluate it at one of the points $1/2$,
$\tau/2$ or $(1+\tau)/2$ where there is no ambiguity associated with
zero modes and we can use operator product 
expansion.
The constant term is determined by the first subleading correction in the
expansion \refb{esstilde}. The $\p\phi$ term does not contribute due to the
vanishing of one point function of $\bar J$ in the matter sector, and the
$\psi^\mu\psi^\nu$ terms does not contribute due to Lorentz invariance
and anti-symmetry under $\mu\leftrightarrow \nu$.
Therefore the result is
\be \label{easinter}
A_s={1\over 2} \ve_{\alpha\beta}\, \bigg\langle  b_\tau \, \bar b_{\bar\tau} \, \bar c(0)\, 
c(0) J(0)\bar J(0) 
\bigg\rangle_{\delta_e}\, ,
\ee
where $\delta_e$ is any of the even spin structures. Since the result does not
depend on the spin structure we can in fact take the average over the three
even spin structures. Furthermore the right hand side of \refb{easinter}
vanishes if we replace $\delta_e$ by the odd spin structure $PP$ due to
zero modes of $\psi^\mu$, and hence we can include in the sum the PP spin
structure as well to write\footnote{Again the
result in the PP sector 
is somewhat ambiguous due to existence of both fermionic and 
bosonic zero modes, but we can define it by inserting an additional 
pair of operators
$e^{-\phi/2} \Sigma_\gamma(x_1) e^{\phi/2} \wt\Sigma^c_\delta (x_2)$
at two arbitrary points $x_1$ and $x_2$ and picking up the coefficient of the
$\ve_{\gamma\delta}(x_1-x_2)^{-1}$ 
term as $x_1\to x_2$. This can be shown to vanish 
following logic similar to the one showing the absence of pole of
\refb{eas1} at $v=0$. In any case, the relevant quantity is really 
\refb{easinter}, and the equality of \refb{easfin} and
\refb{easinter} can be taken as the definition of what appears on the right
hand side of \refb{easfin}.}
\ben \label{easfin}
A_s &=& {1\over 3} \, 
{1\over 2} \ve_{\alpha\beta}\, \sum_\delta 
\bigg\langle  b_\tau \, \bar b_{\bar\tau} \, \bar c(0)\, 
c(0) J(0)\bar J(0) 
\bigg\rangle_{\delta} \nonumber \\
&=& {1\over 3} \, 
 \ve_{\alpha\beta}\,
\bigg\langle  b_\tau \, \bar b_{\bar\tau} \, \bar c(0)\, 
c(0) J(0)\bar J(0) 
\bigg\rangle\, ,
\een
where we have reverted to the earlier notation that correlator without 
a subscript denotes implicit sum over spin structures accompanied by a
factor of 1/2.
Substituting this into \refb{e9.9} we get
\be \label{e9.9b}
\Gamma_P^{(2,2)}(\XX_0\goldc, \Lambda_0) = -\ve_{\alpha\beta}
\, {1\over 6\sqrt 3 \pi i}
\int_\FF d\tau\wedge d\bar\tau \bigg\langle  
b_\tau \, \bar b_{\bar\tau} \, \bar c(0)\, 
c(0) J(0)\bar J(0) 
\bigg\rangle = - \ve_{\alpha\beta} \, \Xi\, ,
\ee
where
\be \label{edefxi}
\Xi = {1\over 3\pi \sqrt 3}
\,  \int_\FF d^2\tau \bigg\langle  
b_\tau \, \bar b_{\bar\tau} \, \bar c(0)\, 
c(0) J(0)\bar J(0) 
\bigg\rangle  \, .
\ee
Note that in \refb{edefxi} we have replaced $d\tau\wedge d\bar\tau$ by
$2i d^2\tau$ in the spirit of \refb{eorifin}. Up to normalization,
the same factor
appears in the expression for the one loop renormalized mass$^2$ of
the scalar $\chi_R$ in the shifted vacuum given in \refb{enewchi}. 

$\Xi$ has been calculated in \cite{ADS,DIS} with non-zero result -- the actual
value depends on the massless field content of the theory. We shall not try
to repeat the analysis here, but express all further 
results in terms of this quantity.

\def\ALPHA{\gamma}

\def\BETA{\beta}

\subsection{Goldstino-Dilatino coupling} \label{sgolddil}

Next we turn to the computation of $\Gamma_P^{(2,2)}(\fermion, \gold)$.
Associated with the choice of $\XX_0 \goldc$ given in
\refb{evop}, we have from \refb{ezaza}
\ben
&& \zeta^a =\goldc = -{4\over \sqrt 3} (\p c+\bar\p \bar c) \bar c\, 
c \, e^{-3\phi/2}\, \wtsp_\beta\, \bar J \, ,
\nonumber \\
&& \zeta_a = \gold ={1\over 4\sqrt 3} 
\bar c \, c \, e^{-\phi/2} \, \Sigma^\beta \bar J\, ,
\een
where $\Sigma^\beta\equiv \ve^{\beta\gamma} \Sigma_\gamma$ and we have used
$\ve^{\beta\gamma}\ve_{\gamma\alpha}=-\delta^\beta{}_\alpha$.

There are two possible candidates for $\scalar$ whose tadpoles may be
generated at this order:
\be
\scalar \propto \bar c c e^{-\phi} \psi_\mu \bar \p X^\mu, \quad c\eta\, ,
\ee
representing the zero momentum vertex operators for the trace of the
graviton and the dilaton. Their fermionic partners have been constructed
in appendix \ref{sap}, yielding the results
\be \label{ecanfer}
\fermion\propto \bar c c e^{-\phi/2} (\gamma_\mu)_\ALPHA{}^\dotalpha
\Sigma^c_{\dotalpha} \bar \p X^\mu, 
\quad c\eta e^{\phi/2} \wtsp_\ALPHA\, ,
\ee
and similar operators related to the above by the exchange of dotted and
undotted indices.

First consider the case where $\fermion=c\eta e^{\phi/2} \wtsp_\ALPHA$.
In this case the computation of the two point function proceeds in a way very
similar to that of the previous subsection. We take the vertex operator
$\zeta_a= \gold$ to be at fixed position 0 and the vertex operator
$c\eta e^{\phi/2} \wtsp_\ALPHA$ at $y$. Then the analog of \refb{e9.6}
will be
\be \label{e9.6old}
\Gamma_P^{(2,2)}(\gold, \fermion) = -{1\over 4\pi^2}
\int d\tau\wedge d\bar\tau \wedge dy \wedge d\bar y  \bigg\langle
\{\XX(v)
b_\tau \bar b_{\bar\tau} b_y \bar b_{\bar y} +\cdots\}
\gold(0) \, \fermion(y)\bigg\rangle
\ee
where $b_\tau, \bar b_{\bar\tau}, b_y, \bar b_{\bar y}$ 
have been defined in \refb{e9.7} and as before
the $\cdots$ inside the curly bracket represent 
sum of four terms, in each
one of the factors $b_\tau$, $\bar b_{\bar\tau}$, $b_y$ and 
$\bar b_{\bar y}$ is replaced by $\p\xi(v) \p_\tau v$, $\p\xi(v) 
\p_{\bar\tau} v$, $\p\xi(v) \p_y v$ and $\p\xi(v) \p_{\bar y} v$
respectively and the $\XX(v)$ factor is dropped.
We now notice that since $\fermion(y)=c\eta e^{\phi/2}\wtsp_\ALPHA(y)$ is
annihilated by $\bar b_{\bar y}$, we must pick the term where 
$\bar b_{\bar y}$ is replaced by $\p\xi(v) \p_{\bar y} v$
and the $\XX(v)$ factor is dropped. This gives the analog of
\refb{e9.6a}:
\be \label{e9.6new}
\Gamma_P^{(2,2)}(\gold, \fermion) = -{1\over 4\pi^2}
\int d\tau\wedge d\bar\tau \wedge dy \wedge d\bar y  \bigg\langle
b_\tau \bar b_{\bar\tau} \, b_y \p\xi(v) \p_{\bar y} v
\gold(0) \, \fermion(y)\bigg\rangle\, .
\ee
Following the same logic as in the previous section we see that if we take
$v$ to be constant on most of the torus outside a small disk containing the
spurious pole and take it to be another constant $w$ inside the disk so that
there is no spurious pole inside the disk, then the integrand vanishes both
inside and outside the disk and picks up non-zero contribution only from the
boundary of the disk where the PCO location jumps.
This leads to the analog of \refb{e9.9a}
\be \label{e9.9anew}
\Gamma_P^{(2,2)}(\gold, \fermion) = -{1\over 4\pi^2}
\int_S d\tau\wedge d\bar\tau \wedge dy \,  \bigg\langle
b_\tau \, \bar b_{\bar\tau}  \, (\xi(v)-\xi(w)) \, 
\gold(0)\, b_y  \, \fermion(y)\bigg\rangle\ee
where $S$ denotes the boundary of the tubular neighborhood in the
$\tau,y$ space enclosing
the spurious pole. 
From this we can reach the analog of \refb{eresi}, \refb{e9.9}, i.e.\
if near the location of the spurious pole $y_s$ we have
\be \label{eresinew}
\wt F(y)\equiv 4\sqrt 3\, \bigg\langle
b_\tau \, \bar b_{\bar\tau} \, (\xi(v)-\xi(w)) \, 
\gold(0) \, b_y   \, \fermion(y)\bigg\rangle = 
{\wt A_s(\tau,\bar\tau)\over y-y_s} +
\hbox{non-singular terms}\, ,
\ee
then we have 
\be \label{e9.9new}
\Gamma_P^{(2,2)}(\gold, \fermion) = {1\over 8\sqrt 3\pi i}
\int_\FF d\tau\wedge d\bar\tau \wt A_s(\tau,\bar\tau)\, .
\ee

Now following the same logic that led to \refb{efy} we can manipulate the
expression for $\wt F(y)$ given in \refb{eresinew} to the form
\ben \label{efynew}
\wt F(y) &=& {1\over 2} \sum_\delta {\vt_\delta(w+v - {3\over 2}y)\over
\vt_\delta(v-{1\over 2} y)\vt_\delta(w -{1\over 2} y)}
{E(w,v)\over E(w,y) E(v,y)} \vt_\delta(y/2) \nonumber \\
&& \times \, \left\langle b_\tau \, \bar b_{\bar\tau}\, \xi(w)\, 
e^{\phi/2} \, \wtsp_\ALPHA(y)
\, \, \bar c\, c \, e^{-\phi/2} \, \Sigma^\BETA \bar J(0)\right\rangle'_\delta\, .
\een
The spurious poles now arise from two sets of points -- at the zeroes
of $\vt_\delta(v -{1\over 2}y)$ and at the zero of $E(v,y)$. First let
us analyze the contribution from the first set of poles. 
For this we
restrict the sum over $\delta$ to be in the PP sector only at the
cost of extending the range of $y$ to a parallelogram of sides
2 and $2\tau$. The spurious pole is at $y=2v$. Evaluating the residue at the
pole we get the analog of \refb{eas1}:
\be \label{eas1newold}
\wt A_s^{(1)} =  \left\langle b_\tau \, \bar b_{\bar\tau}\, e^{\phi/2} \, 
\wtsp_\ALPHA(2v)
\, \, \bar c\, c \, e^{-\phi/2} \, \Sigma^\BETA \bar J(0)\right\rangle_{PP}\, .
\ee
Next we consider the contribution from the pole at $y=v$ arising from the
zero of $E(v,y)$. The residue at this pole gives
\be  \label{eas1newold2}
\wt A_s^{(2)}=-{1\over 2} \sum_\delta \left\langle b_\tau \, \bar b_{\bar\tau}\, \xi(w)\, 
e^{\phi/2} \, \wtsp_\ALPHA(v)
\, \, \bar c\, c \, e^{-\phi/2} \, \Sigma^\BETA \bar J(0)\right\rangle'_\delta\, .
\ee
We can evaluate the correlator in any spin structure and then get the
result for the other spin structures by shifting $v$ by 1, $\tau$ or $1+\tau$.
Now in the PP sector the correlator is similar to the one that appeared
in \refb{eas1}. Following the same logic given below \refb{eas1} one
can show that \refb{eas1newold2} is independent of $v$. Therefore 
\refb{eas1newold2} is also independent of $\delta$ and we can express the 
result as the contribution from the PP sector multiplied by 
a factor of 4. This gives
\be  \label{eas1newold3}
\wt A_s^{(2)}=- 2 \left\langle b_\tau \, \bar b_{\bar\tau} \, 
e^{\phi/2} \, \wtsp_\ALPHA(v)
\, \, \bar c\, c \, e^{-\phi/2} \, \Sigma^\BETA \bar J(0)\right\rangle_{PP}\, .
\ee
Adding this to \refb{eas1newold} and using the fact that both 
expressions are independent of $v$, we get
\be\label{eas1new}
\wt A_s =  -\left\langle b_\tau \, \bar b_{\bar\tau}\, e^{\phi/2} \, 
\wtsp_\ALPHA(2v)
\, \, \bar c\, c \, e^{-\phi/2} \, \Sigma^\BETA \bar J(0)\right\rangle_{PP}\, .
\ee
Following the same logic that led from \refb{eas1} to \refb{easfin} we get
\be \label{easfinnew}
\wt A_s = {1\over 3} \delta^\BETA{}_\ALPHA
\left\langle b_\tau \, \bar b_{\bar\tau}\bar c(0) c(0)
J(0)\bar J(0)
\right\rangle \, .
\ee
Substituting this into \refb{e9.9new} and using $d\tau\wedge d\bar\tau
= 2i d^2\tau$, we get
\be \label{e9.9newer}
\Gamma_P^{(2,2)}(\gold, \fermion) = {1\over 12\sqrt 3\pi } \delta^\BETA{}_\ALPHA
\int_\FF d^2\tau \left\langle b_\tau \, \bar 
b_{\bar\tau}\bar c(0) c(0)
J(0)\bar J(0)
\right\rangle = {1\over 4} \delta^\BETA{}_\ALPHA \Xi\, ,
\ee
where $\Xi$ has been defined in \refb{edefxi}.

Let us now turn to the second candidate for $\fermion$:
\be
\fermion=\bar c c e^{-\phi/2} (\gamma_\mu)_\ALPHA{}^\dotalpha
\Sigma^c_{\dotalpha} \bar \p X^\mu\, .
\ee
Eq.\refb{e9.6old} has identical form
\be \label{e9.6older}
\Gamma_P^{(2,2)}(\gold, \fermion) = -{1\over 4\pi^2}
\int d\tau\wedge d\bar\tau \wedge dy \wedge d\bar y  \bigg\langle
\{\XX(v)
b_\tau \bar b_{\bar\tau} b_y \bar b_{\bar y} +\cdots\}
\gold(0) \, \fermion(y)\bigg\rangle\,.
\ee
However since now both $\gold$ and $\fermion$ carry  factors of
$e^{-\phi/2}$, $\phi$ charge conservation tells us that the contribution
from the $\cdots$ terms inside the curly bracket must vanish. Furthermore
we must pick the $e^\phi T_F(v)$ term inside the PCO $\XX(v)$.
Using the $b_y$, $\bar b_{\bar y}$ factors to remove the $\bar c c$ factor
from $\fermion(y)$ we can express \refb{e9.6older} as
\ben
\Gamma_P^{(2,2)}(\gold, \fermion) &=& -{1\over 16\sqrt 3\, \pi^2}\, 
(\gamma_\mu)_\ALPHA{}^\dotalpha \, 
\int d\tau\wedge d\bar\tau \wedge dy \wedge d\bar y  \nonumber \\
&& \bigg\langle
e^\phi T_F(v) \, 
b_\tau \bar b_{\bar\tau}  \, 
\bar c c e^{-\phi/2} \Sigma^\BETA \bar J(0) \, 
\,  e^{-\phi/2}   
\Sigma^c_{\dotalpha} \bar \p X^\mu (y)\bigg\rangle\,. \nonumber \\
\een
Furthermore in order to get a non-zero correlation function involving the
$X^\mu$ fields we must pick the contribution $-\psi_\nu \p X^\nu$ 
from $T_F$. Using the fact that the two point function $\p X^\nu(v)
\bar \p X^\mu(y)$ on the torus gives a factor of $-\pi\eta^{\mu\nu}/(2\tau_2)$,
we can express this amplitude as
\ben
\Gamma_P^{(2,2)}(\gold, \fermion) &=& -{1\over 32\sqrt 3\, \pi }\, 
(\gamma_\mu)_\ALPHA{}^\dotalpha \, 
\int d\tau\wedge d\bar\tau \wedge dy \wedge d\bar y \, (\tau_2)^{-1}\nonumber \\
&& \bigg\langle
e^\phi \psi^\mu(v) \, 
b_\tau \, \bar b_{\bar\tau} \, 
\bar c \, c \, e^{-\phi/2} \, \Sigma^\BETA \bar J(0) \, 
\,  e^{-\phi/2}   
\Sigma^c_{\dotalpha} (y)\bigg\rangle\,. \nonumber \\
\een
We now notice that the operator $e^\phi \psi^\mu(v)$ has no singularity
near either of the operators inserted at 0 or $y$. In each spin structure $\delta$
it can have at most one spurious pole at the zero of $\vt_\delta(v - y/2)$.
Since it is a doubly periodic function of $v$ in each spin structure and since 
the only doubly periodic function with $\le 1$ poles is a constant, we
conclude that the correlation function is $v$ independent. Therefore we can
evaluate it at any value of $v$ which we shall choose to be $v=y$.
Using the operator product expansion \refb{emainope},
$e^\phi \psi^\mu(v) e^{-\phi/2}   
\Sigma^c_{\dotalpha} (y)$ in the $v\to y$ limit
 reduces to
$(i/2) (\gamma^\mu)_{\dotalpha}{}^\delta e^{\phi/2}
\wt\Sigma^c_\delta(y)$. Therefore
the correlator becomes
\ben
&& \Gamma_P^{(2,2)}(\gold, \fermion) 
\nonumber \\ &=& 
 -i\, 
{1\over 16\sqrt 3\, \pi }\, 
\int d\tau\wedge d\bar\tau \wedge dy \wedge d\bar y \, (\tau_2)^{-1}
\bigg\langle
b_\tau \, \bar b_{\bar\tau} \, 
\bar c \, c \, e^{-\phi/2} \, \Sigma^\BETA \bar J(0) \, 
\,  e^{\phi/2}   
\wt\Sigma^c_\ALPHA (y)\bigg\rangle\,. \nonumber \\
\een
We can get the correlator in any spin structure by starting with the one
in the $PP$ sector and then translating $y$ by 1, $\tau$ and $1+\tau$. We now
note that in the PP sector the correlator is identical in form
to the one that appears
in \refb{eas1new}. Hence it is independent of $y$ and is given by 
\refb{easfinnew} with a sign flip due to different ordering of the two
operators inside the correlator. 
Multiplying this by a factor of 4 due to the
four spin structures, and including a factor of 1/2 by which we need to
multiply the sum over spin structures, we get
\be
\Gamma_P^{(2,2)}(\gold, \fermion) 
=
 -i \, \delta^{\BETA}{}_{\ALPHA}
{1\over 24\sqrt 3\, \pi }\, 
\int d\tau\wedge d\bar\tau \wedge dy \wedge d\bar y \, (\tau_2)^{-1}
\bigg\langle
b_\tau \, \bar b_{\bar\tau} \, 
\bar c \, c \, J\, \bar J(0) \, 
\bigg\rangle\, .
\ee
Expressing $d\tau\wedge d\bar\tau \wedge dy \wedge d\bar y $ as
$-4d^2\tau d^2y$ and noting that the $y$ integral gives a factor of $\tau_2$
we finally arrive at
\be \label{e9.47}
\Gamma_P^{(2,2)}(\gold, \fermion) 
=
 i \, \delta^{\BETA}{}_\ALPHA
{1\over 6\sqrt 3\, \pi }\, 
\int_\FF d^2\tau \, 
\bigg\langle
b_\tau \, \bar b_{\bar\tau} \, 
\bar c \, c \, J\, \bar J(0) \, 
\bigg\rangle = {1\over 2} i \, \delta^{\BETA}{}_\ALPHA \Xi\, ,
\ee
where $\Xi$ has been defined in \refb{edefxi}.

\subsection{Tadpoles} \label{stadpole2}

Finally let us assemble all the pieces to compute the two loop tadpoles.
Our first task will be to compute $\scalar=[\Lambda_0\fermion]_0$ with
precise normalization factors for the candidate $\fermion$'s given in
\refb{ecanfer}. This can be easily done using the fact that up to BRST
exact terms, $[\Lambda_0\fermion]_0$ can be found by computing the
operator product of $\Lambda_0$ and $\fermion$ and then applying the
$b_0^-$ operator on the resulting state. This gives
\ben 
\left[\Lambda_0  \, \left(\bar c c e^{-\phi/2} 
(\gamma_\mu)_\ALPHA{}^\dotalpha
\Sigma^c_{\dotalpha} \bar \p X^\mu\right)\right]_0 &=&
i\, (\gamma_\mu \gamma^\nu)_\gamma{}^\beta \,
\ve_{\beta\alpha} \, \bar c c e^{-\phi} \psi_\nu \bar\p X^\mu\, ,
\nonumber \\
\left[\Lambda_0 \, \left(c\eta e^{\phi/2} \wtsp_\ALPHA\right)\right]_0 &=& 
\ve_{\gamma\alpha} c\eta\, ,
\een
for $\Lambda_0 = c e^{-\phi/2} \Sigma_\alpha$.
Taking the product of \refb{e9.9b} and
\refb{e9.9newer} and a minus sign to account for the fact that in
\refb{e9.9newer} the arguments of $\Gamma^{(2,2)}_P$ are 
exchanged compared to those in \refb{e9.4}, we get
\be 
\ve_{\gamma\alpha} \, \EE_4(c\eta) 
=   {1\over 4} \, g_s^4  \,
\ve_{\alpha\gamma} \, \Xi^2 = - {1\over 4} \, g_s^4  \,
\ve_{\gamma\alpha} \, \Xi^2\, .
\ee
Similarly taking the product of \refb{e9.9b} and
\refb{e9.47} and a minus sign, we get
\be \label{esstt1}
i\, (\gamma_\mu \gamma^\nu)_\gamma{}^\beta \,
\ve_{\beta\alpha} \, \EE_4\left(\bar c c e^{-\phi} \psi_\nu \bar\p X^\mu
\right) = {i\over 2} \, \ve_{\alpha\gamma} \, g_s^4\, \Xi^2
= -{i\over 2} \, \ve_{\gamma\alpha} \, g_s^4\, \Xi^2\, .
\ee
We now use the fact that due to Lorentz invariance of the background,
we must have
\be \label{esstt2}
\EE_4\left(\bar c c e^{-\phi} \psi_\nu \bar\p X^\mu
\right)
={1\over 4} \delta_\nu{}^\mu \, \EE_4\left(\bar c c e^{-\phi} \psi_\rho \bar\p X^\rho
\right)\, .
\ee
Substituting this into \refb{esstt1} we get
\be
i\, \ve_{\gamma\alpha} \, \EE_4\left(\bar c c e^{-\phi} \psi_\rho \bar\p X^\rho
\right)
= -{i\over 2} \, \ve_{\gamma\alpha} \, g_s^4\, \Xi^2\, .
\ee
This gives
\be \label{efintadpole}
\EE_4(c\eta) = - {1\over 4} \, g_s^4 \,
\Xi^2\, , \qquad \EE_4\left(\bar c c e^{-\phi} \psi_\rho \bar\p X^\rho\right)
= -{1\over 2} \, g_s^4\, \Xi^2,
\qquad \EE_4\left(\bar c c e^{-\phi} \psi_\mu \bar\p X^\nu\right)
= -{1\over 8} \, \delta_\mu{}^\nu  \, g_s^4\, \Xi^2\, .
\ee
In appendix \ref{slow} we shall verify that these results and the result 
for the one loop renormalized mass$^2$ of the scalar given in \refb{enewchi}
are all consistent with the predictions of low energy effective supersymmetric
field theory.

\sectiono{Two loop dilaton tadpole in the shifted vacuum} \label{sshifted}

In this short section we shall compute the dilaton tadpole at the shifted vacuum to order 
$g_s^4$ and show that it vanishes.

The relevant formula in this case is \refb{efin2}. Since we have already demonstrated that
the shifted vacuum has unbroken global supersymmetry to order $g_s^2$, the result of
\S\ref{stadpole} shows that all tadpoles vanish to order $g_s^3$. The dilaton tadpole in the
shifted vacuum to order $g_s^4$ can be computed using \refb{efin2} for $q=2$. 
Denoting by $\Gamma_S^{(n,k)}$ the $\Gamma^{(n,k)}$'s in the shifted
vacuum ($\beta\ne 0$), we get
\be \label{eshift1}
\EE_{4} (\scalar)  
=g_s^4\sum_a
\Gamma_S^{(2,1)}(\fermion, \zeta_a) \, \Gamma_S^{(2,3)} 
(\XX_0\zeta^a, \Lambda_2)
+ \OO(g_s^{5})\, .
\ee
As discussed below \refb{efin20}, $\Gamma_S^{(2,1)}(\fermion, \zeta_a)$ is given by
\be \label{ecruc}
\{\fermion \zeta_a \Psi_1\}_{0} = \beta g_s \{\fermion \zeta_a \VV_{\chi_R}\}_0
\ee
where, as in \S\ref{srestore},
the subscript 0 on $\{~\}$ reflects genus zero contribution, 
and we have used the
result $\Psi_1 = \beta g_s\VV_{\chi_R}$. Now for the dilaton tadpole the 
relevant $\fermion$'s have been given in \refb{ecanfer}:
\be \label{ecanferagain}
\fermion\propto \bar c c e^{-\phi/2} (\gamma_\mu)_\beta{}^\dotalpha
\Sigma^c_{\dotalpha} \bar \p X^\mu, 
\quad c\eta e^{\phi/2} \wtsp_\beta\, ,
\ee
and other operators related to the above by the exchange of dotted and
undotted indices.
On the other hand we have $\VV_{\chi_R}=\bar c c e^{-\phi} V_{\chi_R}$. Finally the
candidate goldstino states $|\zeta_a\rangle$
 have been listed in \refb{egoldcan}, and the only candidates that
could possibly contribute satisfying $\bar J$ charge conservation are
\be \label{egoldcannew}
\gold\propto \bar c c e^{-\phi/2} V^{f}_{\chi,\alpha}, 
\quad \bar c c e^{-\phi/2} V^{f}_{\chi^*,\dotalpha}\, .
\ee
It is now easy to see that the sphere 3-point amplitude appearing on the right hand side of
\refb{ecruc} vanishes for all the candidates. For the first candidate for $\fermion$ given in
\refb{ecanferagain} the amplitude vanishes due to the vanishing of one point function of
$\bar \p X^\mu$ on the sphere. For the second candidate for $\fermion$ the amplitude 
vanishes by $\phi$ charge conservation. This shows the vanishing of the dilaton tadpole
in the shifted vacuum to order $g_s{}^4$.

Note that this analysis leaves open the possibility that there may be other
massless fields whose tadpoles do not vanish to this order. Additional work will
be necessary to rule out the existence of such tadpoles.

\bigskip

\noindent {\bf Acknowledgement:}
We thank Roji Pius, Arnab Rudra and Edward Witten
 for useful discussions.
This work  was
supported in part by the 
DAE project 12-R\&D-HRI-5.02-0303 and J. C. Bose fellowship of 
the Department of Science and Technology, India.

\appendix

\sectiono{Glossary of symbols} \label{sglossary}

In this appendix we shall give a glossary of symbols that have been 
used extensively in the paper. This can be used as a quick reference while
reading the paper. For superconformal ghosts and total Virasoro generators
we use standard notation $b$, $c$, $\beta$, $\gamma$, 
$\xi$, $\eta$, $\phi$, $L_n$,
$\bar L_n$ that will not be listed below. Unless stated otherwise, the
description below refers to the heterotic string theory, but can be easily generalized
for type II string theories.

The following symbols refer to the general part of our analysis.
\begin{enumerate}
\item $Q_B$ denotes BRST charge.
\item $b_0^\pm$, $c_0^\pm$ and $L_0^\pm$ denote respectively
$(b_0\pm \bar b_0)$, $(c_0\pm \bar c_0)/2$ and $(L_0\pm \bar L_0)$.
\item $T_m$, $\bar T_m$ denote energy momentum tensors in the matter sector.
\item $T_F$ denotes the world-sheet superpartner of $T_m$.
\item $\XX$ denotes the picture changing operator. For type II string
theory we also have $\bar \XX$ representing picture changing operator
in the left-moving sector.
\item $\XX_0$ is the zero mode of the picture changing operator.
\item For the heterotic string 
$\GG$ is identity in the  NS sector and $\XX_0$ in the R sector.
For type II strings $\GG$ is 
identity in the NSNS sector, 
$\XX_0$ in the NSR sector, $\bar \XX_0$ in the RNS sector and
$\XX_0\bar\XX_0$ in the RR sector.
\item $\HH_T$ is the subspace of states in the Hilbert space of the matter
and ghost CFT annihilated by $b_0^-$ and $L_0^-$. 
\item $\wh\HH_T$ is the subspace of $\HH_T$ in which NS sector states carry
picture number $-1$ and R sector states carry picture number $-1/2$.
\item $\wt\HH_T$ is the subspace of $\HH_T$ in which NS sector states carry
picture number $-1$ and R sector states carry picture number $-3/2$.
\item $\{A_1\cdots A_N\}$ is the 1PI amplitude with external states
$|A_1\rangle,\cdots |A_N\rangle\in\wh\HH_T$, 
normalized so that tree level 
amplitudes do not carry any factor of $g_s$.
\item $[A_1\cdots A_N]$ is a state in $\wt\HH_T$ 
such that $\langle A_0|c_0^-|[A_1\cdots A_N]\rangle$
is $\{A_0A_1\cdots A_N\}$ for all states $\langle A_0|$.
\item $|\Psi\rangle$ denotes the string field which is an arbitrary state in $\wh \HH_T$
of ghost number 2.
\item $|\Psi_{\rm vac}\rangle$  denotes a Lorentz invariant 
classical solution to the 
equations of motion carrying zero momentum, representing a candidate 
vacuum state.
\item $|\Psi_k\rangle$ is the approximation to $|\Psi_{\rm vac}
\rangle$ to order $g_s^k$.
\item $|\psi_k\rangle$ is the projection of $|\Psi_k\rangle$ to $L_0^+=0$
sector. 
\item ${\bf P}$ is the  projection operator to $L_0^+= 0$ states.
\item $|\tilde \psi_k\rangle=|\psi_k\rangle-|\psi_{k-1}\rangle$ 
is the order $g_s^k$ contribution to 
$|\Psi_{\rm vac}\rangle$, projected to the $L_0^+=0$ sector.
\item $|\Phi\rangle$ denotes the shifted field 
$|\Psi\rangle - |\Psi_{\rm vac}\rangle$.
\item $\wh Q_B=Q_B + \GG K$ is the kinetic operator around the shifted 
vacuum where $K$ is the operator defined in \refb{edefK}.
\item $\wt Q_B$ is the operator $Q_B + K \GG $.
\item $\{A_1\cdots A_N\}''$ for $N\ge 3$ 
is the 1PI amplitude with external states
$|A_1\rangle,\cdots |A_N\rangle$ around the shifted vacuum.
\item $[A_1\cdots A_N]''$ for $N\ge 2$ is a state in $\wt\HH_T$ 
such that $\langle A_0|c_0^-|[A_1\cdots A_N]''\rangle$
is $\{A_0A_1\cdots A_N\}''$ for all states $\langle A_0|$.
\item $|\Lambda\rangle$ is the local gauge transformation parameter,
represented by an arbitrary state in $\wh\HH_T$ of ghost number 1.
\item $|\Lambda_{\rm global}\rangle$ is a gauge transformation parameter
satisfying $\wh Q_B |\Lambda_{\rm global}\rangle=0$ and represents
a global symmetry transformation.
\item $|\Lambda_k\rangle$ is a gauge transformation parameter
satisfying
$\wh Q_B|\Lambda_k\rangle=\OO(g_s^{k+1})$. It represents a global
symmetry transformation parameter to order $g_s^k$.
\item $|\lambda_k\rangle$ is the projection of $|\Lambda_k\rangle$ to
the $L_0^+=0$ sector.
\item $|\Phi_{\rm linear}\rangle$ is a solution to the linearized equations
of motion $\wh Q_B |\Phi_{\rm linear}\rangle=0$ around the shifted vacuum.
\item $|\Phi_\ell\rangle$ is a solution to the linearized equation around the
shifted vacuum to order $g_s^\ell$.
\item $L_0^+\simeq 0$ states are those whose $L_0^+$ value for 
on-shell momentum is of order $g_s$ or less. 
\item $P$ denotes projection
operator to $L_0^+\simeq 0$ states.
\item $|\phi_\ell\rangle$ is the projection of $|\Phi_\ell\rangle$ to $L_0^+\simeq 0$
sector.
\item $\Gamma^{(n)}(k_1,a_1;\cdots k_n,a_n)$ denotes the $n$-point
Green's function with external propagators  
truncated, normalized so that the 1PI contribution to
this agrees with $\{A_1,\cdots A_n\}$. $k_1,\cdots k_n$ label the momenta 
carried by the external states and $a_1,\cdots a_n$ denote other quantum numbers.
\item $g_s^\ell \Gamma^{(n,\ell)}(k_1,a_1;\cdots k_n,a_n)$ denotes the
order $g_s^\ell$ contribution to $\Gamma^{(n)}(k_1,a_1;\cdots k_n,a_n)$
\item $|\scalar\rangle$ represents an arbitrary zero momentum Lorentz scalar 
state of ghost number 2, picture number $-1$
satisfying $Q_B|\scalar\rangle=0$.
\item $|\fermion\rangle$ is the superpartner of $|\scalar\rangle$ at the leading
order in $g_s$ defined through \refb{epair0}.
\item $|\gold\rangle$ is a zero momentum, ghost number 2, picture number
$-1/2$ element of the BRST cohomology representing a candidate
for the goldstino state.
\item $|\goldc\rangle$  is a zero momentum, ghost number 3, 
picture number
$-3/2$ element of the BRST cohomology representing dual of
the goldstino state.
\item $\{|\zeta^a\rangle\}$ describe the collection of all candidate states for
$\goldc$ given in \refb{ep2}.
\item $\{|\zeta_a\rangle\}$ describe the collection of all candidate 
states for $\gold$ listed in \refb{egoldcan}.
\end{enumerate}

The following symbols are used specifically in the analysis of SO(32)
heterotic string theory on Calabi-Yau 3-folds.
\begin{enumerate}
\item $\chi$ denotes the chiral multiplet complex scalar whose real part
$\chi_R = (\chi+\chi^*)/\sqrt 2$
condenses at one loop to produce the supersymmetric vacuum.
\item $\sigma$ represents a generic massless
complex scalar field belonging to a chiral multiplet.
\item $J$ is the holomorphic U(1) current associated with the R-symmetry
of the (2,2) supersymmetry algebra associated with the compact
directions.
 \item $\bar J$ is the anti-holomorphic U(1) current associated 
 with the R-symmetry
of the (2,2) supersymmetry algebra associated with the compact
directions. This is also the U(1) current whose associated gauge symmetry
is anomalous.
\item $\VV_\sigma = \bar c c e^{-\phi} V_\sigma$ is the zero momentum
vertex operator for
$\sigma$ in $-1$ picture. 
Here $V_\sigma$ is a matter sector vertex operator of dimension
(1,1/2).
\item $\wt V_\sigma(w)=-\ointop_w dz T_F(z)V_\sigma(w)$ is used for
the construction of the vertex operator for $\sigma$ in the 0 picture. 
Here $T_F$ denotes the matter part of the super-stress tensor.
\item $\VV_\chi$ is the zero momentum
vertex operator for
$\chi$ in $-1$ picture, obtained by taking $\sigma=\chi$ in the above
definition.
\item $\VV_{\chi_R}=(\VV_\chi + \VV_{\chi^*})/\sqrt 2$ is the vertex operator
for the real part of $\chi$.
\item $\Sigma^{(4)}_\alpha$, $\Sigma^{(4)}_\dotalpha$ are spin fields of
dimension $(0,1/4)$ associated with free fermions.
\item $\Sigma^{(6)}$, $\Sigma^{(6)c}$ are spin fields of dimension
$(0, 3/8)$ associated with
the compact directions. 
\item $\Sigma_\alpha$, $\Sigma^c_\dotalpha$, $\wt\Sigma_\dotalpha$
and $\wt\Sigma^c_\alpha$ are various products of $\Sigma^{(6)}$ and
$\Sigma^{(4)}$ given in \refb{edefSigma} and \refb{edefSigmawrong}.
\item $\bar c c e^{-\phi/2} V^f_{\sigma ,\alpha}$ is the zero momentum
vertex
operator for the fermionic partner of $\sigma$ in $-1/2$ picture. 
Schematically $V^f_{\sigma,\alpha}$ can be expressed as 
$V^f_{\sigma}\Sigma^{(4)}_\alpha$ with
$V^f_\sigma$ denoting a matter sector vertex operator of dimension 
(1, 3/8) in the CFT associated with the compact directions.
Similar definitions apply to $V^f_{\sigma,\dotalpha}$, $V^f_{\sigma^*,\alpha}$
and $V^f_{\sigma^*,\dotalpha}$, the details of which can be found in
\refb{edefcomb}.
\item $\bar c c e^{-\phi/2} V^f_{\chi,\alpha}$ is the zero momentum
vertex
operator for the fermionic partner of $\chi$ in $-1/2$ picture.
\item $\gb^{(n,2g)}(0,\chi_R;\cdots 0,\chi_R)$ denotes the full $n$-point, genus 
$g$ amplitude of $n$ external zero momentum $\chi_R$ states  
in the perturbative vacuum.
\item $\AAA^{(4,0)}$ is a particular contribution to 
$\gb^{(4,0)}(0,\chi_R; 0,\chi_R;0,\chi_R;  0,\chi_R)$, defined in \refb{edefa40}.
\item $\wt\AAA^{(4,0)}$ is $\AAA^{(4,0)}$ computed with wrong
PCO arrangement that makes it vanish.
\item $\Gamma_P^{(n,2k)}$ 
denotes $\Gamma^{(n,2k)}$ computed in the perturbative
vacuum in which supersymmetry is broken at one loop order.
\item $\Gamma_S^{(n,2k)}$ denotes 
$\Gamma^{(n,2k)}$ computed in the shifted
vacuum in which supersymmetry is restored at one loop order.
\end{enumerate}

\sectiono{Pairing of physical states between different picture
numbers}  \label{enewappendix}

Eq.\refb{eqexp} gives the linearized equation of motion around the
shifted vacuum
\be  \label{ezz1}
\wh Q_B|\Phi_{\rm linear}\rangle \equiv Q_B|\Phi_{\rm linear}\rangle 
+ \GG\, K |\Phi_{\rm linear}\rangle = 0\, . 
\ee
The solutions to this equation give the possible external states in the S-matrix
element. On the other hand \refb{enewcons} imposes an additional
constraint
\be \label{ezz2}
\GG|\wt \Phi_{\rm linear}\rangle - |\Phi_{\rm linear}\rangle = 0\, .
\ee
The goal of this appendix will be to show that given a perturbative 
solution to 
\refb{ezz1} we can always find a $|\wt \Phi_{\rm linear}\rangle$ satisfying
\refb{ezz2}, possibly after adding a pure gauge term to
$|\Phi_{\rm linear}\rangle$. 
Hence \refb{ezz2} does not impose any additional constraint 
on possible choices of external states. Since in the NS sector $\GG$ is
the identity operator, the result holds trivially there by choosing 
$|\wt \Phi_{\rm linear}\rangle=|\Phi_{\rm linear}\rangle$. Therefore we need to
focus on the Ramond sector.

For this analysis we shall restrict ourselves to the case when 
$|\Phi_{\rm linear}\rangle$ carries non-zero momentum. 
Let us now define $|\wt\Phi_\ell\rangle$ for $0\le \ell\le n$ to be the
solution to the following recursion relations analogous to
\refb{esa3}-\refb{esa5}:
\be \label{esa3zz}
|\wt\Phi_0\rangle = |\wt\phi_n\rangle, \qquad
|\wt\Phi_{\ell+1}\rangle = -{b_0^+\over L_0^+} (1-P)  K \GG |\wt\Phi_\ell\rangle 
+ |\wt\phi_{n}\rangle + 
\OO\left(g_s^{\ell+2}\right)\, ,  \quad \hbox{for} \quad 0\le \ell\le n-1\, ,
\ee
where $|\wt\phi_{n}\rangle$ satisfies
\be \label{esa4zz}
P|\wt\phi_{n}\rangle = |\wt\phi_{n}\rangle\, , 
\ee
\be \label{esa5zz}
Q_B|\wt\phi_{n}\rangle = - P 
K \GG|\wt\Phi_{n-1}\rangle +\OO(g_s^{n+1}) \, .
\ee
Given a solution to \refb{esa3zz}-\refb{esa5zz},
we can find a solution to \refb{esa3}-\refb{esa5} by setting
\be \label{e5zz}
|\phi_n\rangle = \GG |\wt\phi_n\rangle, \qquad 
|\Phi_\ell\rangle = \GG|\wt\Phi_\ell\rangle \, ,  
\quad \hbox{for} \quad 0\le \ell\le n\, .
\ee
Non-trivial solutions to \refb{esa3zz}-\refb{esa5zz} can be found by
starting with a physical state at order $g_s^0$, and then systematically 
correcting the state as well as the momenta carried by the state by
solving the recursion relations.
On the other hand from the result of \cite{9711087}  it follows that 
to order $g_s^0$, there is a one to one correspondence 
between the physical states in picture numbers $-1/2$ and $-3/2$ at
non-zero momentum. Therefore given
a physical state at picture number $-1/2$, we can first construct a physical
state at picture number $-3/2$, then correct it by solving the recursion
relations \refb{esa3zz}-\refb{esa5zz}, and then find the corrected physical
state in picture number $-1/2$ using \refb{e5zz}.  This shows that in
perturbation theory we
do not lose any physical state by imposing the additional condition 
\refb{ezz2}.

\sectiono{Bose-Fermi pairing at zero momentum} \label{sap}

For non-zero momentum supersymmetry pairs NS and R-sector vertex 
operators. However 
at zero momentum some of this pairing may break down since the BRST cohomology
at zero momentum is not always given by the analytic continuation from non-zero 
momentum\cite{9711087}. 
On the other hand, for the analysis of \S\ref{stadpole} we need to assume the
existence of such pairing. In this appendix we shall prove 
the results which
were used in the analysis of \S\ref{stadpole}.
More specifically we shall prove that at zero momentum, 
for every BRST invariant, Lorentz scalar state $| \scalar\rangle\in\wh\HH_T$
of ghost number 2, 
we can find a BRST invariant Ramond sector state $|\fermion\rangle
\in\wh\HH_T$ of ghost number 2, and another BRST invariant Ramond sector
state $|\Lambda_0\rangle
\in\wh\HH_T$ of ghost number 1 -- representing a global supersymmetry
transformation parameter at zeroth order --
such that 
\be \label{egoal}
| \scalar\rangle = [\Lambda_0 \fermion]_0\, ,
\ee
up to addition of BRST trivial states. 
Here $\Lambda_0$ is the leading contribution to a
global supersymmetry generator, and the subscript 0 on $[\cdots]$ denotes
that we only include genus zero contribution in computing $[\cdots]$.
There may be more than one $|\fermion\rangle$ satisfying the desired relation, 
but for us
the existence of one such state will be enough.

To prove \refb{egoal} we take a brute force approach and write down the general form of
possible $| \scalar\rangle$. Up to addition of BRST exact terms,
there are two kinds of BRST invariant scalar vertex operators at zero momentum
(see {\it e.g.} \cite{1408.0571}):
\be \label{ephiform}
 \scalar: \bar c c e^{-\phi} \, W, \quad  c\eta \, ,
\ee 
where $W$ denotes a dimension (1,1/2) superconformal primary in the NS sector
of the matter
CFT. The first set of vertex operators are familiar, the second one gives the
zero momentum dilaton vertex operator in the $-1$ picture. 
Their 
conjugate BRST invariant states 
$ (\scalar)^c$ in the ghost number 3 sector are, up to
normalization and addition of pure gauge states
\be \label{econjugate}
 (\scalar)^c: (\p c + \bar \p\bar c) \bar c  c e^{-\phi} \, W^c, \quad  (\p c + \bar \p\bar c)
 \bar c\,
 \bar \p^2 \bar c \, c\p\xi e^{-2\phi}\, ,
\ee
where $W^c$ denotes the operator in the matter sector conjugate to $W$.

On the other hand
the leading contribution
to the supersymmetry generators $\Lambda_0$ have the form
\be \label{elambdaform}
\Lambda_0: c e^{-\phi/2} \Sigma\, ,
\ee
where $\Sigma$ is 
a dimension $(0,5/8)$ operator in the R sector of the matter CFT.
For example for four dimensional theories preserving $\NN=1$ supersymmetry,
$\Sigma$ is
one of the operators $\Sigma^c_{\dotalpha}$ or
$\Sigma_\alpha$ introduced in \refb{edefSigma}.
Let  $\wt\Sigma$ be the dimension (0,5/8) 
operator  `conjugate' to $\Sigma$
satisfying
\be
\Sigma(z) \wt\Sigma(w) = (z-w)^{-5/4} + \cdots\, .
\ee
For four dimensional theories preserving $\NN=1$ supersymmetry,
$\wt\Sigma$ is
one of the operators $\wts_{\dotalpha}$ or
$\wtsp_\alpha$ introduced in \refb{edefSigma}.

Now space-time supersymmetry at the leading order 
ensures existence of matter primaries $V^f$ of dimension (1,5/8) 
in the Ramond sector
such that for some conjugate pair $\Sigma$, $\wt\Sigma$,
\be \label{extraope}
\wt\Sigma(z) W(w) \sim (z-w)^{-1/2} V^f(w)  + \cdots, \qquad 
\Sigma(z) V^f(w) \sim (z-w)^{-3/4} W(w) + \cdots
\ee
where $\cdots$ denotes less singular terms. The first equation in
\refb{extraope} can be taken to be the defining equation for $V^f$; the second
equation then follows as a consequence.

We now consider the Ramond sector vertex
operators
\be \label{echiform}
\fermion: \bar c c e^{-\phi/2} V^f, \quad c\eta e^{\phi/2} \wt\Sigma\, .
\ee
One can easily show that these represent not-trivial elements of the BRST
cohomology. 
Indeed, these are precisely the candidate goldstino states listed in
\refb{egoldcan}.
Furthermore the three point function of
$\Lambda_0$, $\fermion$ and $ (\scalar)^c$ on a sphere can be shown to be 
non-zero, establishing 
that  $\langle (\scalar)^c| c_0^- [\Lambda_0\fermion]_0\rangle \ne 0$.
Therefore $[\Lambda_0\fermion]_0$ for $\fermion$ given in
\refb{echiform} indeed produces the states $\scalar$ given in \refb{ephiform}
up to addition of BRST exact states. 
This
shows that for every zero momentum scalar vertex operators $ \scalar$ 
in the $-1$ picture
we can find a zero momentum fermionic vertex operator $\fermion$ in the $-1/2$ picture
satisfying \refb{egoal}.

\sectiono{Vanishing of $|\tilde\psi_2\rangle$} \label{svan}

For SO(32) heterotic string theory on a Calabi-Yau manifold, the order $g_s^2$ 
contribution to the vacuum solution, projected to the $L_0^+$ sector, is given by 
$|\tilde\psi_2\rangle$ satisfying \refb{epsi4}:
\be \label{epsi4rep}
Q_B |\tilde\psi_2\rangle = - {\bf P} \left({1\over 2} \beta^2 \, 
g_s^2[\VV_{\chi_R}\VV_{\chi_R}]_0 + [~]_1 \right)
\, .
\ee
In this appendix we shall show that each term on the right hand side of
\refb{epsi4rep} vanishes. This in turn would imply that $|\tilde\psi_2\rangle$
can be taken to be zero.

We begin with the first term. In order to show that it vanishes, we need to
show that
for an arbitrary state $|A\rangle\in\HH_T$ with $L_0^+=0$, ghost number 2 and
picture number $-1$, 
\be \label{e2obs}
\langle A|c_0^-  [\VV_{\chi_R}\VV_{\chi_R}]_0\rangle = \{\VV_{\chi_R}
\VV_{\chi_R} A \}_0
\ee
vanishes. 
Using $\VV_{\chi_R} =\bar c c e^{-\phi} \Vm_{\chi_R}$, expressing 
$\Vm_{\chi_R}$ as $(\Vm_\chi+\Vm_{\chi^*})/\sqrt 2$ and using the 
result \refb{ejbarcharge}
that $\Vm_{\chi}$ and $\Vm_{\chi^*}$ carries $\bar J$ charges 2 and $-2$ respectively,
we see that $A$ must carry $\bar J$
charge 0 or $\pm 4$. It is easy to see that there are no states
in $\HH_T$ with $L_0^+=0$ and $\bar J=\pm 4$ since the $\bar L_0$ eigenvalue is bounded
from below by $-1+\bar j^2/6$ where $\bar j$ is the $\bar J$ eigenvalue. 
Therefore $A$ must
carry $\bar J$ charge 0. This in turn allows us to express \refb{e2obs} as
\be \label{e3obs}
 \{\VV_{\chi}
\VV_{\chi^*} A \}_0\, .
\ee
We also know from Lorentz invariance that $A$ must be a Lorentz scalar for
\refb{e3obs} to be non-zero. Finally if $A$ is a pure gauge then the result vanishes
trivially to order $g_s^2$ since $\VV_{\chi}$ and
$\VV_{\chi^*}$ are BRST invariant operators.
Therefore we need to focus on $\bar J$ neutral, scalar vertex operators $A$ of ghost number
2 and picture number $-1$,
representing physical or unphysical states.
For this we use the classification of possible choices of $A$ given in 
\cite{1408.0571} (section 5.5)
\begin{enumerate}
\item $A$ can be either a chiral or an anti-chiral operator of the form 
$\VV_\sigma=\bar c c e^{-\phi} \Vm_\sigma$ or $\VV_{\sigma^*}=
\bar c c e^{-\phi} \Vm_{\sigma^*}$ as given in
\refb{echiral}. In either case the $\phi$ charge conservation tells us that in 
the single
PCO that has to be inserted in the definition of $\{\VV_{\chi}
\VV_{\chi^*} A \}$ we must pick the $e^{\phi} T_F$ term. Furthermore Lorentz
invariance guarantees that the possible non-vanishing contributions come from 
the internal part $T^{(int)}_F = T^+_F+T^-_F$ of $T_F$. Now consider the 
insertion involving $e^\phi T^+_F(w)$. It follows from the operator product expansion
of the $e^{q\phi}$'s given in \refb{eghope} and the operator product of
$T_F^\pm$ given in \refb{edefwtvm} that as a function of $w$, there is a zero at
the location of $\VV_\chi$ and no poles at the locations of $\VV_{\chi^*}$ and $A$.
Furthermore since $e^\phi T_F^+(w)$ is a dimension zero
primary, as $w\to \infty$ the function goes to a constant. Such a function,
having no poles and one zero in the complex plane, vanishes identically. Similar
argument can be given for $e^\phi T^-_F$, with the roles of $\VV_\chi$ and
$\VV_{\chi^*}$ interchanged.
\item $A$ can be the zero momentum dilaton vertex operators $\bar c c e^{-\phi}\psi_\mu
\bar\p X^\mu$. In this case the correlation function on the sphere vanishes due to the 
presence of the single $\bar\partial X^\mu$ factor. (The PCO can supply factors of
$\p X^\mu$ but no factor of $\bar\p X^\mu$.)
\item $A$ can be $c\eta$ -- the second operator listed in \refb{ephiform}. In this
case by $\phi$ charge conservation we must pick the 
$c\p\xi$ term from the PCO. However the correlator now contains 
4 factors of c and two factors of $\bar c$, and hence vanishes by separate
ghost charge conservation in the holomorphic and anti-holomorphic
sectors.
\item This essentially exhausts all physical state candidates for $|A\rangle$.
So we now turn to the cases when $|A\rangle$ is
unphysical. 
One class of candidates listed  in \cite{1408.0571} 
(section 5.5) have the form
\be 
A = (\p c +\bar \p \bar c) c e^{-\phi} V\, ,
\ee
for some GSO odd matter sector vertex operator $V$ of dimension 
$(0,1/2)$. In the SO(32) heterotic string on a Calabi-Yau 3-fold there are
no such operators that are Lorentz scalars.
\item Among the other class of unphysical operators listed in section 5.5 of
\cite{1408.0571} 
there is only one candidate that is neutral under $\bar J$:
\be \label{ecand}
A = (\p c +\bar\p \bar c) c\p\xi e^{-2\phi}\bar c \bar J\, .
\ee
Note that this operator 
is not BRST invariant but is still a dimension zero matter primary. Hence
$\{\VV_\chi \VV_{\chi^*}A\}$ can be computed without specifying the choice of
local coordinates at the punctures.
Let $(z_3,\bar z_3)$ denote the argument of $A$, $(z_1,\bar z_1)$ denote the
argument of $\VV_\chi$, $(z_2,\bar z_2)$ denote the argument of 
$\VV_{\chi^*}$ and $w$ be the argument of the PCO. 
Then the amplitude is given by
\be \label{eampli}
\left\langle  \XX(w) A(z_3) \VV_{\chi}(z_1) \VV_{\chi^*}(z_2)\right\rangle\, .
\ee
Since $A$, $\VV_\chi$, $\VV_{\chi^*}$ are proportional to
$e^{-2\phi}$, $e^{-\phi}$ and $e^{-\phi}$ respectively,
$\phi$ charge conservation tells us that we must pick the term proportional to
$e^{2\phi}$ from the PCO and hence there is no matter operator coming from the
PCO. This allows us to evaluate the matter part of the correlator easily.
First using \refb{ejbarcharge} and a standard argument in CFT, we can remove the
$\bar J$ from the correlation function at the cost of multiplying the result by a
factor of
\be\label{einter}
2\left( {1\over \bar z_3 - \bar z_1} -  {1\over \bar z_3 - \bar z_2}\right)\, . 
\ee
The rest of the correlation function in the matter sector can be easily evaluated
as it involves a two point function on a sphere of operators of dimension (1,1/2).
This gives  a factor of $(z_1-z_2)^{-1} (\bar z_1 - \bar z_2)^{-2}$.
The net result, after combining with \refb{einter}, is
\be
2(z_1-z_2)^{-1} 
(\bar z_1-\bar z_2)^{-1} (\bar z_1 -\bar z_3)^{-1} (\bar  z_2-\bar z_3)^{-1}\, .
\ee
The important point to note is that this 
is symmetric under the exchange of $(z_1,\bar z_1)$ 
with $(z_2,\bar z_2)$. On the other
hand the rest of the correlator involving the ghost insertions is of the form
\be
\langle \XX(w)  \, (\p c +\bar\p \bar c) c\p\xi e^{-2\phi}\bar c(z_3)\, 
 \bar c c e^{-\phi}(z_1) \, \bar c c e^{-\phi}(z_2)\rangle\, .
\ee
Note now that this is anti-symmetric under the exchange of $z_1$ and $z_2$.
Combining this with the matter sector result we see that the net result is anti-symmetric
under the exchange of  $(z_1,\bar z_1)$ and $(z_2,\bar z_2)$. 
However the vertex is supposed to be
symmetrized under the exchange of the external states, and hence under the
exchange of $z_1,z_2, z_3$ keeping the external states fixed. This makes the
contribution
vanish. Indeed, if we had started with the amplitude \refb{e2obs} so that 
we have the three point amplitude of $A$, $\VV_{\chi_R}$ and $\VV_{\chi_R}$
placed at $z_3$, $z_1$ and $z_2$, and then expressed $\VV_{\chi_R}$ as
$(\VV_\chi+\VV_{\chi^*})/\sqrt 2$, we would have arrived at the average of two
terms -- the amplitude \refb{eampli} and another one related to it by the exchange
of $z_1$ and $z_2$. This vanishes due to the anti-symmetry of \refb{eampli}
under $z_1\leftrightarrow z_2$ exchange.
\end{enumerate}

Let us now turn to the second term on the right hand side of \refb{epsi4rep}. We need
to analyze,  for an arbitrary $L_0^+=0$ state $|A\rangle$, the quantity
\be \label{etwovan}
\langle A| c_0^- |[~]_1\rangle = \{A\}_1\, .
\ee
This is one point function of $A$ on the torus. For physical states
described by chiral scalar vertex operators of the form 
$\VV_\sigma$ or $\VV_{\sigma^*}$, or the dilaton vertex operators
$\bar c c e^{-\phi}\psi_\mu \bar\p X^\mu$ or $c\eta$, 
this vanishes
due to the result of \S\ref{stadpole} and the existence of global supersymmetry
at tree level string theory.
The result also vanishes trivially for pure gauge states due to the identity
\refb{emain}.
The only remaining candidate for $A$ is the operator described in \refb{ecand}. 
We shall now argue that the one point function of
\refb{ecand} on the torus also vanishes. 
Since $A$ given in \refb{ecand} carries picture number $-1$, we have
one PCO insertion on the torus and the $\phi$ charge conservation tells us
that we need to pick up the term involving $e^{2\phi}$ 
from the PCO. Therefore the PCO
insertion involves purely ghost operators.
Let us consider the contribution from a given
spin structure. Since for a given spin structure the correlation function
in the matter and ghost sectors factorize, the one point function of $A$ given
in \refb{ecand} can be regarded as the product of the one point function of
$\bar J$ in the matter sector and an appropriate correlation function 
in the ghost sector.  Due to translation invariance on the torus the one
point function of $\bar J$ in the matter sector is independent of its location
$u$.
Now under $u\to -u$ transformation the spin structure remains unchanged
but $\bar J$ changes sign. This shows that the one point function of $\bar J$
in the matter sector must vanish separately for each spin structure. 

This establishes the vanishing of each term on the right hand side of 
\refb{epsi4rep}, and 
in turn shows that
$|\tilde \psi_2\rangle$ can be taken to vanish.

\sectiono{Uniqueness of the goldstino candidate} \label{sunique}

For SO(32) heterotic string theory on Calabi-Yau manifolds, we derived in
\refb{eL2} the condition for unbroken supersymmetry to order $g_s^2$:
\ben \label{eL2pre}
&& \{(\XX_0\goldc)\lambda_0\}_1 
- \left\{ (\XX_0 \goldc) \lambda_0 \left(
b_0^+ (L_0^+)^{-1} (1 - {\bf P}) [~]_1\right) \right\}_0  \nonumber \\ &&
-{1\over 2} \beta^2 g_s^2 \left\{ (\XX_0\goldc) \lambda_0 
\left({b_0^+} (L_0^+)^{-1} (1 - {\bf P})  
[\VV_{\chi_R} \VV_{\chi_R}]_0\right) \right\}_0 
+{1\over 2}\beta^2 g_s^2 \{(\XX_0\goldc) \lambda_0 \VV_{\chi_R} \VV_{\chi_R}\}_0 
\nonumber \\ &&
-
\beta^2 g_s^2 \left\{ (\XX_0\goldc) \VV_{\chi_R} \left( b_0^+(L_0^+)^{-1}
(1 - {\bf P}) \XX_0[\VV_{\chi_R}\lambda_0]_0 \right)\right\}_0
\nonumber \\ 
&=& \OO(g_s^3)\, .
\een
The condition must hold for each candidate state $\XX_0\goldc$ listed in 
\refb{ep21}.
In this appendix we shall show that there is only one $\XX_0\goldc$ given in
\refb{eunique} for which the different terms in \refb{eL2pre} 
are not zero. This is important for establishing the existence of global
supersymmetry to this order, since otherwise by
adjusting a single constant $\beta$ we may not be able to make 
the left hand side of \refb{eL2pre} vanish to order $g_s^2$ for different choices of 
$\goldc$. 

For our analysis it will be useful to note that since $\VV_{\chi_R}$ carries
picture number $-1$ and $\lambda_0$ and $\XX_0\goldc$ carry picture
number $-1/2$, each of the terms in the left hand side of \refb{eL2pre} 
requires insertion of one PCO.
First let us consider
the second candidate for $\XX_0\goldc$ listed in \refb{ep21}, given by 
$\bar c c \bar \partial^2 \bar c \, e^{-\phi/2}\, \hat\Sigma$. Now in each of the correlation
functions appearing in \refb{eL2pre}, the only vertex operators carrying four dimensional
Lorentz indices are $\lambda_0=c e^{-\phi/2} \Sigma$ 
and $\XX_0\goldc$. Since the Lorentz invariant tensors
are $\ve_{\alpha\beta}$ and $\ve_{\dotalpha\dotbeta}$, $\Sigma$ appearing
in $\lambda_0$ and $\hat\Sigma$ appearing in $\XX_0\goldc$ must either both carry
dotted indices or both carry undotted indices. 
Since both $\Sigma$ and $\hat\Sigma$ are
to be chosen from the list \refb{edefSigma} we see that they must either both
carry $\Sigma^{(6)}$ or both carry $\Sigma^{(6) c}$. Expressing 
$\VV_{\chi_R}$ as $(\VV_\chi+\VV_{\chi^*})/\sqrt 2$ and using the U(1) charge
assignments under the left and right handed U(1) currents $J$ and $\bar J$ given in
\refb{eJ1}, \refb{ejcharge} and \refb{ejbarcharge} it is easy to check that none of the
terms in \refb{eL2pre} can satisfy both charge conservations even after including
the effect of a possible factor of $T_F$ coming from one PCO that needs to be inserted
into the correlation functions. Hence the left hand side of \refb{eL2pre} 
vanishes identically for this choice of $\XX_0\goldc$.

Next we turn to the first candidate for $\XX_0\goldc$ in \refb{ep21} given by
$\bar c c \eta e^{\phi/2} V^f$ where $V^f$ is a dimension (1,5/8) Ramond
vertex operator
in the matter sector. 
Let us for definiteness assume that $\Sigma$ appearing in $\lambda_0$ carries an
undotted index, i.e.\ has the form $c e^{-\phi/2} \Sigma_\alpha$; 
an identical analysis can be carried out in the other case.
In this case by Lorentz invariance, $V^f$ must also carry an 
undotted index. We shall
now consider several possibilities:
\begin{enumerate}
\item
First let us assume that $V^f$ has the form $\Vm^f_{\sigma^*,\alpha}$
where $\sigma^*$ denotes some anti-chiral superfield and 
$\Vm^f_{\sigma^*,\alpha}$ has been
introduced in \refb{edefcomb}. Note that the `wrong' assignment, 
$\Vm^f_{\sigma^*,\alpha}$ (instead of $\Vm^f_{\sigma^*,\dotalpha}$
or $\Vm^f_{\sigma,\alpha}$
as in \refb{evmf1})
is necessary due to GSO projection
rules since $V^f$ is accompanied by the operator $\eta e^{\phi/2}$, which has opposite
GSO parity compared to $e^{-\phi/2}$.  Since according to \refb{eJ1}, \refb{eJ2}
$\Sigma^{(6)}$ carries $J$-charge 3/2
and $\Vm^f_{\sigma^*}$ carries $J$-charge 1/2, the total $J$-charge carried by 
$\lambda_0$ and $\goldc$ is $3/2+1/2=2$.  Since the $\phi$ charges in
each correlator in \refb{eL2pre} add up to the correct value (0 on the torus and $-2$ on the
sphere), we need to pick the $c\partial \xi$ term from the PCO which has no $J$-charge.
Finally among the two $\VV_{\chi_R}$ insertions in the various terms in \refb{eL2pre},
the $\bar J$ charge conservation tells us that one of them must be $\VV_\chi$ and the
other one $\VV_{\chi^*}$ once we express $\VV_{\chi_R}$ as 
$(\VV_{\chi}+\VV_{\chi^*})/\sqrt 2$.\footnote{Had we picked two
$\VV_\chi$'s or two $\VV_{\chi^*}$'s the total $\bar J$ charge carried by
them would be $\pm 4$. To compensate for this $V^f_{\sigma^*,\alpha}$ 
would have
to carry $\bar J$ charge $\mp 4$ which will give a lower bound of $8/3$ to its
$\bar L_0$ eigenvalue.}
Therefore their $J$-charges also cancel according to
\refb{ejcharge}. This tells us that we have a net $J$-charge 2 carried by all the operators
in the correlation function. Hence such a contribution must vanish.
\item Next we consider the case where $V^f$ has the form 
$\wt\Sigma^c_\alpha
\bar J^a$ where $\bar J^a$ is some anti-holomorphic current in the matter SCFT
associated with the compact directions.
Again although this appears to have the `wrong GSO parity' 
($\wt\Sigma^c_\alpha$
instead of $\Sigma_\alpha$), this is what is needed
to compensate for the fact that $\eta e^{\phi/2}$ has opposite GSO charge compared
to $e^{-\phi/2}$. Now we see that the total $J$ charge of $\lambda_0$ and $\goldc$
add up to $3/2-3/2=0$ and hence this matrix element can be non-zero. Indeed, for the choice
$\bar J^a=\bar J$ this precisely gives the operator \refb{eunique} that gives the
non-vanishing contribution to \refb{eL2pre}. However as long as there are no other U(1)
gauge groups in the theory, all other $\bar J^a$'s must be conserved
currents associated with non-abelian groups. 
The matrix elements of such operators
will vanish by the associated non-abelian global symmetry of the world-sheet
theory.
\item Finally consider the case where $V^f = 
\bar\p X^\mu (\gamma_\mu)_\alpha{}^\dotbeta\wt\Sigma_{\dotbeta}$. 
In this case since $\phi$ charge
conservation requires us to pick the $c\partial \xi$ term from the PCO, there is no
other $\p X^\mu$ or $\bar \p X^\mu$ insertions in the correlation function. As a result
the correlator involving $\bar \p X^\mu$ vanishes.
\end{enumerate}
 This shows that the operator given in \refb{eunique} is the only operator that could
 contribute to any of the terms on the left hand side of \refb{eL2pre}.

\sectiono{Calculation of $\gb^{(2,2)}(0,\chi_R; 0, \chi_R)$} \label{srep}

In order to compute the shift $\beta$ of the scalar field using 
\refb{ebetvalue}, we need two quantities: the tree level four point function
$\gb^{(4,0)}(0,\chi_R;0,\chi_R;0,\chi_R;0,\chi_R)$ and the one loop
two point function $\gb^{(2,2)}(0,\chi_R;0,\chi_R)$. The former was
calculated in \S\ref{s4} (eq.\refb{egbvalue}) while the latter was calculated in 
\cite{ADS,DIS,1209.5461,1304.2832,1408.0571}. In this short appendix we
shall review the result for the latter
in the normalization convention of this paper.

We shall use the result of \cite{1408.0571} whose conventions we are using
here. There are however some additional normalization factors we have to
account for. First, using $\chi_R=(\chi+\chi^*)/\sqrt 2$ we can write
\be
\gb^{(2,2)}(0,\chi_R; 0, \chi_R) = \gb^{(2,2)}(0,\chi; 0, \chi^*)\, .
\ee
This is the amplitude given in eq.(7.13) of \cite{1408.0571} with 
$V_1$ identified as $V_\chi$ and $V_2$ identified as $V_{\chi^*}$.
There are however two additional normalization 
factors to be accounted for:
\begin{enumerate}
\item The normalization factor of $(2\pi i)^{-3g+3-n}$ gives a factor of $-1/4\pi^2$
for $g=1$, $n=2$. This was not explicitly included in (7.13) of \cite{1408.0571}.
\item The $d^2\tau d^2y$ in eq.(7.13) of \cite{1408.0571} should actually be
$d\tau \wedge d\bar\tau\wedge dy\wedge d\bar y$. This translates to 
$-4 d^2 \tau d^2 y$. 
\end{enumerate}
Together these give an additional factor of $1/\pi^2$. With this normalization
factor included the final result (7.34) of \cite{1408.0571} translates to
\be 
{q\over 4\pi} \int_\FF d^2\tau \Big\langle b_\tau \bar b_{\bar\tau} 
\bar c(0) c(0) V_D(0)\Big\rangle
\ee
where $q$ denotes the $\bar J$ charge carried by $\chi$ and $V_D$ is a 
dimension (1,1) operator that appears in the operator product of $V_\chi$
and $V_{\chi^*}$ (eq.(7.30) of \cite{1408.0571}):
\be \label{echchd}
V_{\chi}(0) V_{\chi^*}(y) = {q\over \bar y} V_D(0)\, .
\ee
According to \refb{ejbarcharge} we have $q=2$, and so
\be \label{egexp}
\gb^{(2,2)}(0,\chi_R; 0, \chi_R) ={1\over 2\pi}
\int_\FF d^2\tau \Big\langle b_\tau \bar b_{\bar\tau} 
\bar c(0) c(0) V_D(0)\Big\rangle\, .
\ee

The relevant operator $V_D$ is in fact proportional to $J\bar J$. To find the
constant of proportionality, let $V_D = \lambda \, J \bar J$. Consider now
the three point function on the sphere:
\be  \label{eref}
\langle J\bar J(w) V_{\chi}(z) V_{\chi^*}(y)\rangle_{sphere}\, .
\ee
Using the results \refb{ejcharge}, \refb{ejbarcharge} 
that $V_\chi$ and $V_{\chi^*}$ carry $(J,\bar J)$
chargers $(1, 2)$ and $(-1,-2)$ respectively, we can express this correlator as
\ben \label{eref2}
&& 2 \left( {1\over w-z} - {1\over w-y}\right) \left( {1\over \bar w - \bar z}
- {1\over \bar w - \bar y}\right) 
\langle V_{\chi}(z) V_{\chi^*}(y)\rangle_{sphere} \nonumber \\
&=& 2 |z-y|^2 |w-z|^{-2} |w-y|^{-2} (z-y)^{-1} (\bar z - \bar y)^{-2}
\nonumber \\ &=& 2 |w-z|^{-2} |w-y|^{-2} (\bar z - \bar y)^{-1}\, ,
\een
where we have used the normalization \refb{enorm} of $V_\chi$, $V_{\chi^*}$.
On the other hand using \refb{echchd} we see that in the $y\to z$ limit
\refb{eref} has the form
\be
q (\bar y - \bar z)^{-1} \langle J(w)\bar J(\bar w) V_D(z) \rangle
= q\, \lambda (\bar y - \bar z)^{-1} \langle J(w)
\bar J(\bar w) J(z)\bar J(\bar z)\rangle
= 9 \, q \, \lambda (\bar y - \bar z)^{-1}  |w-z|^{-4}\, .
\ee
Using $q=2$ and comparing this with the $y\to z$ limit of \refb{eref2} we
get
\be 
\lambda = -{1\over 9} \, .
\ee
Therefore we can rewrite \refb{egexp} as
\be \label{eg21fin}
\gb^{(2,2)}(0,\chi_R; 0, \chi_R) = -{1\over 18\pi}
\int_\FF d^2\tau \Big\langle b_\tau \bar b_{\bar\tau} 
\bar c(0) c(0) J(0)\bar J(0)\Big\rangle = -{1\over 2\sqrt 3} \Xi\, ,
\ee
where $\Xi$ has been defined in \refb{edefxi}.

Using \refb{emr2} we now get
\be \label{enewchi}
m_B^2
= 8 g_s^2 \gb^{(2,2)}(0,\chi_R;0,\chi_R) 
= -{4\over \sqrt 3}  \, g_s^2 \, \Xi
\, .
\ee

\section{Consistency with low energy effective field theory} \label{slow}

In this appendix we shall verify that in SO(32) heterotic string theory on
Calabi-Yau manifolds, the results for the tadpoles at the perturbative
vacuum given in 
\refb{efintadpole} and the renormalized scalar mass$^2$ given in \refb{enewchi}
are all consistent with the prediction of low energy supersymmetric field
theory. This will be done in several steps.

Our first step will be to identify the metric field in the string field $|\Psi\rangle$.
For this consider the configuration 
\be \label{eins}
\Psi = \tilde h_{\mu\nu} \bar c c e^{-\phi} \psi^\mu \bar\p X^\nu\, ,
\ee
where $\tilde h_{\mu\nu}$ are constants. Then following the analysis of
\S\ref{sorientation} and \S\ref{sreal} 
one can show that to linear order, shifting the background by 
\refb{eins}  will correspond to the insertion of
\be \label{eins1}
-{1\over 2\pi} \int d^2 z \, \tilde h_{\mu\nu} \, \p X^\mu \bar \p X^\nu
\ee
into the world-sheet correlation function. Here the $-1/2$ factor arises from
the picture changing operation given in \refb{epicchange} and 
the $1/\pi$ is the same
factor that appears by combining \refb{eori2} and \refb{eorifin}.
On the other hand the world-sheet action involving the flat space-time
coordinates in the $\alpha'=1$ unit has the form
\ben\label{eworld}
&& S_w = {1\over 4\pi} \int d^2\, \xi \, \p_\alpha X^\mu \p^\alpha X^\nu \eta_{\mu\nu}
=  {1\over \pi} \int d^2 z  \, \p X^\mu \bar\p X^\nu \eta_{\mu\nu}\, ,
\nonumber \\
&& z \equiv \xi^1+i\xi^2, \quad \bar z \equiv \xi^1-i\xi^2, \quad 
d^2 z \equiv d\xi^1 d\xi^2, \quad \p \equiv {\p\over \p z}, \quad
\bar\p \equiv {\p\over \p \bar z}\, .
\een
We now interpret the insertion of \refb{eins1} as the addition of a
term to $-S_w$ that appears in the exponent in the world-sheet path
integral. This amounts to replacing $\eta_{\mu\nu}$ in \refb{eworld}
by $\eta_{\mu\nu}+ h_{\mu\nu}$ where $h_{\mu\nu} =\tilde h_{\mu\nu}/2$. 
Therefore to linear order in $h_{\mu\nu}$, the space-time
metric $g_{\mu\nu}$
can be identified as $\eta_{\mu\nu}+ h^S_{\mu\nu}$ and the 
background 2-form field can be identified as $h^A_{\mu\nu}$ where the
superscripts $S$ and $A$ stand for symmetric and anti-symmetric parts
respectively. For our analysis we shall be interested in the symmetric 
part of $h$ only.

Our next task will be to identify the dilaton field. For this we consider
a more general string field configuration of the form
\ben \label{egenst}
\Psi &=& \bigg[ 2 h_{\mu\nu} \bar c c e^{-\phi} \psi^\mu \bar\p X^\nu 
+ \phi_1 \bar c \bar\p^2\bar c c\p\xi e^{-2\phi} 
+ \phi_2 c\eta + a_\nu (\p c +\bar\p \bar c) \bar c c \p\xi e^{-2\phi}\bar\p X^\nu
\nonumber \\ && 
\quad + b_\mu  (\p c +\bar\p \bar c) c e^{-\phi} \psi^\mu
\bigg] e^{ik\cdot X}\, ,
\een
where $h_{\mu\nu}$, $\phi_1$, $\phi_2$, $a_\nu$ and $b_\mu$ are 
constants. By the analysis of the previous paragraph 
$\eta_{\mu\nu}+h_{\mu\nu}$ represents the metric $g_{\mu\nu}$
to linear order in 
$h_{\mu\nu}$.
Now the linearized equations of motion $Q_B|\Psi\rangle=0$
give
\ben \label{eqb1}
&& {k^2\over 2} h_{\mu\nu} - i k_\nu b_\mu -{i\over 2} k_\mu a_\nu = 0
\nonumber \\ &&
{k^2\over 4} \phi_2 +{i\over 4} k^\mu b_\mu = 0
\nonumber \\ &&
{k^2\over 4} \phi_1 - {i\over 4} k^\nu a_\nu = 0
\nonumber \\ &&
i k_\nu \phi_2 +{i\over 2} k^\mu h_{\mu\nu} +{1\over 2} a_\nu = 0
\nonumber \\ &&
b_\mu +{i\over 2} k^\rho h_{\mu\rho} - {i\over 2} k_\mu \phi_1=0
\, .
\een
Furthermore linearized gauge transformation $\delta|\Psi\rangle
= Q_B|\Lambda\rangle$ leads to the following gauge symmetries
\ben
\delta h_{\mu\nu} &=& i\eps_\mu k_\nu + i k_\mu \xi_\nu
\nonumber \\
\delta \phi_1 &=& i k\cdot \xi +\zeta
\nonumber \\
\delta \phi_2 &=& -{i\over 2} k\cdot \eps +{1\over 2}\zeta
\nonumber \\
\delta b_\mu &=& {k^2\over 2} \eps_\mu +{i\over 2} k_\mu\zeta
\nonumber \\
\delta a_\nu &=& k^2 \xi_\nu - i k_\nu \zeta \, , 
\een
where $\eps_\mu$, $\xi_\nu$ and $\zeta$ are gauge transformation parameters.
We shall from now on take $h_{\mu\nu}$ to be symmetric and 
set $\phi_1$ to 0 using the $\zeta$ gauge transformations. The gauge
symmetries of the resulting theory are generated by $\xi_\mu$, with the
other parameters fixed by the constraints 
$\eps_\mu=\xi_\mu$, $\zeta = -ik\cdot \xi$.
Eliminating
$a_\nu$ and $b_\mu$ using the last two equations in \refb{eqb1}
and combining other equations we get
\ben \label{eqb2}
&& {k^2\over 2} h_{\mu\nu} -{1\over 2} k_\nu k^\rho h_{\mu\rho}
 -{1\over 2} k_\mu k^\rho h_{\rho\nu} - k_\mu k_\nu \phi_2 = 0\, ,
 \nonumber \\ &&
 k^2 \phi_2+{1\over 2} k^\rho k^\sigma h_{\rho
 \sigma} = 0\, .
 \een
 We now compare this with the linearized equations of motion derived from
 the low energy effective action involving the metric and the dilaton. Up to
 a constant of proportionality this action takes the form
 \be
 \int d^4 x \, \sqrt{-\det g} \, e^{-2\Phi} (R + 4 D_\mu\Phi D^\mu\Phi)
 \ee
 leading to the following linearized equations of motion around 
 $g_{\mu\nu}=\eta_{\mu\nu}$, $e^{2\Phi}=g_s^2$:
 \ben \label{elin2}
&&{k^2\over 2}  h_{\mu\nu} -  {1\over 2} k_\nu k^\rho h_{\mu\rho} 
- {1\over 2} k_\mu k^\rho h_{\rho\nu}  + {1\over 2} k_\mu k_\nu h^\rho{}_\rho
- 2 k_\mu k_\nu \wt\Phi = 0\, ,
\nonumber \\ &&
k^2 h^\rho{}_\rho - k^\rho k^\sigma h_{\rho\sigma} - 4 k^2 \wt\Phi = 0\, .
\een
Here $h_{\mu\nu}=g_{\mu\nu}-\eta_{\mu\nu}$ and $\wt\Phi=\Phi -\ln g_s$.
Comparing 
\refb{eqb2} and \refb{elin2} we arrive at the identification
\be\label{eid}
\phi_2 = 2\wt\Phi -{1\over 2} h^\rho{}_\rho\, .
\ee

We shall now consider a zero momentum string field configuration 
describing change in the background metric and dilaton fields by 
$h_{\mu\nu}$ and 
$\wt\Phi$ respectively. 
Since at zero momentum $a_\mu$ and $b_\mu$ vanish by the
last two
equations of motion in \refb{eqb1} and we have chosen to work in
the $\phi_1=0$ gauge, we see from \refb{egenst} and \refb{eid} that
this corresponds to the state
\be
\Psi_0 =  2 h_{\mu\nu} \bar c c e^{-\phi} \psi^\mu \bar\p X^\nu 
+ \left(2\wt\Phi - {1\over 2} h^\rho{}_\rho\right) c\eta \, .
\ee
Using   \refb{efintadpole} we get
\be \label{ecomp1}
\EE_4(\Psi_0) = 2  {1\over 4} h^\rho{}_\rho \left( -{1\over 2} g_s^4 \, \Xi^2\right)
+  \left(2\wt\Phi - {1\over 2} h^\rho{}_\rho\right)  \left( -{1\over 4} g_s^4 \, \Xi^2
\right) = -{1\over 4} g_s^4 \, \Xi^2 \left(2\wt\Phi + {1\over 2} h^\rho{}_\rho\right)\, .
\ee

We shall now compare this with the term in the action linear in fields 
at the perturbative vacuum
of the low energy effective supersymmetric field theory. The relevant fields
for our analysis will be a U(1) gauge field $A_\mu$, its associated field
strength $F_{\mu\nu}$ and a scalar field $\chi$ carrying charge $\q$ under this
gauge field.
We shall also keep the metric $g_{\mu\nu}$ and $\Phi$ as background
fields.
The action takes the form
\be \label{elowact}
-{1\over 4} \int d^4 x \, \sqrt{-\det g} \,  e^{-2\Phi}\left[ (\p_\mu\chi^*  - i \q A_\mu\chi^*)
(\p^\mu\chi+i\q A^\mu\chi) + {1\over 4} F_{\mu\nu} F^{\mu\nu} 
+{1\over 2} (\q \chi^* \chi - c \, e^{2\Phi})^2 \right]\, .
\ee
The overall factor of $-1/4$ ensures that the scalar kinetic terms in the action
are given by $-k^2 /4$ as in the string field theory action.  $c$ is a constant
giving the Fayet-Iliopoulos term. 
Since it is generated at one loop, we have a factor of $e^{2\Phi}$ multiplying
it.
Now if we expand the action around the 
perturbative vacuum $\chi=0$, $A_\mu=0$,
$g_{\mu\nu}=\eta_{\mu\nu}$, 
$e^{2\Phi}=g_s^2$ 
in powers of
$h_{\mu\nu}=g_{\mu\nu}-\eta_{\mu\nu}$ and $\wt\Phi=\Phi 
- \ln g_s$, we get, to linear order
\be \label{ecomp2}
-{1\over 8} \, g_s^2 \, \left(2\wt\Phi + {1\over 2} h^\rho{}_\rho\right) c^2 \, .
\ee
We have used the convention stated below \refb{evacnormprime} 
that the volume
of space-time has been normalized to 1.
This has to be compared with $g_s^{-2}\EE_4(\Psi_0)$ given in \refb{ecomp1},
with the $g_s^{-2}$ factor representing the overall normalization factor of 
$g_s^{-2}$ multiplying the action \refb{eactpsi} which was not reflected in the 
definition \refb{eeom} of the equation of motion $\EE$. This gives
\be \label{ec2c2}
c^2 = 2 \, \Xi^2\, .
\ee
Note that at this stage we already have a non-trivial consistency check between
the 1PI effective string field theory result and low energy effective field
theory result since both \refb{ecomp1} and \refb{ecomp2} have 
the combination
$\left(2\wt\Phi+ {1\over 2} h^\rho{}_\rho\right)$.

By expanding the action \refb{elowact}
around the shifted vacuum $\q\chi^*\chi=ce^{2\Phi}$
we see that the tadpoles of $h^\rho{}_\rho$ and $\Phi$ vanish there,
in agreement
with the result of \S\ref{sshifted}. Furthermore by expanding the action around
this vacuum to quadratic order in $\chi$ we learn that the mass of $\chi$ at
the  shifted vacuum is given by
\be \label{ecomp3}
m_B{}^2 = 2 c\, g_s^2\, \q \, .
\ee
This has to be compared with \refb{enewchi}, but for this we need to determine
the value of $\q$. To do this, we note from \refb{elowact} that  
in momentum space, the three point
coupling of an external $\chi$ and a $\chi^*$ carrying momenta $\pm k$ and
a zero momentum gauge field of polarization $\eps^\mu$
is given by
\be \label{eqmukmu}
-{1\over 2} \q\, k^\mu \eps_\mu\, .
\ee
On the other hand, this can be computed in string theory from the
3-point function of a pair of $\chi$, $\chi^*$ vertex operators and the
zero momentum gauge field vertex operator
\be 
i\sqrt{2\over 3}\, \bar c c  e^{-\phi} \psi^\mu \bar J\, ,
\ee
where the $i\sqrt{2\over 3}$ factor has been included to ensure that the vertex
operator satisfies the normalization condition \refb{eno2}. Therefore the
required 3-point function is given by
\be
A_3= i\sqrt{2\over 3} \, \eps_\mu \, 
\left\langle  \bar c c e^{-\phi} V_\chi e^{ik\cdot X} (z_1) 
\, \bar c c e^{-\phi} V_\chi^* e^{-ik\cdot X} (z_2) \, 
\bar c c  e^{-\phi} \psi^\mu \bar J(z_3)  \, \XX(w)\right\rangle\, .
\ee
The PCO location $w$ can be chosen arbitrarily. We shall take the limit
$w\to z_3$. This gives
\be
A_3 =
 -i \sqrt{1\over 6} \, \eps_\mu \, \left\langle  \bar c c e^{-\phi} 
V_\chi e^{ik\cdot X} (z_1) 
\, \bar c c e^{-\phi} V_\chi^* e^{-ik\cdot X} (z_2) \, 
\bar c c  \p X^\mu \bar J(z_3)  \right\rangle\, .
\ee
We can now use \refb{ematterope}, \refb{ejbarcharge} to
remove the $\p X^\mu(z_3)$ and $\bar J(\bar z_3)$
factors at the cost of picking up the factors 
\be
-{i k^\mu\over 2} \left({1\over z_3-z_1} - {1\over z_3-z_2}
\right) \times 2 \left({1\over \bar z_3-\bar z_1} - {1\over \bar z_3-\bar z_2}
\right) 
= -i k^\mu |z_3-z_1|^{-2} |z_3-z_2|^{-2} |z_1-z_2|^2\, .
\ee
This gives
\ben
A_3 &=& -\sqrt{1\over 6} \, \eps_\mu k^\mu \, \left\langle  \bar c c e^{-\phi} 
V_\chi e^{ik\cdot X} (z_1) 
\, \bar c c e^{-\phi} V_\chi^* e^{-ik\cdot X} (z_2) \, 
\bar c c  (z_3)  \right\rangle 
\, |z_3-z_1|^{-2} |z_3-z_2|^{-2} |z_1-z_2|^2
\nonumber \\
&=& -\sqrt{1\over 6} \, \eps_\mu k^\mu \, ,
\een
where in the last step we have used \refb{enorm} and the normalization
condition \refb{evacnorm} to evaluate the correlation function. 
Comparing this with \refb{eqmukmu} we get
\be 
\q = \sqrt{2\over 3} \, .
\ee

\refb{ecomp3} and \refb{ec2c2} now gives
\be 
(m_B^2)^2  = 4 c^2 \, g_s^4\, \q^2 =  4 \times 2 \, g_s^4\, \Xi^2 \times {2\over 3} 
= {16\over 3} g_s^4\, \Xi^2 \, .
\ee
This agrees with the result given in \refb{enewchi}.

\end{document}